\documentclass[12pt]{article}

\usepackage{scicite}
\usepackage{amsmath}
\usepackage{graphicx}
\usepackage{times}
\usepackage{authblk}
\usepackage{float}
\usepackage{amssymb}
\usepackage{url}
\usepackage[usenames,dvipsnames]{color}

\usepackage{xcolor} % to access the named colour LightGray
\definecolor{LightGray}{gray}{0.9}

\newenvironment{sciabstract}{%
\begin{quote} \bf}
{\end{quote}}

\title{A Retrofit Sensing Strategy for Soft Fluidic Robots}

\author[1]{Shibo Zou}
\author[1,3]{Sergio Picella}
\author[1]{Jelle de Vries}
\author[2]{Vera Kortman}
\author[2]{Aimée Sakes}
\author[1,3]{Johannes T. B. Overvelde$^\ast$}

\affil[1]{\normalsize{Autonomous Matter Department, AMOLF,}\protect\\
\normalsize{Amsterdam, 1098 XG, The Netherlands}}

\affil[2]{\normalsize{Bio-Inspired Technology Group, Department of BioMechanical Engineering,}\protect\\
\normalsize{Delft University of Technology, Delft, 2628 CD, The Netherlands}}

\affil[3]{\normalsize{Institute for Complex Molecular Systems and Department of Mechanical Engineering,}\protect\\
\normalsize{Eindhoven University of Technology, Eindhoven, 5600 MB, The Netherlands}\newline
\newline
\normalsize{$^\ast$To whom correspondence should be addressed; E-mail:  overvelde@amolf.nl.}}

\date{}

\begin{document} 


\baselineskip24pt


\maketitle 

\graphicspath{{Figures/}}


\begin{sciabstract}
%Perception is indispensable for robots to achieve autonomy. 
Soft robots are intrinsically capable of adapting to different environments by changing their shape in response to interaction forces with the environment. However, sensing and feedback are still required for higher level decisions and autonomy.  
Most sensing technologies developed for soft robots involve the integration of separate sensing elements in soft actuators, which presents a considerable challenge for both the fabrication and robustness of soft robots due to the interface between hard and soft components and the complexity of the assembly. To circumvent this, here we present a versatile sensing strategy that can be retrofitted to existing soft fluidic devices without the need for design changes. We achieve this by measuring the fluidic input that is required to activate a soft actuator and relating this input to its deformed state during interaction with the environment. We demonstrate the versatility of our sensing strategy by tactile sensing of the size, shape, surface roughness and stiffness of objects. Moreover, we demonstrate our approach by retrofitting it to a range of existing pneumatic soft actuators and grippers powered by positive and negative pressure. Finally, we show the robustness of our fluidic sensing strategy in closed-loop control of a soft gripper for practical applications such as sorting, fruit picking and ripeness detection. Based on these results, we conclude that as long as the interaction of the actuator with the environment results in a shape change of the interval volume, soft fluidic actuators require no embedded sensors and design modifications to implement useful sensing. We believe that the relative simplicity, versatility, broad applicability and robustness of our sensing strategy will catalyze new functionalities in soft interactive devices and systems, thereby accelerating the use of soft robotics in real world applications.

%such that no further design modifications are needed to turn any soft fluidic actuator into a sensor.

%such that no additional components are necessary to increase the autonomy of soft robots.  

%Adapting the fluidic sensing strategy to soft actuators with different shapes, pressure- or vacuum- powered demonstrates the universality of our approach.
\end{sciabstract}

% In setting up this template for *Science* papers, we've used both
% the \section* command and the \paragraph* command for topical
% divisions.  Which you use will of course depend on the type of paper
% you're writing.  Review Articles tend to have displayed headings, for
% which \section* is more appropriate; Research Articles, when they have
% formal topical divisions at all, tend to signal them with bold text
% that runs into the paragraph, for which \paragraph* is the right
% choice.  Either way, use the asterisk (*) modifier, as shown, to

%INTRODUCTION

The intrinsic compliance of soft robots provides adaptability to unknown environments\cite{pfeifer2012challenges, rus2015design,laschi2016soft}. For example, a soft robotic gripper passively adapts its body shape, making it possible to grasp various objects without the need for active sensing\cite{brown2010universal,shintake2018soft}. However, when it comes to more advanced tasks such as identifying and sorting objects, sensory feedback from the gripper becomes essential to achieve closed-loop control in gripping and manipulation\cite{hughes2016soft}. %as the large deformation and infinite degrees of freedom of soft robots during interactions with unknown environments make them difficult to comply with these more traditional sensing technologies for rigid robots. 
Benefiting from advances in soft materials, soft robotic sensing has been enabled by embedding flexible or stretchable sensors made from piezoresistive and piezocapacitive polymer composites\cite{amjadi2016stretchable}, liquid metals\cite{ren2020advances}, electrically and ionically conductive hydrogels\cite{wang2020stretchable}, and polymeric optical waveguides\cite{guo2019soft}. Both proprioception (sensing of self-deformation) and exteroception (sensing of external stimuli) of soft robots have been successfully demonstrated with embedded sensors. Moreover, multimodal sensing, i.e., the simultaneous perception of multiple physical parameters, has been achieved by machine learning\cite{thuruthel2019soft,shih2020electronic} and embedding various sensors into the soft actuator\cite{zhao2016optoelectronically,truby2018soft,shih2019design}. A common feature in all these sensing strategies for soft robotic applications is the separation of actuation and sensing elements \cite{polygerinos2017soft,wang2018toward}. This is a result of the compliance of the soft systems, which complicates integration and reduces reliability of the sensors that need to be embedded in the soft actuator, therefore placing considerable constraints on the design of both the sensors and actuators.

%Importantly, a common denominator in all these previously developed sensing strategies is the separation of sensing and actuation tasks into different components.   

%To our best knowledge, inherently integrated actuation-sensing systems which use the actuation media directly for sensing remain vacant in soft robotics.

As fluidic actuation represents a plurality in soft robotics\cite{jumet2021data}, sensing strategies based on fluidic media, either gas or liquid, have been investigated to reduce the integration difficulties of actuation and sensing elements, such as fluidic resistance sensing\cite{kusuda2007fluid,koivikko2022integrated}, fluidic pressure sensing\cite{fishel2008robust, slyper2012prototyping,farrow2015soft,scharff2017towards,tawk2019soft,wang2020mechanoreception,hughes2020simple,hughes2020sensorization,tawk2021force,block2022arms,truby2022fluidic,he20223d} and electrical resistance sensing\cite{helps2018proprioceptive}. 
Most fluidic sensing strategies incorporate an additional cavity to the soft actuator\cite{fishel2008robust,scharff2017towards,tawk2019soft,wang2020mechanoreception,hughes2020sensorization,tawk2021force}. Since the enclosed cavity contains a fixed amount of fluid, deformation of the actuator or contact with the environment changes the volume of the cavity and thus increases or decreases the internal pressure. Interestingly, this pressure response can be measured remotely by connecting the cavity and electronic pressure sensor via a tube, such that no electronic components need to be embedded in the soft actuator. 

A particularly interesting yet simple method uses the cavity of the soft actuator itself to sense external force by measuring and analyzing the fluidic pressure of the soft actuator\cite{lazeroms1996hydraulic,joshi2023sensorless}. The benefit of such a self-sensing approach has also been demonstrated in dielectric elastomer actuators\cite{keplinger2008capacitive, gisby2013self, rosset2013self} and electrohydraulic actuators\cite{acome2018hydraulically,ly2021miniaturized, yoder2022soft}. In these systems the electrical characteristics of the actuator can be measured to infer the mechanical deformation while it is being actuated, hence no additional sensors and associated electronics are needed\cite{gisby2013self}. While fluidic self-sensing has originally been demonstrated in a potential medical application \cite{lazeroms1996hydraulic}, a natural question to ask is how widely applicable, versatile and robust such an approach is. To answer this question, we need to gain a better understanding of the underlying principles that allow for fluidic self-sensing, and determine if we can infer the interaction of a wide variety of soft actuator with their environment by measuring and analyzing the fluidic response of the enclosed cavity. And if so, we want to determine how easy it is to integrate and retrofit such a sensing approach, and if interactions with the environment can be robustly measured.

To achieve this, in this work we will first experimentally show how the response of a typical soft fluidic bending actuator changes when interacting with the environment. We next introduce several strategies to sense these interactions without the need to embed additional sensing elements in the soft actuator. We demonstrate how to apply our fluidic sensing strategy to a soft gripper, and to enable a versatile range of sensing applications such as size, shape, surface roughness and stiffness sensing of objects. To demonstrate that the sensing approach can be retrofitted, we apply the sensing strategy to a filament actuator, a McKibben actuator, a thermoplastic polyurethane (TPU) actuator, a soft suction gripper specifically designed for medical applications and two commercially available soft grippers. We furthermore developed a basic model based on a linear extension actuator to study the underlying factors that determine the sensing resolution. Finally, we show that our fluidic sensing strategy is robust enough to implement closed-loop control in gripping and sorting applications.

%A particularly interesting, yet simple strategy, for fluidic sensing incorporates an additional cavity to the soft actuator\cite{fishel2008robust,tawk2019soft,wang2020mechanoreception,tawk2021force}. Here, since the enclosed cavity contains a fixed amount of fluid, deformation of the actuator or contact with the environment changes the volume of the cavity and in return increases or decreases the pressure inside the cavity. Interestingly, this pressure can be measured at a distance by connecting the cavity and electronic pressure sensor via a tube, such that no electronic components need to be embedded in the soft actuator. 
%history%However, the addition of sensing cavity affects the response of the actuator (e.g., becomes stiffer or more difficult to bend), and requires custom design of the sensor arrangements based on the shape of the actuator and sensing applications, making it difficult to transfer the sensing strategy from one actuator to another.
%history%In this work, our aim is to develop a fluidic sensing strategy that is universal for any type of soft fluidic device without the need for embedded sensors. 
%%With the realization that any soft fluidic actuator is also an enclosed cavity, a natural question to ask is if we can sense the state of the actuator during the environment interaction by measuring and analyzing the fluidic response of the soft actuator. 

%Figure 1 - pv responses

\begin{figure}
    \centering
    \resizebox{160mm}{!}{\includegraphics{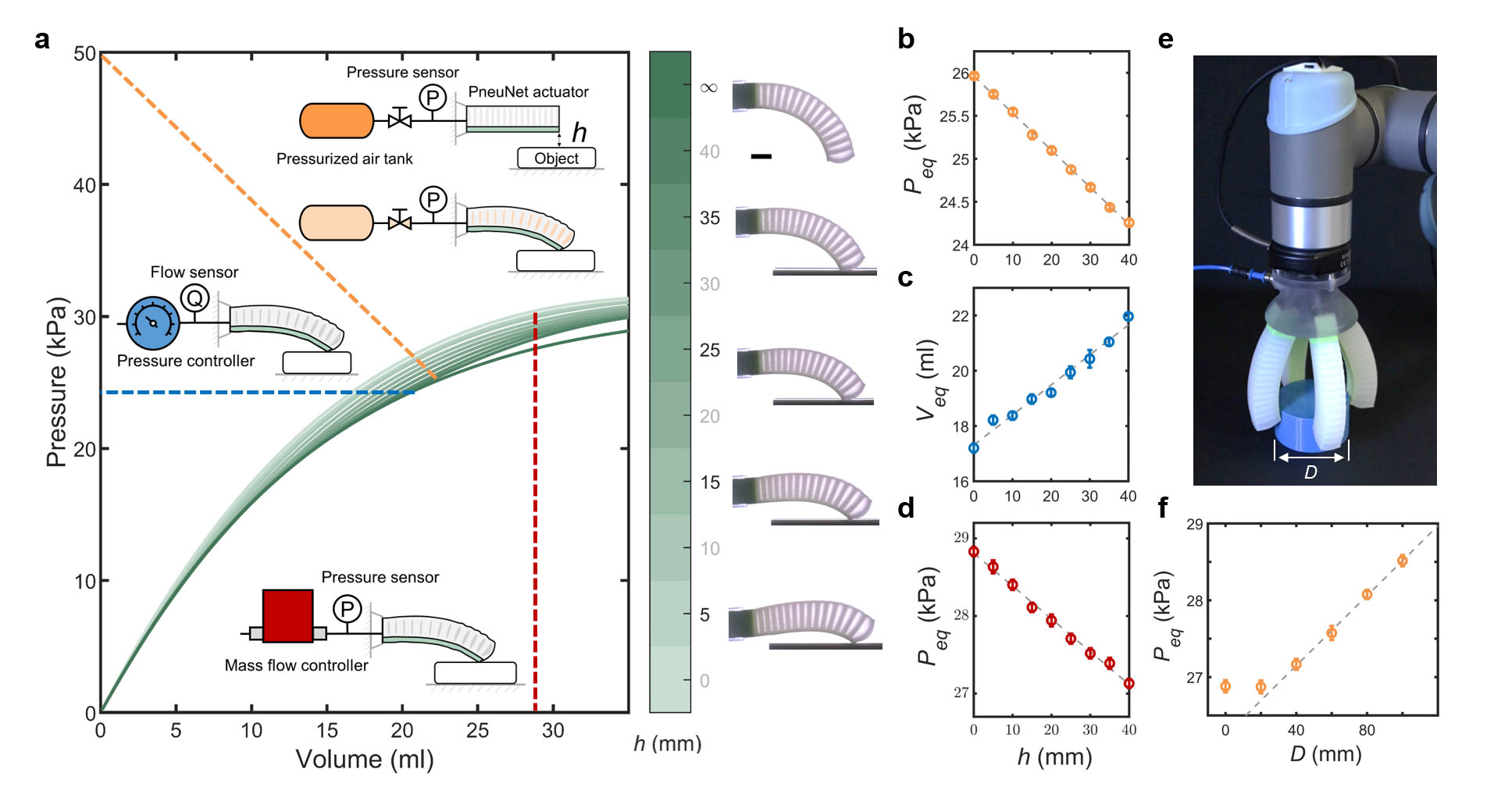}}
    \caption{\textbf{Fluidic sensing of the soft robot-environment interaction. a,} Overview of the fluidic sensing methodology. Inflating a PneuNet actuator onto a rigid plate from a height $h$ changes the pressure-volume response of the soft actuator, which can be characterized by three different fluidic sensing methods: pressurized air tank and solenoid valve (orange dashed line), pressure control (blue dashed line), and flow control (red dashed line). For clarity, the presented date is fitted based on measured results shown in Fig.~\ref{FigS_pv_raw}. The snapshots represent the equilibrium state of the soft bending actuator with an input volume of 28.7 ml at $h=5, 15, 25, 35$ and $\infty$ mm. Scale bar: 20 mm. \textbf{b-d,} Calibrations of the height $h$ based on sensing with the three fluidic sensing methods (colors match sensing method). The dashed lines represent the linear fits of all the test data. \textbf{e,} A soft gripper with four PneuNet actuators gripping a cylindrical object with a diameter $D$. \textbf{f,} Calibration of cylinder diameter $D$ sensing with the soft gripper. The dashed line represents a linear fit of the test data with $D=40, 60, 80, 100$ mm.}
    \label{Fig1}
\end{figure}

%TEXT ABOUT FIGURE 1
\subsubsection*{Fluidic sensing of the soft robot-environment interaction}

We start by looking into the characteristic behavior of a typical soft PneuNet bending actuator\cite{mosadegh2014pneumatic} when interacting with the environment.
We inflate a soft actuator onto a rigid plate from different heights $h$, and characterize the pressure-volume response for each height, i.e., the pressure $P$ as a function of supplied air volume $V$, at different heights (fitting curves in Fig.~\ref{Fig1}a, test results in Fig.~\ref{FigS_pv_raw}, test procedure in Methods section \textbf{Distance Sensing with PneuNet Bending Actuator} and Supplementary Video 1). Interestingly, the interaction with the plate influences the pressure-volume response of the soft actuator. This influence originates from the compliance of the soft actuator and the effect that external forces have on the internal geometric volume of the inflated actuator. According to the ideal gas law, the difference in internal geometric volume gives a direct fluidic response in the form of a variation in pressure, if the temperature is constant and the amounts of fluid substances at different heights are equal. %Hence, when the rigid plate is placed closer to the soft actuator, the pressure-volume curve shifts up.
Importantly, by definition any physical interaction with the environment leads to a change in the internal geometric volume of a soft actuator because of the compliance of the soft body. 

In order to effectively sense these differences in the pressure-volume response, and with that the interaction of the soft actuator with the environment, we connect the soft actuator to a pressurized air tank via a solenoid valve and measure the equilibrium pressure using an external pressure sensor after opening of the valve (orange line in Fig.~\ref{Fig1}a). Depending on the initial distance $h$ of the actuator to the surface, the equilibrium pressure $P_{\mathrm{eq}}$ will be slightly different. Interestingly, for this specific actuator design and interaction, the relationship between the equilibrium pressure $P_{\mathrm{eq}}$ and height $h$ can be approximated by a linear relationship. Therefore, the initial distance between the actuator and plate can be inferred from the fluidic signal based on the calibration curve (Fig.~\ref{Fig1}b and Fig.~\ref{FigS_pt_raw}a) with an accuracy of $\pm1.7$ mm (Fig.~\ref{FigS_sensing_accuarcy}a). The force applied by the actuator on the plate can also be inferred from the equilibrium pressure $P_{\mathrm{eq}}$ (Fig.~\ref{FigS_force_calibration}). Note that since the steel air tank has a linear pressure-volume relationship, varying the tank size effectively changes the slope of the tank's pressure-volume curve. This changes the intersection points with the actuator's pressure-volume curves in Fig.~\ref{Fig1}a, making it possible to tune the sensing resolution (Fig.~\ref{FigS_resolution_2caps}). 

While here we connect the actuator to a steel tank and solenoid valve to implement sensing, depending on the available equipment and precision requirement, the interaction of a soft actuator with the environment can also be characterized by controlling the pressure and measuring the volume flow input (blue line in Fig.~\ref{Fig1}a, Fig.~\ref{Fig1}c and Fig.~\ref{FigS_pt_raw}b) with a sensing accuracy of $\pm4.7$ mm (Fig.~\ref{FigS_sensing_accuarcy}b), or controlling the volume flow input and measuring pressure (red line in Fig.~\ref{Fig1}a, Fig.~\ref{Fig1}d and Fig.~\ref{FigS_pt_raw}c) with a sensing accuracy of $\pm3.4$ mm (Fig.~\ref{FigS_sensing_accuarcy}c).

Our sensing strategy can also be directly applied to a soft gripper to sense the size of objects. We demonstrate size sensing of cylindrical objects using a soft gripper consisting of four PneuNet bending actuators (Fig.~\ref{Fig1}e, Fig.~\ref{FigS_gripper_pt_raw} and Supplementary Video 1). To enable sensing, the actuators are jointly connected to the external system that contains an air tank, a solenoid valve and a pressure sensor. Fig.~\ref{Fig1}f shows that when gripping larger objects, a higher equilibrium pressure $P_{\mathrm{eq}}$ is reached. %  objects have  the inflation of the actuators, shifting up the pressure-volume curves of the soft gripper (Fig.S2). 
%Thereby after the valve is opened, the system equilibrates at a higher pressure when grasping larger objects (Fig.1f). 
Interestingly, the relationship between the equilibrium pressure $P_{\mathrm{eq}}$ and cylinder diameter $D$ can also be fitted with a linear function within the grasping range of the gripper ($d\gtrapprox20$ mm). Repeated tests on a similar gripper showed the same results, where we found that the pressure measurements vary within $\pm$ 0.08 kPa over 100 cycles (Fig.~\ref{FigS_cycling}).
%Since the width of the PneuNet actuator is around 20 mm, the gripper has a minimum opening around 20 mm even when fully inflated, making it incapable to differentiate objects smaller than 20 mm (Fig.1g).

\begin{figure}
    \centering
    \resizebox{110mm}{!}{\includegraphics{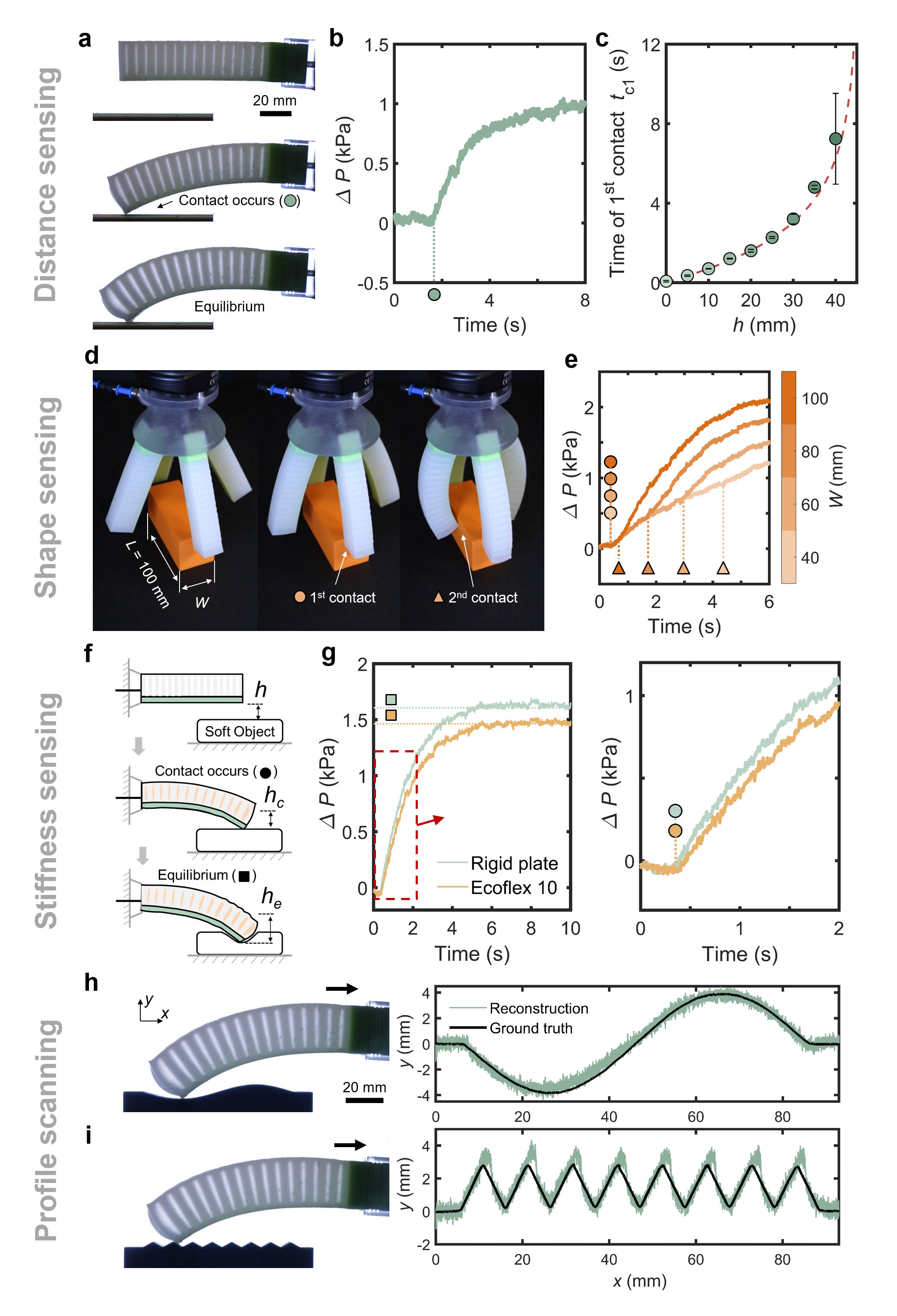}}
    \caption{\textbf{Time-enabled sensing versatility. a-c,} Inferring the initial distance $h$ between the actuator and plate from the time of first contact $t_{\mathrm{c1}}$. $h = 20$ mm in \textbf{a} and \textbf{b}. The dashed line in \textbf{c} represents the vertical displacement-time curve of the tip of actuator in the case of free actuation. \textbf{d, e,} Shape sensing of rectangular objects by measuring both the time of first and second contact of the soft gripper. \textbf{f, g,} Stiffness sensing by measuring both the time of first contact and the equilibrium pressure, which indicate the vertical displacement of the actuator at the first contact ($h_c = h$) and equilibrium ($h_e \geq h$ depending on the object stiffness), respectively. $h = 5$ mm in \textbf{g}. \textbf{h, i,} Surface scanning and profile reconstruction of two different surface profiles.}
    \label{Fig2}
\end{figure}

%TEXT ABOUT FIGURE 2
\subsubsection*{Time-enabled sensing versatility}

Having demonstrated the basic principles of our sensing approach, we next show that versatile sensing applications can be achieved by measuring the pressure response of the soft actuator over time. To show how we can extract more information from the pressure-time response, we first revisit height sensing where so far we only considered the equilibrium pressure at a specific moment in time (Fig.~\ref{Fig1}). Instead, if we correlate the pressure-time response of the actuator to a reference response, i.e., free actuation without interacting with the environment, we can determine the moment of contact (Fig.~\ref{Fig2}a-b). In Fig.~\ref{Fig2}b
%Assuming all components from the fixed mass method in Fig.1a remain constant, 
%Before the actuator touches the plate, the pressure-time response of the actuator is the same as the reference. As soon as the actuator touches the plate, the pressure-time response deviates from the reference. Therefore, 
we evaluate $\Delta P = P - P_{\mathrm{ref}}$ over time from the onset of actuation and determine the time of first contact $t_{\mathrm{c1}}$ when $\Delta P > 0$. We can then use $t_{\mathrm{c1}}$ to infer the initial distance $h$ between the actuator and plate (Fig.~\ref{Fig2}c) based on the tip displacement-time curve of the actuator in the  reference response, with an accuracy of -2.9 to 3.8 mm (Fig.~\ref{FigS_sensing_accuarcy}d). Note that the sensing speed of this strategy is dominated by the actuation speed, which is determined by the flow resistance between the air tank and actuator (Fig.~\ref{FigS_responsetime}). Interestingly, a higher sensing response speed can be achieved by measuring the time of contact $t_{\mathrm{c1}}$ (Fig.~\ref{Fig2}c) compared to the equilibrium $P_{\mathrm{eq}}$ (Fig.~\ref{Fig1}b), because the measurement of $t_{\mathrm{c1}}$ does not require the system to reach equilibrium. 

Based on this approach, we show how we can sense \emph{(i)} the shape of objects, \emph{(ii)} the stiffness of a soft substrate and \emph{(iii)} the profile of a surface. As a first demonstration of the sensing versatility, we use our previously introduced gripper to sense the aspect ratio of rectangular objects. Fig.~\ref{Fig2}d-e and Supplementary Video 2 show that two contact events occur in the pressure-time response when the soft gripper grips a rectangular object, indicating the length and width of the object, respectively. While this could also be achieved by individually addressing each actuator, which would likely make it easier to extract shape information from the soft gripper, it would also require additional hardware that might not be needed or available in specific applications. 

As a second demonstration of the sensing versatility, we achieve multimodal sensing of distance and stiffness when the actuator interacts with a soft plate (Fig.~\ref{Fig2}f). Here, the time of first contact $t_{\mathrm{c1}}$ extracted from the pressure-time response (Fig.~\ref{Fig2}g, Supplementary Video 2) indicates the initial distance between the actuator and plate, while the equilibrium pressure $P_{\mathrm{eq}}$ indicates the final vertical displacement of the actuator. By comparing these two displacements, we can extract the indentation depth of the soft actuator, which can be correlated to the stiffness of the plate when considering the stiffness of the soft actuator. Therefore, measuring $t_{\mathrm{c1}}$ and $P_{\mathrm{eq}}$ together makes it possible to compare the stiffness values of objects (Fig.~\ref{Fig2}g). 

As a final demonstration of the sensing versatility, we use the actuator as a profilometer by considering the variations in the equilibrium pressure when moving the actuator along a surface (Fig.~\ref{Fig2}h-i and Supplementary Video 2). To show this, we move the actuator horizontally along a surface with a robotic arm and measure the pressure response continuously. The profile of the object can be reconstructed using a calibration curve (Fig.~\ref{Fig1}b) and a reference pressure response, which rules out the influence of system leakage or other variations over time. Note that in this sensing application the sharpness of the tip of the soft actuator will determine the resolution of the sensing signal.

\begin{figure}
    \centering
    \resizebox{160mm}{!}{\includegraphics{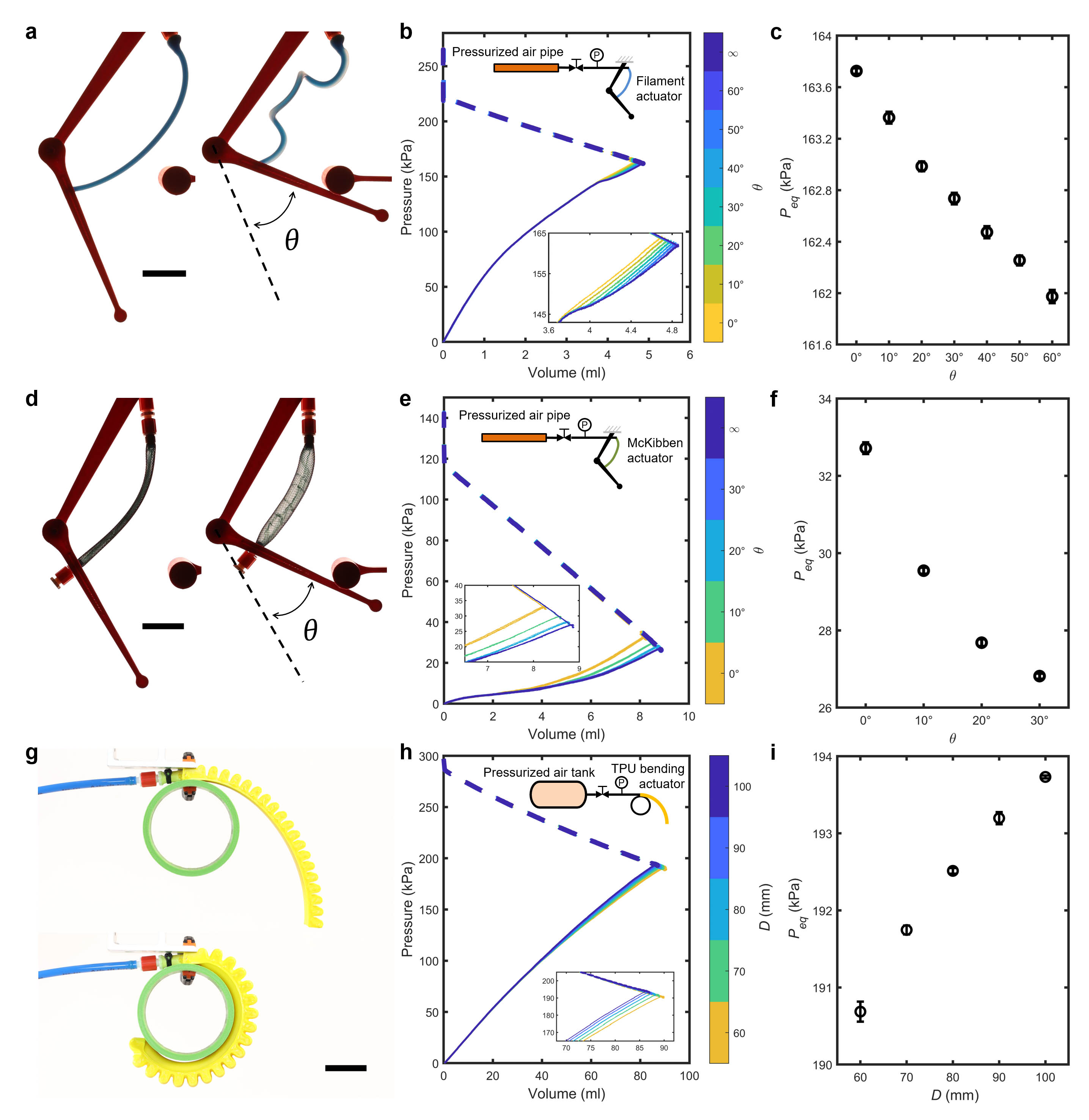}}
    \caption{\textbf{Retrofitting the fluidic sensing approach to a filament actuator (a-c), a McKibben actuator (d-f) and a 3D-printed bending actuator (g-i).} The filament (\textbf{a}) and McKibben (\textbf{d}) actuator are used  as a muscle to rotate an arm towards a stopper. TPU bending actuator (\textbf{g}) wrapping around a cylinder with a diameter $D$. Corresponding pressure-volume relation for the soft actuator (solid) and tank (dashed) (\textbf{b, e, h}) and equilibrium pressure in the system (\textbf{c, f, i}) for different positions of the stopper or cylinder diameter $D$. Experimental results from five tests are shown for each $\theta$ (\textbf{b, e}) and each $D$ (\textbf{h}). Scale bars, 30 mm.}
    \label{Fig_retrofit_1}
\end{figure}

\begin{figure}
    \centering
    \resizebox{155mm}{!}{\includegraphics{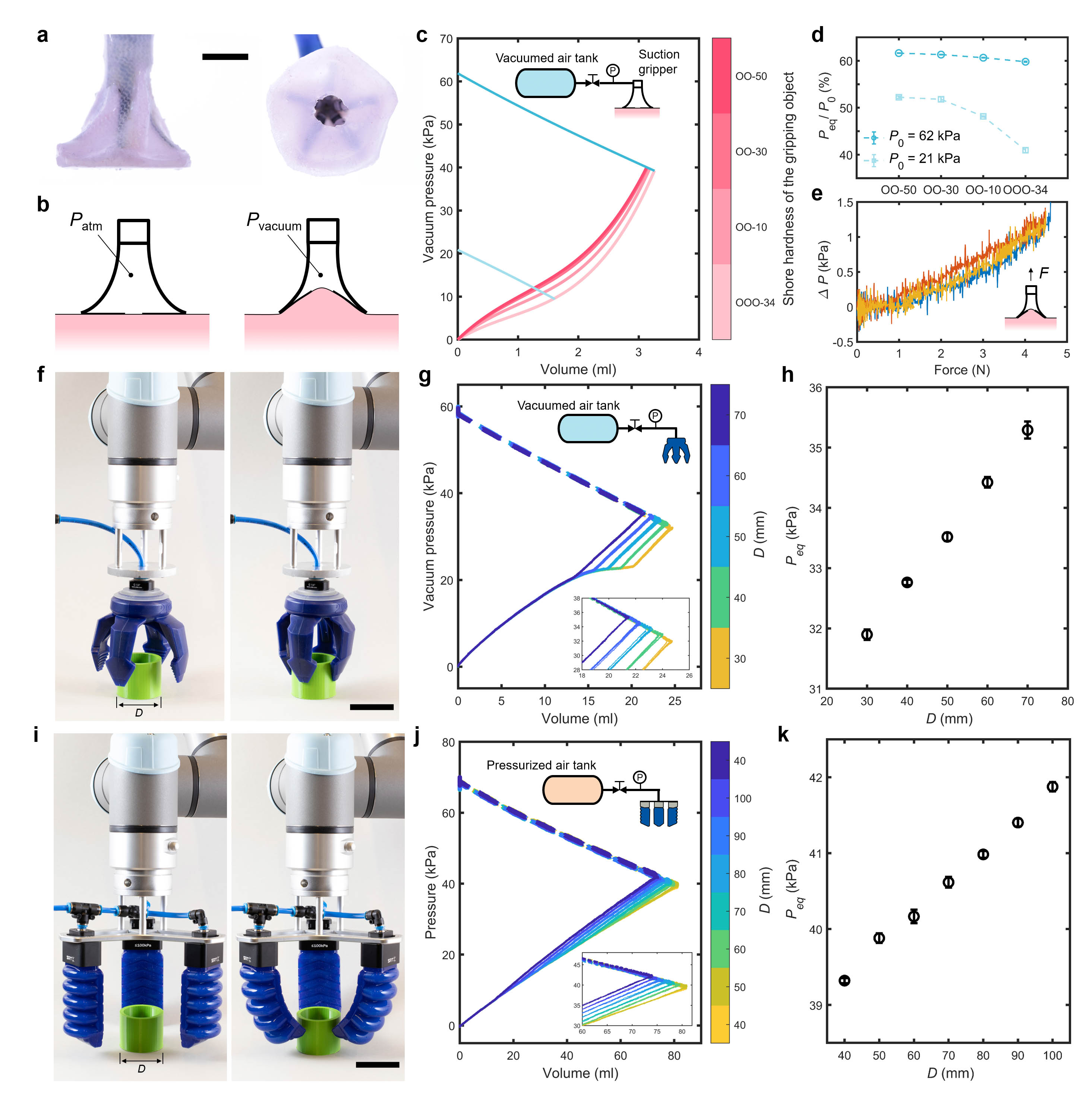}}
    \caption{\textbf{Retrofitting the fluidic sensing approach to a suction cup (a-e) and two commercial soft grippers (f-k).} \textbf{a,} Front and bottom views of the suction gripper. Scale bar, 10 mm. \textbf{b,} Schematic of the suction gripper attaching to a soft object. \textbf{c,} Smoothened pressure-volume responses of the suction gripper (pink) and air tank (blue) when attaching to silicone samples with different shore moduli. \textbf{d,} Experimental sensing results obtained using two different initial pressures $P_{\mathrm{0}}$ in the air tank. \textbf{e,} Force-pressure responses from three pulling tests on the suction gripper when attached to a silicone sample with a shore hardness of OO-30. Commercial vacuum (\textbf{f}) and pressurized (\textbf{i}) grippers gripping a cylindrical object with a diameter $D$. Scale bar, 50 mm. Corresponding pressure-volume relation for the soft gripper (solid) and tank (dashed) (\textbf{g, j}) and equilibrium pressure in the system (\textbf{h, k}) for objects. Experimental results from five tests are shown for each $D$ (\textbf{g, j}).}
    \label{Fig_retrofit_2}
\end{figure}

%TEXT ABOUT FIGURE 3 and 4
\subsubsection*{Retroffiting the fluidic sensing approach}

In all demonstrations so far, we used one or more identical soft bending actuator. However, our sensing approach can also be retrofitted to a broad range of fluidic actuators without the need for any design changes. To demonstrate the wide applicability, using our approach we sensorize a filament actuator, a McKibben actuator, a 3D-printed bending actuator\cite{yap2016high}, a suction cup, and two commercial soft grippers (Fig.~\ref{Fig_retrofit_1}, Fig.~\ref{Fig_retrofit_2} and Supplementary Video 3). 

We start by retrofitting our sensing strategy to a filament actuator \cite{becker2020mechanically, jones2021bubble,doi:10.1073/pnas.2209819119} and a McKibben actuator \cite{daerden2002pneumatic, faudzi2017soft} to sense the angular displacement of a joint in the artificial muscle demonstration \cite{acome2018hydraulically, jones2021bubble}. In both cases (Fig.~\ref{Fig_retrofit_1}a-f), we can correlate measured equilibrium pressure to the angular displacement of the joint. We do observe that for the filament actuator, less sensing resolution is achieved compared to the McKibben actuator. This is likely due to the fact that the deformation of the filament gripper is less well-defined, especially when comparing it to the McKibben actuator where the environment has a strong influence on the internal volume. We hypothesize that this stronger influence is the results of the wires that to some extent limit the degrees of freedom.

To determine if our sensing approach can also be used for higher actuation pressures, we next retrofit our sensing strategy to a 3D-printed TPU bending actuator that requires an actuation pressure around $200$ kPa \cite{yap2016high}. In previous tests with the bending actuator, we only consider a single contact between the soft actuator and the environment. Since the TPU bending actuator forms a circular shape at higher pressures \cite{yap2016high}, we tested our sensing strategy with conformal grasping \cite{tawk20223d}, where the soft actuator interacts with the cylindrical object at multiple contact points (Fig.~\ref{Fig_retrofit_1}g). We find that the conformal grasping of cylindrical objects with various diameters results in different pressure-volume responses of the soft actuator (Fig.~\ref{Fig_retrofit_1}h) and that we can also correlate the equilibrium pressure with the diameter of the grasping object, even for these higher pressure ranges (Fig.~\ref{Fig_retrofit_1}i).

To test our retrofitting approach with soft grippers, we first apply it to a suction cup (Fig.~\ref{Fig_retrofit_2}a) that was specifically designed for tissue gripping in Minimal Invasive Surgery (MIS) \cite{kortman2023bio}. The requirements for the foldability, adaptability and biocompatibilty of the tissue gripper make it challenging to embed sensors in the gripper itself to obtain sensory feedback during operation. With our fluidic sensing strategy, the pressure sensor can be connected remotely to the gripper outside of the human body and no additional design change of the gripper is needed.   
%To investigate the sensing capacity, we first study the interaction between the soft gripper and different silicone samples by characterizing the pressure-volume responses of the gripper (Fig.3). Once we apply vacuum to the gripper, the silicone sample is sucked into the gripper, reducing the internal volume of the gripper and thus slowing down the increase of vacuum pressure. Therefore, the pressure-volume curve shifts down when gripping softer silicone samples. Then we apply the same setup from the third method in Fig.1a to the tissue gripper. Instead of applying positive pressure to the air tank, we vacuumed the tank. Then we open the solenoid valve and let the tank and gripper equilibrate at the same pressure.
Once vacuum is applied to the soft gripper, the connected surface gets pulled into the gripper, reducing its internal geometric volume (Fig.~\ref{Fig_retrofit_2}b). The surface stiffness influences the pressure-volume response of the gripper through the amount of reduced internal geometric volume of the gripper (fitting curves in Fig.~\ref{Fig_retrofit_2}c, and test results in Fig.~\ref{FigS_suctioncup_pv}). The final equilibrium pressure $P_{\mathrm{eq}}$ can be used to infer the stiffness of the surface that is attached to the gripper, where the sensing resolution can be tuned by the initial pressure $P_{\mathrm{0}}$ in the air tank (Fig.~\ref{Fig_retrofit_2}d and Fig.~\ref{FigS_suctioncup_softness_pt}). Furthermore, we found that the equilibrium pressure in the gripper changes almost linearly with the pulling force applied on the gripper (Fig.~\ref{Fig_retrofit_2}e), making it possible to predict when the gripper would detach from the surface which in our experiments occured at $\Delta P \approx 1.2$ kPa with an average detaching force of 4.53 N. %Therefore the doctors can estimate the pulling force and predict the detachment of the tissue with an externally-connected pressure sensor during MIS.

Similarly, we demonstrate that we can retrofit sensing to commercial soft grippers (Fig.~\ref{Fig_retrofit_2}f-k), both to grippers powered by vacuum pressure \cite{piSOFTGRIP} and by positive pressure \cite{SRTgripper}. Even though the pressure-volume relation for both grippers is relatively different, in both cases we find a linear correlation between the size of cylindrical objects and the measured equilibrium pressure. Note that because the internal volumes of the actuators and grippers are different in these examples, we had to replace some of the external hardware. For example, the internal volume of the air tank, which determines the slope of the tank's pressure-volume response, is chosen based on a compromise between initial tank pressure and sensing resolution (Fig.~\ref{FigS_resolution_2caps}).

\subsubsection*{Characterizing the sensing resolution}

Fig.~\ref{Fig_retrofit_1} and Fig.~\ref{Fig_retrofit_2} provide a general picture of how the variations in pressure-volume curves between the soft actuators and grippers lead to different sets of equilibrium points based on the conservation of air mass. The absolute pressure change observed during robot-environment interactions ranges from 1.3 kPa to 5.9 kPa among the soft actuators and grippers we tested (Table 1). Even though the relative pressure difference (pressure difference due to interaction with the environment in comparison to maximum pressure obtained in the actuator) might be smaller for actuators that require higher inflation pressure (e.g., TPU and filament actuators), the absolute pressure change for all actuators we tested is in the same order of magnitude (Table 1). Note that the average sensing resolution can be determined by dividing the absolute pressure change by the tested range of sensing target and is therefore not affected by a lower relative pressure difference.

%However, there are also many other factors involved in this comparison, such as the internal geometric volume of the air tank and soft actuator. 
In order to find out the underlying factors that determine the sensing resolutions of the soft actuators and grippers, we consider both the initial and final states of the system. At the initial state, the air tank with an internal geometric volume $v_{\mathrm{tank}}$ is pressurized at $p_{\mathrm{tank}} = p_0$, and the actuator with an internal geometric volume $v_{\mathrm{act}} = v_{\mathrm{0}}$ is at atmosphere pressure $p_{\mathrm{act}} = p_{\mathrm{atm}}$. At the final state, the air tank and the actuator reach the same pressure $p_{\mathrm{tank}} = p_{\mathrm{act}} = p_1$, and the internal geometric volume of the actuator becomes $v_{\mathrm{act}} = v_1$. Assuming constant temperature, since the total amount of air mass inside the system (air tank and the actuator) stays constant, according to the ideal gas law, we have
    \setcounter{equation}{0}
    \begin{equation} \label{eq:1}
    p_{\mathrm{0}}v_{\mathrm{tank}} + p_{\mathrm{atm}}v_{\mathrm{0}} = p_{\mathrm{1}}v_{\mathrm{tank}} + p_{\mathrm{1}}v_{\mathrm{1}}.
    \end{equation}
    % which can be rewritten as 
    % \begin{equation} \label{eq:2}
    % p_{\mathrm{1}} = \frac{p_{\mathrm{0}}v_{\mathrm{tank}} + p_{\mathrm{atm}}v_{\mathrm{0}}}{v_{\mathrm{tank}}+v_{\mathrm{1}}}.
    % \end{equation}
Note that the absolute pressure here is indicated in lowercase letter to distinguish it from the relative pressure (with respect to atmospheric pressure) that is used elsewhere in the manuscript. When the total amount of air inside the system remains constant, $v_{\mathrm{1}}$ only depends on the interaction of the soft actuator with the environment, i.e., the sensing target $\xi$. Therefore, the sensing resolution $dp_{\mathrm{1}}/d\xi$ can be written as
    \begin{equation} \label{eq:2}
    \frac{dp_{\mathrm{1}}}{d\xi} = -\frac{p_{\mathrm{0}}v_{\mathrm{tank}} + p_{\mathrm{atm}}v_{\mathrm{0}}}{(v_{\mathrm{tank}}+v_{\mathrm{1}})^2}\cdot\frac{dv_{\mathrm{1}}}{d\xi},
    \end{equation}
where, for example for the gripping test of the cylinders  $\xi = D$, i.e., the diameter of the cylindrical objects in Fig.~\ref{Fig_retrofit_2}f and i. Moreover, $dv_1/d\xi$ represents the sensitivity of the internal geometric volume of the gripper to gripping cylindrical objects with different diameters. Equation \ref{eq:2} shows that essentially, the variation of internal geometric volume when the soft actuator interacts with the environment in different ways causes the pressure change, which can be used to infer the interaction.

    \begin{table}[h]
    \centering
    \resizebox{150mm}{!}{\includegraphics{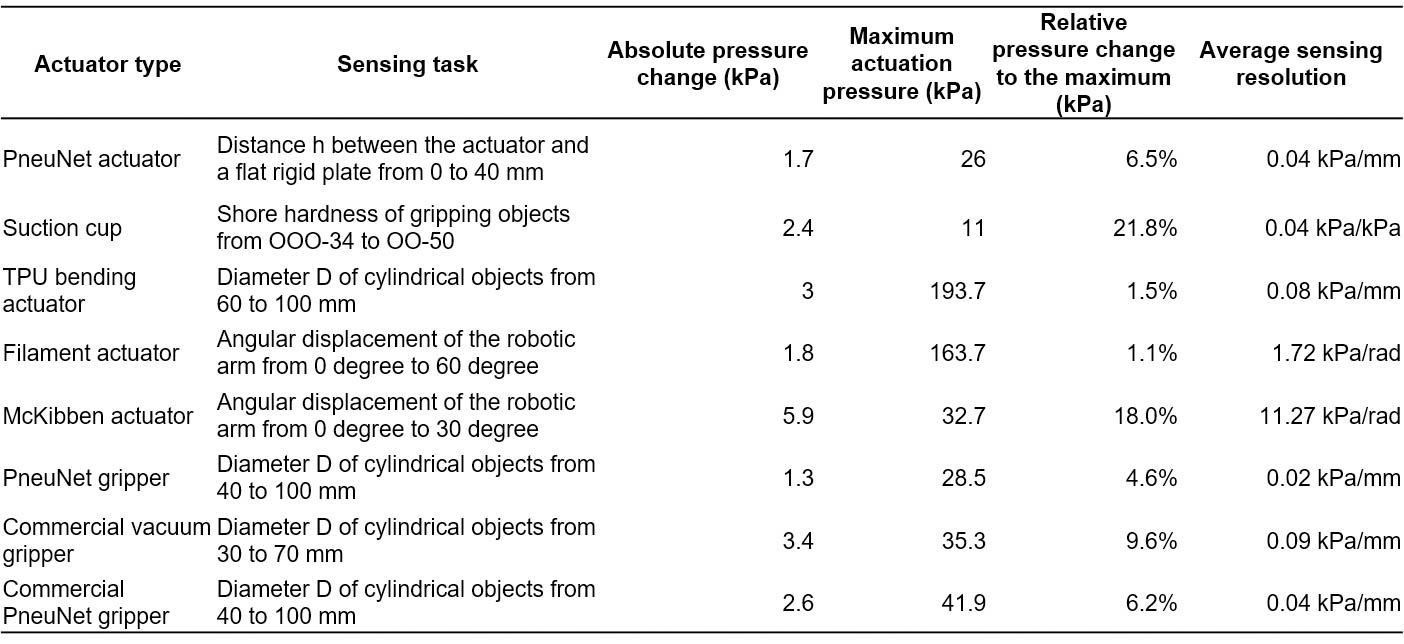}}
    \caption{\textbf{Overview of the sensing performance of the actuators and grippers tested in this work.} The modulus (at 100\% strain) of Ecoflex OO-50 is 82.7 kPa according to the product sheet. The modulus (at 100\% strain) of OOO-34 is estimated from that of OOO-35 (17 kPa) from the reference\cite{byrne2021silico}.}
    \label{TableS_outputrange}
    \end{table}

%The absolute pressure change observed during robot-environment interactions ranges from 1.3 kPa to 5.9 kPa among the soft actuators and grippers we tested (Table 1). Equation \ref{eq:2} shows that this pressure change results from the variation of internal geometric volume when the soft actuator interacts with the environment in different ways. It is important to note that even though the relative pressure difference (pressure difference due to interaction with the environment in comparison to maximum pressure obtained in the actuator) might be smaller for actuators that require higher inflation pressure (e.g., TPU and filament actuators), the absolute pressure change for all actuators we tested is in the same order of magnitude (Table 1). Luckily, there are ample (relatively cheap) pressure sensors on the market that span this full range of pressures with high accuracy (Table S3), so that the sensing resolution is not affected by a lower relative pressure difference.

However, it is not trivial to compare the sensing resolutions of the soft actuators and grippers in Fig.~\ref{Fig_retrofit_1} and Fig.~\ref{Fig_retrofit_2}, because these actuators vary in actuation pressure, internal volume and sensing targets. To give an example, since the initial internal geometric volumes $v_{\mathrm{0}}$ of both commercial grippers in Fig.~\ref{Fig_retrofit_2}f and i are known, we can determine $v_{\mathrm{1}}$ at equilibrium from equation \ref{eq:1} based on experimental measurements of $p_{\mathrm{1}}$ (Table S1 and Fig.~\ref{FigS_commercialgrippers}), from which we can then obtain $dp_1/dD$ according to equation \ref{eq:2}. We find that the sensitivity of the internal geometric volume of the gripper to gripping different cylindrical objects equals $|dv_1/dD|=0.16$ ml/mm for the vacuum gripper and $|dv_1/dD|=0.13$ ml/mm for the PneuNet gripper, while the magnitude of the sensing resolution $|dp_1/dD|=0.08$ kPa/mm of the vacuum gripper is twice that of the PneuNet gripper ($|dp_1/dD|=0.04$ kPa/mm). According to equation \ref{eq:2}, the smaller term $(v_{\mathrm{tank}}+v_{\mathrm{1}})^2$ that is related to internal volumes in the case of the vacuum gripper contributes to the higher magnitude of sensing resolution when compared to the PneuNet gripper, even though the term $p_{\mathrm{0}}v_{\mathrm{tank}} + p_{\mathrm{atm}}v_{\mathrm{0}}$ that is related to the total amount of air in the system is lower in the case of the vacuum gripper.

Equation \ref{eq:2} indicates that the sensing resolution is determined by the total amount of air $p_{\mathrm{0}}v_{\mathrm{tank}} + p_{\mathrm{atm}}v_{\mathrm{0}}$ in the system, the internal geometric volume of the tank $v_{\mathrm{tank}}$ and soft actuator $v_{\mathrm{1}}$ at equilibrium, and the sensitivity of the internal geometric volume of the actuator to the sensing target $dv_1/d\xi$. The effects of system parameters, such as the stiffness and initial geometric volume of the soft actuator, on the sensing resolution depend on how these system parameters affect $p_{\mathrm{0}}v_{\mathrm{tank}} + p_{\mathrm{atm}}v_{\mathrm{0}}$, $v_{\mathrm{tank}}$, $v_{\mathrm{1}}$ and $dv_1/d\xi$ in equation \ref{eq:2}, which should be analyzed case by case. To give an example, we develop a basic model based on the interaction of an extension actuator (with a linear stiffness $k$) with a rigid wall in Methods section \textbf{Modeling the Fluidic Sensing Approach} and show the effects of the linear actuator's initial length, cross section, stiffness and initial tank pressure on the sensing resolution in Fig.~\ref{FigS_modelV4}. For example, for the modeled linear extension actuator, an increase in length (increase in initial volume) reduces the sensing resolution, while an increase in area (also an increase in initial volume) first increases and then decreases the sensing resolution. Despite the simplifications made in the model, we believe that it provides a framework for choosing available parameters for improving the sensing resolution.

Finally, it should be noted that the overall sensing accuracy is determined by both the sensing resolution of the actuator or gripper and the sensing accuracy of the pressure sensor used, and luckily, there are ample (relatively cheap) pressure sensors on the market that span various pressure ranges with high enough accuracy for our purpose (Table S2). For example, in the tests with the PneuNet actuator in Fig.~\ref{Fig1}b, we used a $\pm 34.5 \mathrm{kPa}$ pressure sensor with an accuracy of $\pm 0.25\%$ (of Full Scale Span). By dividing the pressure sensor error ($\pm 0.1725 \mathrm{kPa}$) by the average sensing resolution ($0.04 \mathrm{kPa/mm}$), we can obtain an overall sensing accuracy of $\pm 4.31 \mathrm{mm}$ for the PneuNet actuator. This sensing accuracy is valid for one-time measurement with the pressure sensor. In this work, however, we always do the pressure measurement over a period of time (which we were able to reduce to 0.05 seconds in Fig.~\ref{FigS_actuationspeed}) and calculate the average value, which gives a higher overall sensing accuracy, e.g., $\pm 1.7 \mathrm{mm}$ with a 95\% confidence interval (Fig.~\ref{FigS_sensing_accuarcy}a) for the PneuNet actuator.

%These examples show that our sensing strategy can be retrofitted to soft actuators whose pressure-volume responses change while interacting with the environment, which is true for a wide range of actuators in soft robotics. The differences in the pressure-volume responses of a soft actuator after interacting with the environment determines its maximum potential sensing resolution and is related with the deformation mode of the actuator. For example, for a sensing target ranging from $0^{\circ}$ to $30^{\circ}$ in Fig.~\ref{Fig3}f-k, the sensing output span is 1 kPa for the filament actuator and 6 kPa for the McKibben actuator. This is because once the robotic arm hits the stopper, the McKibben acutator can only deform through radial expansion, while the filament actuator is free to deform in any dimensions other than the longitudinal direction. A more constrained deformation of the actuator after interacting with the environment leads to a lager divergence of the pressure-volume responses and therefore a larger span of the sensing output.

%TEXT ABOUT FIGURE 4
\subsubsection*{Closed-loop control with fluidic sensing}

\begin{figure}
    \centering
    \resizebox{160mm}{!}{\includegraphics{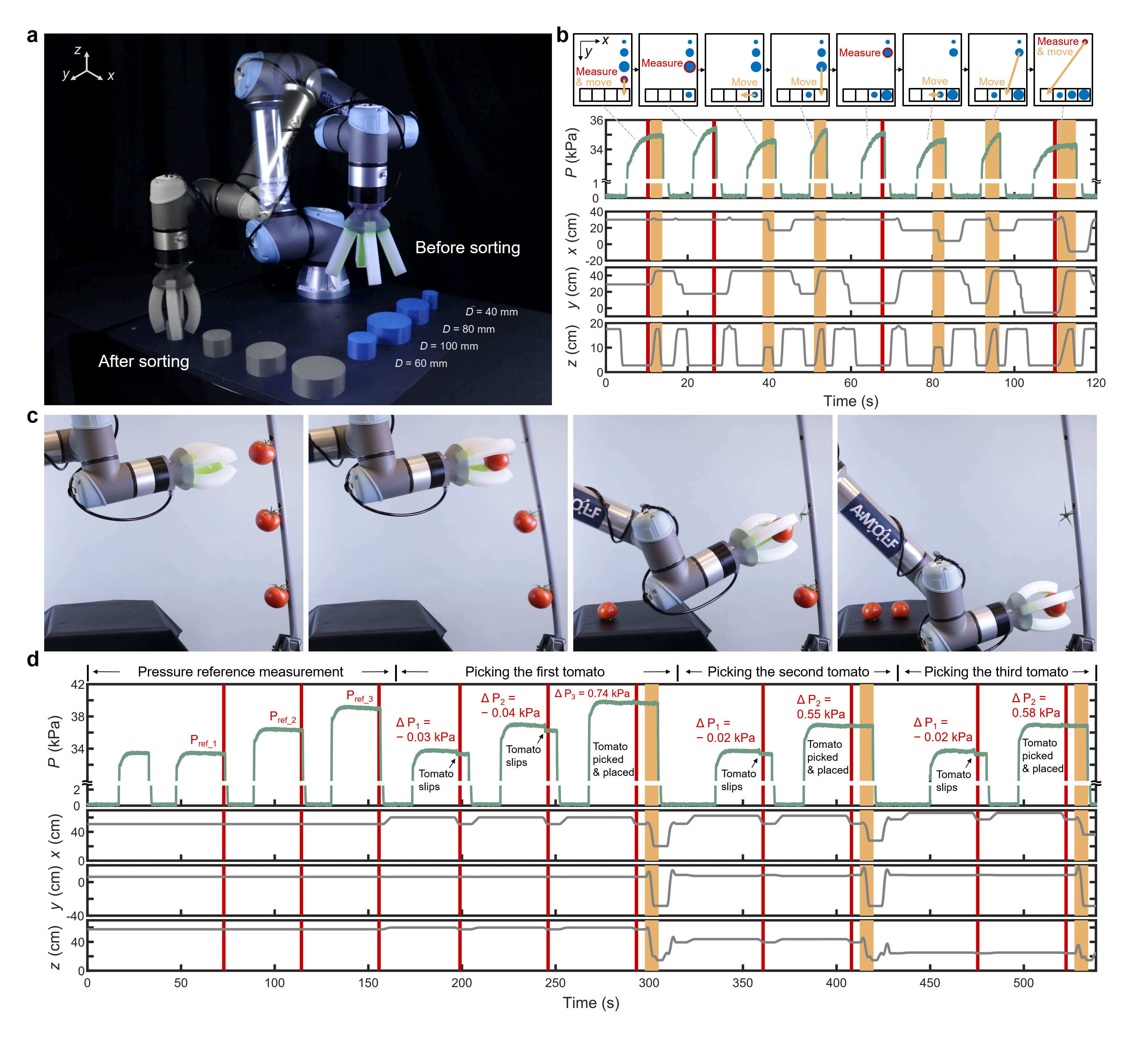}}
    \caption{\textbf{Closed-loop control with fluidic sensing. a,} Superimposed pictures taken at the beginning and ending of the size sorting experiment. The input order of the cylindrical objects was randomly selected. \textbf{b,} gripper actuation pressure and coordinates of the Tool Center Point (TCP) over time during the sorting. The manipulations during each actuation cycle are explained in the corresponding schematics. The red band in the plot represents the pressure feedback measurement, and the yellow band represents the object movement. The four pressure feedback measurements (average $\pm$ standard deviation) are $34.80 \pm 0.07$ kPa, $35.36 \pm 0.07$ kPa, $35.05 \pm 0.08$ kPa, $34.13 \pm 0.07$ kPa, respectively. \textbf{c,} Snapshots of the tomato picking experiment representing the pressure reference measurement and picking of the three tomatoes, respectively. \textbf{d,} gripper actuation pressure and TCP coordinates over time during the tomato picking. The red band represents the pressure feedback measurement, and the yellow band represents the tomato placement on the table.}
    \label{Fig4}
\end{figure}

With the insights gained on sensing performance, we finally demonstrate that the fluidic sensing strategy is robust for closed-loop control in three applications: size sorting, tomato picking and ripeness detection. In the first closed-loop control experiment (Fig.~\ref{Fig4}a-b and Supplementary Video 4), we used a modified insertion sort algorithm which inserts cylinders of different diameters one by one at the correct position in a sorted array based on the fluidic sensing feedback. Every time a new object is gripped, its size is measured via the feedback pressure in the gripper that is averaged over a fixed time window (red band in Fig.~\ref{Fig4}b) after actuation. When the feedback pressure is smaller than the pressures in the sorted array, the robotic arm directly moves the new object to the end of the queue, e.g., the first and last actuation cycles in Fig.~\ref{Fig4}b. Otherwise, the robotic arm moves objects in the sorted array to make the correct position available for the new object, e.g., the third and sixth actuation cycles in Fig.~\ref{Fig4}b. The closed-loop control makes it possible to successfully sort the four cylindrical objects with random input order (Fig.~\ref{FigS_sorting_1234}, Fig.~\ref{FigS_sorting_4321}, Fig.~\ref{FigS_sorting_4123} and Supplementary Video 4). The fluidic sensing strategy also works for sorting random objects with irregular shape (Fig.~\ref{FigS_sorting_random1}, Fig.~\ref{FigS_sorting_random2} and Supplementary Video 4). Note that no calibration curve is needed here since the comparison between the equilibrium pressure values is sufficient for sorting.

In the second closed-loop control experiment (Fig.~\ref{Fig4}c-d and Supplementary Video 5), we perform a tomato picking experiment. We artificially increase the pressure in every step to demonstrate that the proposed sensing approach can provide feedback for picking automation. Note that in a more realistic setting, one would have likely used the highest pressure immediately. During each picking cycle, the algorithm compares the equilibrium pressure $P_{\mathrm{eq}}$ in the gripper to a corresponding reference measurement $P_{\mathrm{ref}}$ in free space, and evaluates $\Delta P = P_{\mathrm{eq}} - P_{\mathrm{ref}}$ to determine a successful or unsuccessful picking. When $\Delta P$ is smaller than a threshold, the algorithm regards it as an unsuccessful picking as the actuators are not deformed by the tomato. It then starts the next picking cycle with a higher actuation pressure. Once $\Delta P$ is larger than a threshold (0.2 kPa), the algorithm regards it as a successful picking. The gripper places the tomato on the table and moves to the next tomato in line. Note that the slip of the tomato out of the gripper in an unsuccessful picking can also be detected from the abrupt decrease in the pressure-time curve, which could potentially provide additional sensing feedback. We tested a total of nine tomatoes in three runs, out of which six tomatoes were successfully picked and placed, one tomato was not picked by the gripper, the other two were successfully picked but not recognized because the tomato slipped into the palm of the gripper after being picked from the stem, resulting in a $\Delta P$ smaller than the threshold (Fig.~\ref{FigS_tomatofailure} and Supplementary Video 5). As our soft gripper was not specifically designed for picking tomatoes, the design of the gripper should be optimized for this task. Importantly, this would not affect the retrofit implementation of our sensing approach.

\begin{figure}
    \centering
    \resizebox{110mm}{!}{\includegraphics{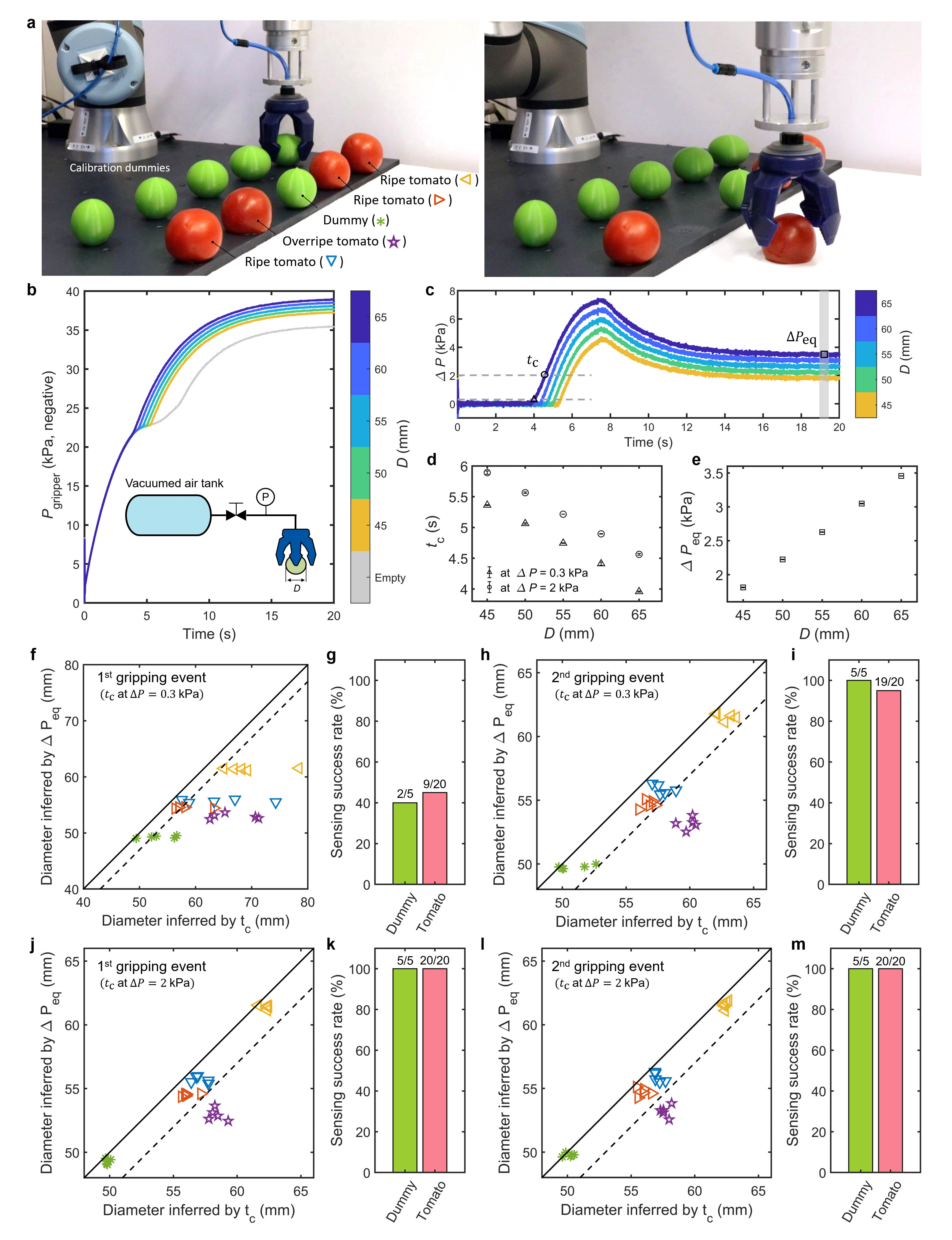}}
    \caption{\textbf{Picking out an overripe tomato with fluidic sensing.} \textbf{a,} Snapshots of the closed-loop control demonstration. Five 3D-printed rigid dummies ($D =$ 45, 50, 55, 60, 65 mm) are used for calibration. The other five objects are tested for sensing, including three ripe tomatoes, one overripe tomatoes and one dummy with a diameter $D =$ 50 mm. The whole demonstration includes one cycle of calibration and five cycles of sensing. In each sensing cycle, two gripping events are carried out for each object. The positions of the objects are shuffled randomly between cycles. \textbf{b-e,} Calibration results. Experimental results from three measurements are plotted for each $D$ in \textbf{b,c}. The error bars in \textbf{d,e} represent the standard deviation of three measurements. \textbf{f-m,} Sensing results. The solid line in \textbf{f,h,j,l} represents that the object diameter inferred by $t_{\mathrm{c}}$ equals that inferred by $\Delta P_{\mathrm{eq}}$, the dashed line represents the object diameter inferred by $\Delta P_{\mathrm{eq}}$ is 3 mm smaller than that inferred by $t_{\mathrm{c}}$. For the ripe tomatoes and rigid dummy, a measurement above the dashed line is considered as a successful sensing event. Conversely, for the overripe tomato, a measurement below the dashed line indicates sensing success.}
    \label{Fig_ripe}
\end{figure}

In the third closed-loop control experiment (Fig.~\ref{Fig_ripe} and Supplementary Video 6), we detect the ripeness of tomatoes by applying the method mentioned in Fig.~\ref{Fig2}f and g to estimate the indentation depth. We demonstrate that the proposed sensing approach can provide feedback for automated sorting of an overripe tomato from ripe tomatoes. For versatility, we select the commercial vacuum gripper for this demonstration, also as it has the largest sensing resolution (Table 1). The demonstration includes one cycle of calibration and five separate cycles of sensing to determine repeatability. The positions of the four tomatoes (including one overripe tomato) and one dummy are shuffled randomly between sensing cycles. All size predictions are based on one calibration process with 3D-printed rigid dummies, where the gripper pressure $P_{\mathrm{gripper}}$ is compared to a reference response $P_{\mathrm{ref}}$ to determine $\Delta P = P_{\mathrm{gripper}} - P_{\mathrm{ref}}$ over time for different diameters of the object that is being gripped (Fig.~\ref{Fig_ripe}b and c). As explained by the stiffness sensing method in Fig.~\ref{Fig2}f and g, the size of the tomato upon gripping can be inferred by the time of first contact $t_{\mathrm{c}}$ (Fig.~\ref{Fig_ripe}d), and the size at equilibrium can be inferred by the $\Delta P$ at equilibrium (Fig.~\ref{Fig_ripe}e).
    
It is important to note that the method that uses the time of first contact is strongly affected by the alignment of the tomato inside the gripper, leading to an early rise of $\Delta P$-time curve and inaccurate size predictions of the tomato upon gripping and sensing success rates of 40\% and 45\% (Fig.~\ref{Fig_ripe}f and g) for the rigid dummy and tomatoes, respectively. To increase the sensing success rate, we can perform the gripping event twice, the first to center the object inside the gripper, and the second gripping event to extract sensing feedback (Fig.~\ref{Fig_ripe}h and i). Alternatively, we can choose $t_{\mathrm{c}}$ at higher $\Delta P$ values for both calibration and sensing (Fig.~\ref{Fig_ripe}j and k), so that the object has been effectively centered during sensing. With $t_{\mathrm{c}}$ at $\Delta P = 2 \mathrm{kPa}$, sensing during either the first or second gripping event gives 100\% success rates for both rigid dummy and tomatoes (Fig.~\ref{Fig_ripe}j-m). We can also infer the initial size of the tomato by $\Delta P_{\mathrm{max}}$ (Fig.~\ref{FigS_tomato_ripeness_DPMAX} and Supplementary Video 6) instead of $t_{\mathrm{c}}$ to avoid the influence of misalignment, which gives 100\% and 90\% success rate for rigid dummy and tomatoes, respectively. While we were able to pick out the rotten tomato consistently, since the gripper squeezes tomatoes for ripeness detection in this method, post-harvest studies should be performed in the future to avoid extra damage to the produce when applying the method in practical applications.

It should be noted that all the closed-loop control demonstrations above were performed under quasistatic conditions. We tested the sensing strategy at different actuation speeds (Fig.~\ref{FigS_actuationspeed}) and find that the effectiveness of the sensing strategy is not affected by the actuation speed, as long as the sensing feedback is collected after the actuator comes into contact with the environment. To speed up the sensing process for real-world applications, it is important to ensure an actuation speed that is high enough for the interaction with the environment to happen before the collection of fluidic sensing feedback. To prove the feasibility, we successfully implemented fast sensing in the size sorting demonstration, where pressure values are collected 0.5 s (instead of 5 s) after the opening of the valve between the air tank and gripper (Fig.~\ref{FigS_sortingfaster} and Supplementary Video 4).

\subsubsection*{Conclusion}

% discussions about limitations to be added
% 1. single contact v.s. multiple contacts, irregularly shaped objects
% 2. about scaling: multiple fluidic inputs, bulkiness of current systems, potential in combining with soft vavles.
% 3. situations that the strategy wouldn't work, nonunique readings (inhomogeneous stiffness), jamming gripper, stachastic gripping. Also make it clear that any 'bad' actuator influences sensing too.

In conclusion, we present a versatile fluidic sensing strategy that relies on measuring the fluidic input response instead of embedded sensing elements into the soft actuators. The soft robot-environment interaction can be accurately interpreted with one pressure sensor that is connected remotely. We show that the proposed strategy can be retrofitted to a wide range of soft devices, and implemented in closed-loop control of gripping applications. %\indent \updated{The absolute pressure change observed during robot-environment interactions ranges from 1.3 kPa to 5.9 kPa among the soft actuators and grippers we tested (Table 1). We show in equation \ref{eq:2} that this pressure change results from the variation of internal geometric volume when the soft actuator interacts with the environment in different ways. It is important to note that even though the relative pressure difference (pressure difference due to interaction with the environment in comparison to maximum pressure obtained in the actuator) might be smaller for actuators that require higher inflation pressure (e.g., TPU and filament actuators), the absolute pressure change for all actuators we tested is in the same order of magnitude (Table 1). Luckily, there are ample (relatively cheap) pressure sensors on the market that span this full range of pressures with high accuracy (Table S3), so that the sensing resolution is not affected by a lower relative pressure difference. It should be noted that the overall sensing accuracy is determined by both the sensing resolution of the actuator/gripper and the sensing accuracy of the pressure sensor used. For example, in the tests with the PneuNet actuator, we used a $\pm 34.5 \mathrm{kPa}$ pressure sensor with an accuracy of $\pm 0.25\%$ (of Full Scale Span). By dividing the pressure sensor error ($\pm 0.1725 \mathrm{kPa}$) by the average sensing resolution ($0.04 \mathrm{kPa/mm}$), we can obtain an overall sensing accuracy of $\pm 4.31 \mathrm{mm}$ for the PneuNet actuator. This sensing accuracy is valid for one-time measurement with the pressure sensor. In this work, we always do the pressure measurement over a period of time (which we were able to reduce to 0.05 seconds in Fig.~\ref{FigS_actuationspeed}) and calculate the average value, which gives a higher overall sensing accuracy, e.g., $\pm 1.7 \mathrm{mm}$ with a 95\% confidence interval (Fig.~\ref{FigS_sensing_accuarcy}a) for the PneuNet actuator.} 
We believe that this relatively straightforward integration of sensing capabilities makes it readily available for other soft robotic devices and applications, including wearable assistive devices \cite{shveda2022wearable} and soft locomotive robots \cite{faudzi2017soft,vasios2020harnessing}, without the need to alter the design of the soft device itself. 

%Finally, while our goal is to demonstrate the potential for retrofit sensing in soft actuators and grippers, higher sensing resolutions could also be obtained by redesigning the actuators as demonstrated by the model we introduced in this work. Yet, we believe that} this relatively straightforward integration of sensing capabilities makes is readily available for integration with other soft robotic devices and applications, including wearable assistive devices \cite{shveda2022wearable} and soft locomotive robots \cite{faudzi2017soft,vasios2020harnessing}, without the need to alter the design of the soft device itself.

While we were able to retrofit our sensing approach to a range of soft actuators and grippers, it should be noted that the air tank method in Fig.~\ref{Fig1}b is not directly applicable to soft actuators driven by incompressible liquid, as the method depends on the compressibility of air and the final pressure balance between the tank and actuator. This could be solved by using a flexible tank (e.g., a balloon), such that the compressibility of the air is replaced by the elasticity of the tank. A simpler approach could instead be to use the pressure control (Fig.~\ref{Fig1}c) or flow control (Fig.~\ref{Fig1}d) method to obtain sensory feedback from the liquid-driven actuator's interaction with the environment.

%Since all the methods depend on the differences observed in the pressure-volume curves of a soft actuator under different interactions with the environment, the larger differences these pressure-volume curves show, the larger the sensing accuracy could be. 

According to equation \ref{eq:2}, the sensing resolution is influenced by the final internal volume $v_{\mathrm{1}}$ of the actuator after the interaction with the environment and its derivative over the sensing target $dv_{\mathrm{1}}/d\xi$. Yet, there is no trivial relationship between the initial volume of the actuator and the sensing resolution.
Even though we demonstrate that we can retrofit sensing to pneumatic actuators, optimizing the sensing resolution by for example changing the tank size and the initial tank pressure should be done on a case by case basis, and can best be done by experimentally obtaining the relation between pressure and volume for specific interactions with the environment.
For example, if we want to apply our strategy to pneumatic actuators that can generate complex motions with multiple degrees of freedom\cite{hu2020bioinspired}, the sensing resolution depends on how the interaction of the actuator with the environment affects $v_{\mathrm{1}}$ and $dv_{\mathrm{1}}/d\xi$ in equation \ref{eq:2}, which is not straightforward to predict beforehand and depends on the application.

Moreover, we did find relatively small variations over time during cyclic gripping. These variations may be due to the performance of the soft actuator, pressure regulator and pressure sensor, or environment variations like the temperature. Comparing the sensing response to a reference helps reduce the influence of long-term system and environment variations, as also demonstrated in the closed-loop control examples.  Still, any non-unique pressure-volume response of a soft actuator would cause inaccurate sensing. Moreover, since the proposed sensing principle uses the soft actuator itself as a sensor, any fragility or unreliability of the soft actuator would have a direct influence on the sensing performance. Additionally, the soft actuation of more complex devices should be designed such that it achieves sensitivity to the sensing task. 

%The proposed sensing principle can be applied to many fluid-actuated soft devices to sense the interaction with the environment without altering the design of the soft devices themselves. Still, as for most soft robotic sensing applications, information about the sensing task is typically needed for calibration or relating the input response to the environment. For example, in order to make use of the versatility of our approach, information is needed about whether we are evaluating the shape of an object or the stiffness, or the orientation of a surface. Importantly, for most applications the task environment can be clearly defined.

Moving forward, machine learning can potentially be applied to read the higher-order difference in the pressure-time curves of the soft robot for various interactions with the environment to achieve more complex sensing applications \cite{shih2020electronic,chin2020machine}. Although our sensing strategy removes the need of embedding or attaching sensors to the soft actuator, the hardware of the system is still bulky and may not be suitable (yet) for small-scale untethered soft robotic system \cite{wehner2016integrated}. To further reduce the bulkiness of the system, the fluidic sensory feedback from the soft robot-environment interaction can be potentially read out by soft pneumatic valves \cite{rothemund2018soft, van2022fluidic, decker2022programmable} to build electronic-free soft robots \cite{drotman2021electronics, van2022fluidic, lee2022buckling} that can sense and respond to their environment. The basic yet powerful principles studied in this work make it possible to bring (some) sensing capabilities to most soft fluidic devices without the need for design changes, and paves the way towards new functionalities in soft interactive devices and systems for real world applications.
%facilitate the transition from soft robotic research towards real-world applications.

%We come up with four practical methods to extract the sensing information: 1) pressure control with pressure measurement; 2) pressure control with volume measurement; 3) flow control with pressure measurement; 4) pressure-time measurement. Methods 1) to 3) measure the fluidic signal at the equilibrium state, which is related with the final shape of the soft actuator after the interaction with the environment, while method 4) can give us more information about when and how many contact events happened. 
%To implement this sensing strategy in practice, each method requires different hardware which may scale up with the number of fluidic sources. Method 1) requires an air tank and a valve for each fluidic source and a pressure sensor for each actuator. Method 2) and 3) requires no additional devices for each fluidic source but a flow sensor and pressure sensor for each actuator, respectively. Method 4) can be done with any of the systems above.

% Your references go at the end of the main text, and before the
% figures.  For this document we've used BibTeX, the .bib file
% scibib.bib, and the .bst file Science.bst.  The package scicite.sty
% was included to format the reference numbers according to *Science*
% style.

%BibTeX users: After compilation, comment out the following two lines and paste in
% the generated .bbl file. 
\section*{Methods}
Details on the methods are provided in Supplementary Materials.

\section*{Data Availability}
All experimental data that support the findings of this work and computer algorithms necessary for running the analysis will be uploaded to Zenodo before publication.

\section*{Code availability}
The numerical data in the Methods section \textbf{Modeling the Fluidic Sensing Approach} and computer algorithms necessary for running the model will be uploaded to Zenodo before publication.

\section*{Acknowledgments}
This work is part of the Dutch Research Council (NWO) and supported by European Research Council Starting Grants (grant agreement ID: 948132).

\bibliography{FluidicSensing}

\bibliographystyle{Science}

%Here you should list the contents of your Supplementary Materials -- below is an example. 
%You should include a list of Supplementary figures, Tables, and any references that appear only in the SM. 
%Note that the reference numbering continues from the main text to the SM.
% In the example below, Refs. 4-10 were cited only in the SM.  

\newpage
\setcounter{page}{1}
\section*{Supplementary Materials for\newline
A Retrofit Sensing Strategy for Soft Fluidic Robots}
Shibo Zou, Sergio Picella, Jelle de Vries, Vera Kortman, Aimée Sakes, Johannes T. B. Overvelde
Corresponding author: Johannes T. B. Overvelde, overvelde@amolf.nl
\renewcommand{\theequation}{\arabic{equation}}
\setcounter{equation}{0}
\renewcommand\thefigure{S\arabic{figure}}  
\setcounter{figure}{0}

\subsection*{Fabrication of PneuNet actuators and soft gripper}
The PneuNet actuator was molded in a two-step process (Fig.~\ref{FigS_actuator_fabrication}). In the first step, the extensible layer that consists of air chambers was molded with Dragon Skin (DS) 10 Medium silicone (Smooth-On). Since the DS silicone cures in 5 hours, we waited 3.5 hours for the DS silicone to be almost cured. In the second step, we molded the Elite Double (ED) 32 silicone (Zhermack) around the almost-cured DS silicone to form the inextensible layer and the inlet unit. This way, the ED silicone could cure together with the DS silicone, leading to a strong adhesion between the extensible and inextensible layers. A grid fabric (Penelope 70/10, Garenenzo) was embedded during the molding to further increase the axial stiffness of the inextensible layer. We used one sacrificial inner mold and one reconfigurable outer mold containing six parts for the two-step molding process. The inner mold was printed with butene-diol vinyl alcohol (BVOH) on a Fused Filament Fabrication (FFF) 3D printer (Ultimaker 3). The outer mold parts were printed with an acrylic-like photopolymer (VeroClear, Stratasys) on a PolyJet 3D printer (Eden260VS, Stratasys). The VeroClear mold tends to inhibit the curing of silicone, especially those freshly printed. Therefore, we brushed a thin layer of anti cure inhibition coating (Inhibit X, Smooth-On) onto the inner surface of the VeroClear mold to avoid the cure inhibition. The outer molds can be assembled into two different configurations corresponding to the two steps of the molding process. In each step, the silicone was first loaded into a two-component cartridge (AF 400-01-10-01, Sulzer), degassed and then injected into the mold through a static mixing nozzle (MFQ 05-24L) using a pneumatic extrusion gun. Before the molding of each silicone, a thin layer of release agent (Ease release 200, Smooth-On) was sprayed and then brushed evenly over the molding surfaces. The silicones were cured at room temperature. After curing, the BVOH inner mold was dissolved and flushed out by connecting the actuator to a water pump (H5750010, FLOJET) in parallel with a tunable flow resistor for venting. The pneumatic connectors for the single PneuNet actuator and the soft grippper were printed with VeroClear on the Stratasys 3D printer. Vacuum grease was brushed into the connector, and the PneuNet actuator was gently pressed into the connector. %The computer-aided design (CAD) files of all the molds and connectors can be found in Supplementary Information.

\subsection*{Distance sensing with PneuNet bending actuator}
The pneumatic connector containing a single PneuNet actuator was mounted on an aluminium frame to ensure that the bottom edge of the actuator at rest stays horizontal under gravity. The actuator was inflated onto a horizontal aluminium bar from a distance $h$. We adjusted $h$ from 0 mm to 40 mm in 5 mm increments by manually moving the aluminium bar with a caliper. To control and measure the fluidic tests, we used a data acquisition card (NI-DAQ USB-6212, National Instruments) with custom software developed at AMOLF. The measurement setup contains (depending on the sensing method used) a proportional pressure regulator (-100 kPa to 100 kPa, VEAB-L-26-D13-Q4-V1-1R1, Festo), mass flow controller (SLA5850, Brooks Instrument), 3/2-way solenoid valve (VDW250-5G-2-01F-Q, SMC), air tanks (CRVZS-0.1, CRVZS-0.75, Festo), bidirectional flow sensor (HAFBLF0750CAAX5, Honeywell), $\pm$ 34.5 kPa and $\pm$ 103.4 kPa pressure sensors (SSCDRRN005PDAA5, SSCDRRN015PDAA5, Honeywell). All measurements were done with a data acquisition frequency of 1000 Hz. All the tests conducted with the data acquisition card in this paper started with 15 s blank measurement for sensor offset, and the data after offset were smoothed in MATLAB using the loess method (local regression using weighted linear least squares and a second degree polynomial model) with a span of 20 data points to remove high frequency noise before any further analysis.

To obtain the pressure-volume curves in Fig.~\ref{Fig1}a, we measured both the flow input and pressure response over time as the actuator was inflated onto the aluminium bar from a distance $h$. The actuation system (Fig.~\ref{FigS_setup_flowcontrol}) included, in the direction of air flow, a mass flow controller and a 3/2-way solenoid valve. The input port of the mass flow controller was connected to the compressed air source that was set at 150 kPa by an independent wall-mounted pressure regulator (LRP-1/4-2.5, Festo). We added a 0.75 L air tank as a buffer in between the wall-mounted pressure regulator and mass flow controller. The normally closed port of the valve was connected to the mass flow controller, the normally open port of the valve was connected to the atmosphere, and the inlet port of the valve was connected to the PneuNet actuator through a bidirectional flow sensor. We added a flow resistor made from Teflon tube and nozzles (920050-TE, Metcal) with an equivalent flow resistance around $1.8 \times 10^9 \: \mathrm{Pa} \cdot \mathrm{s}/\mathrm{m}^3$ in between the valve and flow sensor to ensure that the flow rate was within the sensor range. We used a $\pm$ 103.4 kPa pressure sensor to measure the pressure in the actuator. A total of 8 actuation cycles were performed at each $h$ by programming the custom software for the data acquisition system. Each actuation cycle started by switching the valve on, increasing the flow rate from 0 to 0.6 SLPM (standard liter per minute) in 5 s, holding the flow rate at 0.6 SLPM for 2 s, then decreasing the flow rate to 0 SLPM, holding at 0 SLPM for 30 s, and finally switching the valve off and discharging the actuator for 30 s. For each $h$, the pressure-volume curve was fitted from the measurements of the last 5 actuation cycles using a six-order polynomial with a fixed zero y-intercept.

We come up with four practical methods to extract the sensing information and demonstrate here by distance sensing with a soft PneuNet bending actuator: 1) pressure control with pressure measurement; 2) pressure control with flow measurement; 3) flow control with pressure measurement; 4) pressure-time measurement (Time of first contact in Section \textbf{Time-enabled sensing versatility}). Methods 1) to 3) measure the fluidic signal at the equilibrium state, which is related with the final shape of the soft actuator after the interaction with the environment, while method 4) can give us more information about when and how many contact events happened. To implement this sensing strategy in practice, each method requires different hardware which may scale up with the number of fluidic sources. Method 1) requires an air tank and a valve for each fluidic source and a pressure sensor for each actuator. Method 2) and 3) requires no additional devices for each fluidic source but a flow sensor and pressure sensor for each actuator, respectively. Method 4) can be done with any of the systems above.

\subsubsection*{Pressure control with pressure measurement}
The actuation system (Fig.~\ref{FigS_setup_masscontrol}) included, in the direction of air flow, a proportional pressure regulator, a 3/2-way solenoid valve, a 0.1 L air tank, and another 3/2-way solenoid valve. The input port of the proportional pressure regulator was connected to the compressed air source that was set at 150 kPa by an independent wall-mounted pressure regulator. For the solenoid valve, the inlet port was connected to the proportional pressure regulator and the normally closed port was connected to the air tank, the normally open port was blocked by a cap (FTLLP-6005, Nordson). For the second valve, the inlet port was connected to the PneuNet actuator and the normally closed port was connected to the air tank, the normally open port was connected to the atmosphere. We used a $\pm$ 103.4 kPa pressure sensor and a $\pm$ 34.5 kPa pressure sensor to measure the pressure in the 0.1 L air tank and actuator, respectively. A flow resistor with an equivalent flow resistance around $7.3 \times 10^9 \: \mathrm{Pa} \cdot \mathrm{s}/\mathrm{m}^3$ was added between the second solenoid valve and the actuator to avoid any dynamic effect during the charging of the actuator. A total of 8 actuation cycles were performed at each $h$ by programming the custom software for the data acquisition system. Each actuation cycle started by setting the proportional pressure regulator at 49.9 kPa and waiting for 10 s, then switching on the first valve to charge the 0.1 L air tank for 10 s, switching off the first valve and waiting for 10 s, then switching on the second valve to charge the actuator for 60 s, and finally switching off the second valve to discharge the actuator for 30 s. To obtain the calibration curve in Fig.~\ref{Fig1}b, an average pressure was first taken for each cycle over the last 10 s during the 60 s actuator charging period, then those average pressure values from the last 7 cycles were used to calculate the final average $P_{\mathrm{eq}}$ for each $h$. The error bars in Fig.~\ref{Fig1}b represent the standard deviation of those average pressure values from the last 7 cycles.

\subsubsection*{Pressure control with flow measurement}
The actuation system (Fig.~\ref{FigS_setup_pressurecontrol}) included, in the direction of air flow, a proportional pressure regulator and a 3/2-way solenoid valve. The input port of the proportional pressure regulator was connected to the compressed air source that was set at 100 kPa by an independent wall-mounted pressure regulator. The normally closed port of the valve was connected to the output port of the proportional pressure regulator. We added a 0.75 L air tank as a buffer in between the proportional pressure regulator and the valve. The normally open port of the valve was connected to the atmosphere, and the inlet port of the valve was connected to the PneuNet actuator through a bidirectional flow sensor. We added a flow resistor with an equivalent flow resistance around $1.8 \times 10^9 \: \mathrm{Pa} \cdot \mathrm{s}/\mathrm{m}^3$ in between the valve and flow sensor to ensure that the flow rate lies within the sensor range. We used a 0-100 kPa pressure sensor (MPX5100DP, NXP) to measure the pressure in the actuator. A total of 8 actuation cycles were performed at each $h$ by programming the custom software for the data acquisition system. Each actuation cycle started by setting the proportional pressure regulator at 23.8 kPa and waiting for 10 s, then switching the valve on and charging the actuator for 30 s, and finally switching the valve off and discharging the actuator for 30 s. To obtain the calibration curve in Fig.~\ref{Fig1}c, the total volume of the air going into the actuator during each cycle was calculated by integrating the flow rate over the 30 s charging period. The total volume values of the last 7 cycles were averaged to obtain the $V_{\mathrm{eq}}$ for each $h$. The error bars in Fig.~\ref{Fig1}c represent the standard deviation of the total volume values from the last 7 cycles.

\subsubsection*{Flow control with pressure measurement}
The actuation system (Fig.~\ref{FigS_setup_flowcontrol}) was the same as the one for the characterization of pressure-volume curves mentioned above, except that 
%included, in the direction of air flow, a mass flow controller and a 3/2-way solenoid valve. The input port of the mass flow controller was connected to the compressed air source that was set at 150 kPa by an independent wall-mounted pressure regulator. We added a 0.75 L air tank as a buffer in between the wall-mounted pressure regulator and mass flow controller. The normally closed port of the valve was connected to the mass flow controller, the normally open port of the valve was connected to the atmosphere, and the inlet port of the valve was connected to the PneuNet actuator through a bidirectional flow sensor, which was to verify the flow going into the actuator. We added a flow resistor made from Teflon tube and nozzles (920050-TE, Metcal) with an equivalent flow resistance around $1.8 \times 10^9 \: \mathrm{Pa} \cdot \mathrm{s}/\mathrm{m}^3$ in between the valve and flow sensor to ensure that the flow rate lies within the sensor range. 
we used a $\pm$ 34.5 kPa pressure sensor to measure the pressure in the actuator. A total of 8 actuation cycles were performed at each $h$ by programming the custom software for the data acquisition system. Each actuation cycle started by switching the valve on, increasing the flow rate from 0 to 0.3 SLPM (standard liter per minute) in 5 s, holding the flow rate at 0.3 SLPM for 4 s, then decreasing the flow rate to 0 SLPM, holding at 0 SLPM for 30 s, and finally switching the valve off and discharging the actuator for 30 s. To obtain the calibration curve in Fig.~\ref{Fig1}d, an average pressure was first taken for each cycle over the last 5 s during the 30 s holding period, then those average pressure values from the last 7 cycles were used to calculate the final average $P_{\mathrm{eq}}$ for each $h$. The error bars in Fig.~\ref{Fig1}d represent the standard deviation of those average pressure values from the last 7 cycles.

\subsection*{Size sensing with soft gripper}
Five cylinders with the same height of 40 mm and diameters of 20 mm, 40 mm, 60 mm, 80 mm, and 100 mm were printed with polylactic acid (PLA) filament on a fused filament fabrication (FFF) 3D printer (Ultimaker 3). The pneumatic connector that contains the soft gripper was mounted on a robotic arm (Universal Robots UR5e) and moved to a position above the cylinder that ensures gripping with actuator tips upon inflation (Fig.~\ref{Fig1}e). The robotic arm stayed still during test and each cylinder was placed under the gripper manually without accurate alignment. The actuation system (Fig.~\ref{FigS_setup_gripper}) included, in the direction of air flow, a proportional pressure regulator (-100 kPa to 100 kPa, VEAB-L-26-D13-Q4-V1-1R1, Festo), a 3/2-way solenoid valve, three 0.1 L air tanks that were connected in series and another 3/2-way solenoid valve. The input port of the proportional pressure regulator was connected to the compressed air source that was set at 150 kPa by an independent wall-mounted pressure regulator. We added a 0.75 L air tank as a buffer in between the proportional pressure regulator and the first valve. For the first valve, the inlet port was connected to the proportional pressure regulator and the normally open port was connected to the air tank, the normally closed port was blocked by a cap (FTLLP-6005, Nordson). For the second valve, the inlet port was connected to the PneuNet actuator and the normally closed port was connected to the air tank, the normally open port was connected to the atmosphere. We used two $\pm$ 103.4 kPa pressure sensors to measure the pressure in the air tanks and soft gripper, respectively. We added a small flow resistor in between the second valve and the gripper to avoid any dynamic effect during the actuation of the gripper. All control and measurement were done on the data acquisition card with custom software. For each cylinder, the proportional pressure regulator was set at 63 kPa at the beginning of the test. A total of 6 actuation cycles were performed for each cylinder by programming the custom software for the data acquisition system. Each actuation cycle started by switching off the first valve to charge the three 0.1 L air tanks for 10 s, then switching on the first valve to disconnect the tanks from the proportional pressure regulator and waiting for 10 s, switching on the second valve to charge the gripper for 20 s, then switching off the second valve to discharge the gripper. To obtain the calibration curve in Fig.~\ref{Fig1}f, an average pressure was first taken for each actuation cycle over the measurements between 4 s and 5 s after switching on the second valve. Then those average pressure values from the last 5 cycles were used to calculate the final average $P_{\mathrm{eq}}$ for each cylinder. The error bars in Fig.~\ref{Fig1}f represent the standard deviation of those average pressure values from the last 5 cycles. 

\subsection*{Time-enabled sensing versatility}

\subsubsection*{Time of first contact}
The analysis in this section was performed with the test results obtained from section \textbf{Pressure Control with Pressure Measurement}. An additional test of free actuation ($h = \infty$) was done using the same test setup and the actuator pressure-time curve from the last cycle out of the 8 cycles in total was used as a reference response for the following analysis. The actuator tip vertical displacement-time curve (dashed line in Fig.~\ref{Fig2}c) from the reference test was obtained by applying a point-tracking MATLAB algorithm on the test footage. To find the time of first contact $t_{\mathrm{c1}}$, we first obtained the $\Delta P$-time curve of each cycle by subtracting the pressure response of each cycle with the reference response, then we used a script that detects abrupt changes in slope to find $t_{\mathrm{c1}}$ on the $\Delta P$-time curve. The average $t_{\mathrm{c1}}$ over the last 7 cycles out of the 8 cycles in total at each $h$ was plotted in Fig.~\ref{Fig2}c, with the error bar representing the standard deviation.

\subsubsection*{Shape sensing}
Four rectangular objects with the same height of 40 mm and the same length of 100 mm, and widths of 40 mm, 60 mm, 80 mm and 100 mm were printed with PLA filament on a FFF 3D printer. The test setup was the same as that in section \textbf{Size Sensing with Soft Gripper}, except that we used a 0-100 kPa pressure sensor (MPX5100DP, NXP) and a $\pm$ 34.5 kPa pressure sensor to measure the pressure in the air tanks and soft gripper, respectively. All control and measurement were done on the data acquisition card with custom software. One actuation cycle was performed for each rectangular object. For each object, the actuation cycle started by setting the proportional pressure regulator at 80 kPa in 20 s, waiting for 15 s to charge the three 0.1 L air tanks, switching on the first valve to disconnect the tanks from the proportional pressure regulator and waiting for 5 s, switching on the second valve to charge the gripper for 60 s, then switching off the second valve to discharge the gripper, and finally switching off the first valve. The $\Delta P$-time curves in Fig.~\ref{Fig2}e were obtained by subtracting the pressure response of each object with the reference response of empty gripping.

\subsubsection*{Stiffness sensing}
To prepare the soft samples, the pre-polymer components of silicone (A and B, Ecoflex OO-10, Smooth-On) were mixed with a ratio of 1A:1B and degassed. The silicone mixture was then molded in a plastic Petri dish (diameter: 86 mm) to which a thin layer of release agent was applied (Ease release 200, Smooth-On). The silicone was cured at ambient temperature for 4 hours and then taken out of the Petri dish. The silicone sample had a thickness of 17 mm. A laser-cut acrylic plate with a thickness of 8 mm was used as the rigid sample. For both the soft and rigid samples, the single PneuNet actuator was placed horizontally and 5 mm above the sample, with the bottom surface of the actuator parallel to the top surface of the sample. The test setup was the same as that in section \textbf{Pressure Control with Pressure Measurement}. One actuation cycle was performed for each sample. The actuation cycle started by setting the proportional pressure regulator at 49.9 kPa, waiting for 5 s to charge the 0.1 L air tank, switching on the first valve to disconnect the tank from the proportional pressure regulator and waiting for 10 s, switching on the second valve to charge the actuator for 60 s, then switching off the second valve to discharge the actuator. The $\Delta P$-time curves in Fig.~\ref{Fig2}g were obtained by subtracting the pressure response of each sample with the reference response of free actuation.

\subsubsection*{Profile scanning}
Two plates (100 mm $\times$ 25 mm $\times$ 15 mm, length $\times$ width $\times$ height) with the top surface in the shape of sine and triangle waves, respectively, were printed with PLA filament on a FFF 3D printer. The single PneuNet actuator was mounted on a robotic arm (Universal Robots UR5e) through a 3D-printed adapter. The actuation system included, in the direction of air flow, a proportional pressure regulator, a 3/2-way solenoid valve, a 0.1 L air tank, and another 3/2-way solenoid valve. The input port of the proportional pressure regulator was connected to the compressed air source that was set at 150 kPa by an independent wall-mounted pressure regulator. The connections of the valves were the same as those described in section \textbf{Pressure Control with Pressure Measurement}. The inlet port of the second valve was connected directly to the actuator through a 2 m long polyurethane air hose (PUN-6X1-BL, Festo). We used a 0-100 kPa pressure sensor (MPX5100DP, NXP) and a 0-50 kPa pressure sensor (MPX5050DP, NXP) to measure the pressure in the air tank and actuator, respectively. All control and measurement were done on the data acquisition card with custom software. We connected the analog input of the robotic arm control board to the 5 V digital output of our data acquisition system and wrote a program in PolyScope that starts the movement of the robotic arm when the analog input changes from 0 V to 5 V. The robotic arm moves horizontally in the mode of MoveL for a total distance of 100 mm at a speed of 2 mm/s. To begin the test, we first manually ran the PolyScope program on the robotic arm. Since the 5 V digital output of our data acquisition system was initially off, the robotic arm first stayed still. Then we ran a script in our custom software that started the actuation process by setting the proportional pressure regulator at 49.9 kPa and waiting for 5 s, then switching on the first valve to charge the 0.1 L air tank for 10 s, switching off the first valve and waiting for 10 s, then switching on the second valve to charge the actuator for 60 s, then switching on the 5 V digital output to initiate the robotic arm movement and waiting for 90 s, and finally switching off the second valve to discharge the actuator for 30 s. A reference pressure-time response was also measured using the same test procedure without scanning any object. We obtained the $\Delta P$-time curve by substracting the pressure-time response of the scan with the reference pressure-time response. To reconstruct the profile, a calibration $\Delta P$-$h$ curve was first obtained using the same setup. With the calibration curve, the $\Delta P$-time curve from the scan could then be converted into a tip-plate distance-time curve. Finally the time data was converted to x coordinates through the robotic arm moving speed. The ground truth of the profiles were obtained by digital image analysis of flat-bed scanned pictures of the printed objects.

\subsection*{Retroffiting the fluidic sensing approach}

\subsubsection*{Suction cup}
%The fabrication of the suction-based soft gripper is based on previous work \cite{kortman2023bio}. 
To prepare the gripping samples with different stiffness, we molded Ecoflex OO-50, OO-30, OO-10, and GEL 2 (OOO-34) in Petri dishes using the same procedure described in section \textbf{Stiffness Sensing}. The actuation system (Fig.~\ref{FigS_setup_suctioncup_softness}) for the suction gripper included, a proportional pressure regulator (-100 kPa to 100 kPa, VEAB-L-26-D13-Q4-V1-1R1, Festo), a 3/2-way solenoid valve, a 15 ml air tank made from an empty CO\textsubscript{2} canister and another 3/2-way solenoid valve. The vacuum input port of the proportional pressure regulator was connected to a vacuum source that was set at -65 kPa by an independent vacuum pressure regulator. We added a 0.75 L air tank as a buffer in between the proportional pressure regulator and the first valve. For the first valve, the inlet port was connected to the proportional pressure regulator and the normally closed port was connected to the 15 ml air tank. The normally open port was blocked by a cap (FTLLP-6005, Nordson). For the second valve, the inlet port was connected to the suction gripper and the normally closed port was connected to the 15 ml air tank. The normally open port was connected to the atmosphere. We used two $\pm$ 103.4 kPa pressure sensors to measure the pressure in the 15 ml air tank and suction gripper, respectively. We added a flow resistor with an equivalent flow resistance around $7.3 \times 10^9 \: \mathrm{Pa} \cdot \mathrm{s}/\mathrm{m}^3$ in between the second valve and the gripper to avoid any dynamic effect during the actuation of the gripper. All control and measurement were done on the data acquisition card with custom software. For each gripping object, two sets of initial vacuum pressure (-62 KPa and - 21 kPa) in the 15 ml air tank were tested. A total of 5 actuation cycles were performed for each test. Each actuation cycle started by setting the proportional pressure regulator and waiting for 15 s, switching on the first valve to vacuum the 15 ml air tank for 15 s, switching off the first valve and waiting for 15 s, then switching on the second valve to vacuum the suction gripper for 60 s, and finally switching off the second valve and waiting for 10 s. At the beginning of each test, the suction gripper was gently pressed on top of the gripping sample to ensure successful gripping upon vacuum. For each actuation cycle, the equilibrium pressure was averaged between 8 s and 10 s after switching on the second valve. For each gripping sample, the equilibrium pressure was averaged over the last 4 actuation cycles and the standard deviation was used to plot the error bar in Fig.~\ref{Fig_retrofit_2}d. The pressure-volume curves (Fig.~\ref{Fig_retrofit_2}c  and Fig.~\ref{FigS_suctioncup_pv}) of the suction-based soft gripper when gripping different Ecoflex samples were characterized using the experimental setup in Fig.~\ref{FigS_setup_suctioncup_PV}. A total of 5 actuation cycles were performed for each gripping sample. The test started by setting the proportional pressure regulator and waiting for 15 s. Each actuation cycle included switching on the valve to vacuum the gripper for 30 s, then switching off the valve and waiting for 30 s.

To correlate the pressure response of the suction gripper with the pulling force, we started pulling the suction gripper away from the gripping sample (Ecoflex OO-30) on a custom tensile test setup, after the suction gripper and the 15 ml air tank reached equilibrium pressure. The actuation system for the suction gripper was the same as mentioned above and the suction gripper was connected to a load cell (FLLSB200, FUTEK) through a nylon wire on the custom tensile test setup. The actuation cycle was also the same as mentioned above, except that the second valve was switched off 15 s after it was switched on, disconnecting the suction gripper from the 15 ml air tank and maximizing the sensing resolution, then the load cell moved up vertically on the custom setup at a speed of 1 mm/s. Since the fluidic and tensile test systems were separated, we synchronized both test data based on the detaching moment of the suction gripper.

\subsubsection*{TPU bending actuator}
The TPU bending actuator was designed based on dimensions described in reference \cite{yap2016high} and printed with TPU filament (Filaflex 82A) on a Fused Filament Fabrication 3D printer (Felix Tec). The actuation system (Fig.~\ref{FigS_setup_TPU_actuator}) included, in the direction of air flow, a proportional pressure regulator (-100 kPa to 500 kPa, VEAB-L-26-D18-Q4-V1-1R1, Festo), a 3/2-way solenoid valve with the normally closed port closed with a cap, a 0.1 L air tank, a second 3/2-way solenoid valve with the normally open port venting to air and the inlet port connected to a flow resistor. In this experiment, we used a different model of the solenoid valve (VDW250-5G-1-01F-Q, SMC) that can handle twice higher pressure than the valves used in other experiments. The input port of the proportional pressure regulator was connected to the compressed air source that was set at 400 kPa by an independent wall-mounted pressure regulator. To reduce the pressure fluctuations during regulation, we added a 0.75 L air tank (CRVZS-0.75, Festo) as a buffer between the proportional pressure regulator and the 3/2-way valve. We used a 0-700 kPa pressure sensor (MPX5700DP, NXP) and a $\pm$ 206.8 kPa pressure sensor (SSCDRRN030PDAA5, Honeywell) to measure the pressure in the air tank and TPU actuator, respectively. A bidirectional flow sensor (HAFBLF0750CAAX5, Honeywell) was added to measure the flow from the air tank to the TPU actuator during actuation and a flow resistor was used to restrict the flow within the range of the flow sensor. A total of 8 actuation cycles were performed for each test. Each actuation cycle started by setting the proportional pressure regulator at 300 kPa and waiting for 20 s, switching on the first valve to isolate the tank from the pressure regulator and waiting for 10 s, then switching on the second valve to inflate the TPU actuator for 60 s, finally switching off the second valve to deflate the TPU actuator and waiting for 60 s before switching off the first valve for recharging the air tank. For each actuation cycle, the equilibrium pressure was averaged between 40 s and 41 s after switching on the second valve. For each cylindrical object, the equilibrium pressure was averaged over the last 5 actuation cycles and the standard deviation was used to plot the error bar in Fig.~\ref{Fig_retrofit_1}i. Note that the sensor offset of the $\pm$ 206.8 kPa pressure sensor was recalculated for each actuation cycle based on the average of the last 1 s measurements before switching on the second valve.

\subsubsection*{Filament actuator}

The filament actuator was fabricated with the gravity-assisted molding approach described in reference \cite{becker2020mechanically}. A 400 mm long steel rod with a diameter of 1.5 mm was coated with liquid silicone (Elite Double 32 fast) and suspended at a 5 degree angle relative to the vertical direction during curing. The filament actuator was cut into a length of 150 mm after curing. One end of the filament actuator was dip coated with extra silicone to form a thicker part that could be mounted on the robotic forearm by piercing with a plastic pin (Fig.~\ref{FigS_setup_muscledemo_filactuator}). The other end of the filament actuator was connected to a tube fitting (FTLL004, Nordson) and coated with extra silicone to ensure airtightness. The angle of the hinge between two arms was measured with a 3D-printed angle ruler to adjust the position of the cylindrical stopper at different angular displacements $\theta$ in Fig.~\ref{Fig_retrofit_1}a-c. The actuation system (Fig.~\ref{FigS_setup_muscledemo_filactuator}) included, in the direction of air flow, a proportional pressure regulator (-100 kPa to 500 kPa, VEAB-L-26-D18-Q4-V1-1R1, Festo), a 3/2-way solenoid valve with the normally open port closed with a cap, a 500 mm long air pipe (PUN-6X1-BL, Festo), a second 3/2-way solenoid valve with the normally open port closed with a cap and a third 3/2-way solenoid valve with both the inlet port and normally open port venting to air. The input port of the proportional pressure regulator was connected to the compressed air source that was set at 400 kPa by an independent wall-mounted pressure regulator. The 500 mm long air pipe was used as the actuation tank. To reduce the pressure fluctuations during regulation, we added a 0.75 L air tank (CRVZS-0.75, Festo) as a buffer between the proportional pressure regulator and the 3/2-way valve. We used a 0-700 kPa pressure sensor (MPX5700DP, NXP) and a $\pm$ 206.8 kPa pressure sensor (SSCDRRN030PDAA5, Honeywell) to measure the pressure in the air pipe and filament actuator, respectively. A bidirectional flow sensor (HAFBLF0750CAAX5, Honeywell) was added to measure the flow from the air pipe to the filament actuator during actuation and a flow resistor was used to restrict the flow within the range of the flow sensor. A total of 8 actuation cycles were performed for each test. Each actuation cycle started by setting the proportional pressure regulator at 265 kPa and waiting for 10 s, switching on the first valve to pressurize the air pipe for 20 s, switching off the first valve and waiting for 10 s, then switching on the second valve to inflate the filament actuator for 30 s, finally switching off the second valve and switching on the third valve to deflate the filament actuator and waiting for 20 s before switching off the third valve. Two solenoid valves instead of one were arranged between the air pipe and filament actuator to avoid any leakage, because only the inlet port of the valve allows pressure up to 700 kPa while the other two ports only allow up to 100 kPa. For each actuation cycle, the equilibrium pressure was averaged between 28 s and 29 s after switching on the second valve. For each angular displacement $\theta$, the equilibrium pressure was averaged over the last 5 actuation cycles and the standard deviation was used to plot the error bar in Fig.~\ref{Fig_retrofit_1}c. Note that the sensor offset of the $\pm$ 206.8 kPa pressure sensor was recalculated for each actuation cycle based on the average of the last 0.5 s measurements before switching on the second valve.

\subsubsection*{McKibben actuator}

The McKibben actuator was fabricated with thermoplastic polyurethane (TPU) bladder and polyethylene terephthalate braided sleeve (PTO0.25BK, Techflex PET Overexpanded). The bladder (130 $\times$ 23.6 mm, length $\times$ width at the deflated state) was made by heat sealing two layers of TPU film (Airtech Stretchlon 200, thickness 38 $\mu m$) on a modified 3D printer (Fleix Tec). The PET sleeve was cut into a length of 130 mm with a heat cutter (D-65396 Walluf, HSGM GmbH). The actuation system (Fig.~\ref{FigS_setup_muscledemo_McKibben}) was the same as that for the filament actuator. We used a $\pm$ 206.8 kPa pressure sensor (SSCDRRN030PDAA5, Honeywell) and a $\pm$ 103.4 kPa pressure sensor (SSCDRRN015PDAA5, Honeywell) to measure the pressure in the air pipe and McKibben actuator, respectively. A bidirectional flow sensor (HAFBLF0750CAAX5, Honeywell) was added to measure the flow from the air pipe to the McKibben actuator during actuation and a flow resistor was used to restrict the flow within the range of the flow sensor. A total of 8 actuation cycles were performed for each test. Each actuation cycle started by setting the proportional pressure regulator at 140 kPa and waiting for 10 s, switching on the first valve to pressurize the air pipe for 20 s, switching off the first valve and waiting for 10 s, then switching on the second valve to inflate the filament actuator for 30 s, finally switching off the second valve and switching on the third valve to deflate the filament actuator and waiting for 60 s before switching off the third valve. The data processing was the same as that of filament actuator.

\subsubsection*{Vacuum-powered commercial soft gripper}

We mounted the vacuum-powered commercial soft gripper (SG.S74DS70.SDS70.F.G14F.00, Piab) on the robotic arm and kept the robotic arm stationary throughout the gripping test. The actuation system (Fig.~\ref{FigS_setup_gripper_piab}) was similar to that described in section \textbf{Size Sensing with Soft Gripper}, except that a vacuum pressure source and 0.1 L air tank were used. We added a bidirectional flow sensor (HAFBLF0750CAAX5, Honeywell) to measure the flow between the air tank and the soft gripper and a flow resistor to restrict the flow within the range of the flow sensor. A total of 8 actuation cycles were performed for each test. Each actuation cycle started by setting the proportional pressure regulator at -60 kPa and waiting for 10 s, switching on the first valve to pressurize the air pipe for 20 s, switching off the first valve and waiting for 10 s, then switching on the second valve to inflate the gripper for 30 s, finally switching off the second valve to deflate the filament actuator and waiting for 30 s to fully discharge. For each actuation cycle, the equilibrium pressure was averaged between 24 s and 25 s after switching on the second valve. For each gripping object, the equilibrium pressure was averaged over the last 5 actuation cycles and the standard deviation was used to plot the error bar in Fig.~\ref{Fig_retrofit_2}h. Note that the sensor offset of the $\pm$ 103.4 kPa pressure sensor was recalculated for each actuation cycle based on the average of the last 1 s measurements before switching on the second valve. 

\subsubsection*{Commercial soft PneuNet gripper}

We mounted the commercial soft PneuNet gripper (SFG-FNC3-N5087-S, Soft Robot Technology) on the robotic arm and kept the robotic arm stationary throughout the gripping test. The actuation system (Fig.~\ref{FigS_setup_gripper_SRT}) was similar to that described in section \textbf{Size Sensing with Soft Gripper}, and we added a bidirectional flow sensor (HAFBLF0750CAAX5, Honeywell) to measure the flow between the air tank and the soft gripper and a flow resistor to restrict the flow within the range of the flow sensor. A total of 8 actuation cycles were performed for each test. Each actuation cycle started by setting the proportional pressure regulator at 70 kPa and waiting for 10 s, switching on the first valve to pressurize the air pipe for 20 s, switching off the first valve and waiting for 10 s, then switching on the second valve to inflate the gripper for 60 s, finally switching off the second valve to deflate the filament actuator and waiting for 60 s to fully discharge. For each actuation cycle, the equilibrium pressure was averaged between 50 s and 51 s after switching on the second valve. For each gripping object, the equilibrium pressure was averaged over the last 5 actuation cycles and the standard deviation was used to plot the error bar in Fig.~\ref{Fig_retrofit_2}k. Note that the sensor offset of the $\pm$ 103.4 kPa pressure sensor was recalculated for each actuation cycle based on the average of the last 1 s measurements before switching on the second valve.

\subsection*{Modeling the fluidic sensing approach}
We develop a basic model based on the interaction of an extension actuator (with a linear stiffness $k$) with a rigid wall (Fig.~\ref{FigS_modelV4}), to characterize and better understand the dynamics between the air tank and actuator in the proposed sensing strategy. The linear actuator is initially at atmosphere pressure $p_{\mathrm{act}} = p_{\mathrm{atm}}$, while the tank with compressed air starts at $p_{\mathrm{tank}} = p_{\mathrm{0}}$. We take the sensing target as the initial distance $L$ between the tip of the actuator and the wall. The valve between the air tank and actuator is opened at $t = 0$, so that the system will reach the equilibrium pressure $p_1$. Assuming incompressible and laminar flow between the air tank and actuator, the Hagen-Poiseuille equation that characterizes the flow rate between air tank and actuator can be written as
\setcounter{equation}{2}
\begin{equation} \label{eq:3}
\frac{dV}{dt} = \frac{p_{\mathrm{tank}} - p_{\mathrm{act}}}{r},
\end{equation}
where $V$ represents the volume of air that transfers from the air tank to the actuator, $r$ represents the flow resistance between the air tank and actuator, and $p_{\mathrm{tank}}$ and $p_{\mathrm{act}}$ indicate the absolute pressure in the air tank and actuator, respectively. According to the ideal gas law, we have
\begin{equation} \label{eq:4}
p_{\mathrm{tank}}v_{\mathrm{tank}} = n_{\mathrm{tank}}RT,
\end{equation}
and
\begin{equation} \label{eq:5}
p_{\mathrm{act}}v_{\mathrm{act}} = n_{\mathrm{act}}RT,
\end{equation}
where $R$ represents the ideal gas constant, $T$ represents the absolute temperature of the air, and $v_{\mathrm{tank}}$ and $v_{\mathrm{act}}$ represent the internal geometrical volume of the air tank and actuator, respectively. The volume $v_{\mathrm{tank}}$ is constant, while $v_{\mathrm{act}}$ depends on the interaction of the actuator with its environment. Assuming the internal volume of the actuator does not change anymore after the actuator comes into contact with the wall, we have
\begin{equation} \label{eq:6}
v_{\mathrm{act}} =
    \begin{cases}
        A(l_{\mathrm{0}}+\dfrac{(p_{\mathrm{act}}-p_{\mathrm{atm}})A}{k}) & \text{if $\dfrac{(p_{\mathrm{act}}-p_{\mathrm{atm}})A}{k} < L$},\\
        A(l_{\mathrm{0}}+L) & \text{if $\dfrac{(p_{\mathrm{act}}-p_{\mathrm{atm}})A}{k} \ge L$},\\
    \end{cases}       
\end{equation}
where $A$ and $l_{\mathrm{0}}$ represent the cross section and initial length of the actuator, respectively. By substituting equation \ref{eq:6} into equation \ref{eq:5}, we have
\begin{equation} \small \label{eq:7}
p_{\mathrm{act}} =
    \begin{cases}
        \dfrac{-(Al_{\mathrm{0}} - \dfrac{p_{\mathrm{atm}}A^2}{k})+
        \sqrt{(Al_{\mathrm{0}} - \dfrac{p_{\mathrm{atm}}A^2}{k})^2 + 4n_{\mathrm{act}}RT \dfrac{A^2}{k}} }
        {\dfrac{2A^2}{k}} & \text{if $\dfrac{(p_{\mathrm{act}}-p_{\mathrm{atm}})A}{k} < L$},\\
        \dfrac{n_{\mathrm{act}}RT}{A(l_{\mathrm{0}}+L)} & \text{if $\dfrac{(p_{\mathrm{act}}-p_{\mathrm{atm}})A}{k} \ge L$}.\\
    \end{cases}       
\end{equation}

Combining equations \ref{eq:3}, \ref{eq:4} and \ref{eq:5}, we obtain two differential equations that describe the change of the amount of air mass $n_{\mathrm{tank}}$ in the air tank and $n_{\mathrm{act}}$ in the actuator over time
\begin{equation} \label{eq:8}
\frac{Mdn_{\mathrm{tank}}}{\rho dt} = -\frac{\dfrac{n_{\mathrm{tank}}RT}{v_{\mathrm{tank}}} - p_{\mathrm{act}}}{r},
\end{equation}
\begin{equation} \label{eq:9}
\frac{Mdn_{\mathrm{act}}}{\rho dt} = \frac{\dfrac{n_{\mathrm{tank}}RT}{v_{\mathrm{tank}}} - p_{\mathrm{act}}}{r},
\end{equation}
% \begin{equation} \label{eq:5}
% \frac{Mdn_{\mathrm{tank}}}{\rho dt} = -
% \frac{p_{\mathrm{tank}} - p_{\mathrm{act}}}{r},
% \end{equation}
% \begin{equation} \label{eq:6}
% \frac{Mdn_{\mathrm{act}}}{\rho dt} = 
% \frac{p_{\mathrm{tank}} - p_{\mathrm{act}}}{r},
% \end{equation}
where $M$ and $\rho$ represent the molar mass and density of air, respectively, and $p_{\mathrm{act}}$ is given by equation \ref{eq:7}. With initial conditions $n_{\mathrm{tank}}\Bigr\rvert_{t = 0}=p_{\mathrm{0}}v_{\mathrm{tank}}/RT$ and $n_{\mathrm{act}}\Bigr\rvert_{t = 0}=p_{\mathrm{atm}}Al_{\mathrm{0}}/RT$, equations \ref{eq:8} and \ref{eq:9} can be solved numerically. 

For the parameter values presented in Table S3, we determine the pressure-time and pressure-volume responses, calibration and sensing resolution points as shown in Fig.~\ref{FigS_modelV4}b-e. Note that for the basic interaction of the linear extension actuator with a rigid wall, assuming that the actuator comes into contact with the wall at the equilibrium state, the initial and the equilibrium internal volume of the actuator are known under different environment settings, i.e., $v_{\mathrm{0}} = Al_{\mathrm{0}}$, $v_{\mathrm{1}} = A(l_{\mathrm{0}}+L)$, then we also have $dv_{\mathrm{1}}/dL = A$. Therefore, we can also determine the analytical solution of the equilibrium pressure $p_{\mathrm{1}}$ and sensing resolution $dp_{\mathrm{1}}/dL$ from equation \ref{eq:1} and \ref{eq:2} ($\xi = L$)
\begin{equation} \label{eq:10}
p_{\mathrm{1}} = \frac{p_{\mathrm{0}}v_{\mathrm{tank}} + p_{\mathrm{atm}}Al_{\mathrm{0}}}{v_{\mathrm{tank}}+A(l_{\mathrm{0}}+L)},
\end{equation}
\begin{equation} \label{eq:11}
\frac{dp_{\mathrm{1}}}{dL} = -\frac{p_{\mathrm{0}}v_{\mathrm{tank}} + p_{\mathrm{atm}}Al_{\mathrm{0}}}{(v_{\mathrm{tank}}+A(l_{\mathrm{0}}+L))^2} A.
\end{equation}
% Therefore, even though the numerical simulation results shown in Fig.~4d appears to be linear, it is actually a reciprocal of a linear function with respect to the sensing target according to equation \ref{eq:11}. The sensing resolution $dp_{\mathrm{1}}/dL$ is a reciprocal of a quadratic function with respect to the sensing target according to equation \ref{eq:11}. \\

With the numerical simulation and analytical solution, we can qualitatively study the influence of each individual system parameter on the sensing resolution. 
The initial internal volume of the actuator can be changed by varying either the initial length $l_0$ (Fig.~\ref{FigS_modelV4}f), or the cross section $A$ (Fig.~\ref{FigS_modelV4}g), which interestingly have different effects on the sensing resolution. To understand this difference, we can first look at an extreme case when the volume of the actuator is infinitely large. In this extreme case, the air mass in the air tank is too small to extend the actuator, then the sensing resolution is $dp_1/dL = 0$. This vanishing of sensing resolution with an extremely large actuator can be seen from the numerical simulation results when varying either $l_0$ or $A$ ($dp_{\mathrm{1}}/dL$ becomes zero after $l_{\mathrm{0}}$ and $A$ exceed a certain value in Fig.~\ref{FigS_modelV4}f and g), but not from the analytical solutions. This is because the analytical solution assumes that the actuator always comes into contact with the wall at the equilibrium state. However, the occurrence of this contact also depends on the initial conditions. Therefore, the numerical simulation results agree with the analytical solution in Fig.~\ref{FigS_modelV4}f and g until the actuator can not reach the wall anymore at given initial conditions. Still, the analytical solution in equation \ref{eq:11} provides theoretical insights on how the sensing resolution is affected by $l_0$ and $A$.    
% This difference can be explained by equation \ref{eq:11}. As $l_{\mathrm{0}}$ increases, the absolute value of $dp_{\mathrm{1}}/dL$ decreases in equation \ref{eq:11}, while since $dv_{\mathrm{1}}/dL = A$ for the extension actuator in Fig.~4a, as $A$ increases, the absolute value of $dp_{\mathrm{1}}/dL$ increases in equation \ref{eq:11}. 

Interestingly, the analytical solution also indicates that the stiffness $k$ (Fig.~\ref{FigS_modelV4}h) does not influence the sensing resolution, because $k$ does not affect any of the parameters in equation \ref{eq:11}. However, when $k$ is extremely large, the actuator becomes too stiff to extend a distance of $L$ with given initial conditions, such that the sensing resolution becomes $dp_1/dL = 0$ as no contact occurs, which can be seen from the numerical simulation results in Fig.~\ref{FigS_modelV4}h. Therefore, a higher stiffness does require a higher tank and actuation pressure. On the other hand, the magnitude of sensing resolution increases with the initial tank pressure $p_0$ when the other parameters in equation \ref{eq:11} stay constant(Fig.~\ref{FigS_modelV4}i). 

It should be noted that, in this simplified model with the linear extension actuator, we assume that the actuator volume stops changing after it hits the wall. However, in reality, the interaction with the environment does not fully constrain the deformation and thus the internal volume of the actuator, which can complicate the analysis and needs to be analyzed case by case. We believe that, despite the simplifications made in our model, it provides a framework for choosing available parameters for improving the sensing resolution.

\subsection*{Sorting experiment}
Four cylinders with the same height of 40 mm and diameters of 40 mm, 60 mm, 80 mm, and 100 mm were printed with PLA filament on a FFF 3D printer. The objects were placed into the input array manually. The soft gripper was mounted on a robotic arm (Universal Robots UR5e) through a 3D-printed adapter and connected to the actuation system via a 3.3 m long polyurethane air hose (TIUB07, SMC) that was winded around the robotic arm. The actuation system (Fig.~\ref{FigS_setup_sorting}) included, in the direction of air flow, a proportional pressure regulator (-100 kPa to 100 kPa, VEAB-L-26-D13-Q4-V1-1R1, Festo), a normally open 2/2-way solenoid valve (VX243AZ3AAXB, SMC), a 0.4 L air tank (CRVZS-0.4, Festo) and a 3/2-way solenoid valve (VDW250-5G-2-01F-Q, SMC). The input port of the proportional pressure regulator was connected to the compressed air source that was set at 150 kPa by an independent wall-mounted pressure regulator (LRP-1/4-2.5, Festo).  The 0.4 L air tank was used as the actuation tank. To reduce the pressure fluctuations during regulation, we added a 0.75 L air tank (CRVZS-0.75, Festo) as a buffer between the proportional pressure regulator and the 2/2-way valve. We used a 0-250 kPa pressure sensor (MPX4250DP, NXP) and a $\pm$ 34.5 kPa pressure sensor (SSCDRRN005PDAA5, Honeywell, uncompensated pressure up to 40.7 kPa) to measure the pressure in the 0.4 L air tank and soft gripper, respectively. The proportional pressure regulator was connected to the 24 VDC digital output and 10 VDC analog output of the control board of the robotic arm for power and control, respectively. The solenoid valves were connected to the 24 VDC digital outputs of the control board. The pressure sensors were connected to the analog output and analog input of the control board for power and reading, respectively. To establish the connection with the robotic arm (Universal Robots UR5e), we used the urx and pymodbus libraries in Python and ran the code in Jupyter Notebook. The proportional pressure regulator was set at 64 kPa. Since the 2/2-way valve was normally open, the 0.4 L air tank was initially pressurized to 64 kPa. At the beginning of each actuation cycle, the 2/2-way valve was first switched on to isolate the 0.4 L air tank from the proportional pressure regulator. After 0.5 s, the 3/2-way valve was switched on to connect the 0.4 L air tank to the soft gripper. The pressure measurement of the soft gripper started 5 s after the activation of the 3/2-way valve and continued for 0.75 s at a data acquisition frequency of 30 Hz. The algorithm associated the average value of the pressure measurement with the size of the object and used it for the insertion sort afterwards. Compared to the conventional insert sort algorithm, our algorithm inserted objects into a separated sorted array instead of the same input array. All the coordinates of the input and sorted array were preprogrammed in the algorithm. To open the gripper, the 3/2-way valve was first switched off, connecting the gripper to the atmosphere. After 0.5 s, the 2/2-way valve was switched off, reconnecting the 0.4 L air tank to the proportional pressure regulator for charging.

\subsection*{Tomato picking experiment}
The tomatoes with stems were bought from a local supermarket. Each tomato was cut off with a remaining piece of the thick stem, which was then inserted into the small slot along a plastic tube without any extra fixture. The plastic tube was clamped vertically at the bottom end to simulate the hanging tomatoes. The soft gripper, actuation system and their connections were the same as those described in section \textbf{Sorting Experiment}. The hanging positions of the three tomatoes and corresponding placing positions were preprogrammed in the python code. The experiment included a dummy actuation cycle, three reference actuation cycles at initial tank pressure $P_{\mathrm{0}} = 62 \mathrm{kPa}, 70 \mathrm{kPa}, 78 \mathrm{kPa}$, respectively, and three intended actuation cycles for each tomato at $P_{\mathrm{0}} = 62 \mathrm{kPa}, 70 \mathrm{kPa}, 78 \mathrm{kPa}$, respectively. During the reference actuation cycle, the gripper closed in the air without moving. During the picking actuation cycle, the robotic arm first moved to the preprogrammed picking position and closed the gripper around the tomato. The gripper rotated the tomato 180 degrees and then the robotic arm retracted 8 - 11 cm. For both reference actuation and picking actuation cycles, the pressure measurement started 25 s after the activation of the 3/2-way valve and continued for 1 s at a data acquisition frequency of 30 Hz. The algorithm used the average pressure values at the same $P_{\mathrm{0}}$ to calculate the $\Delta P$. The $\Delta P$ threshold for successful picking was set at 0.2 kPa. 

\subsection*{Tomato ripeness sensing experiment}
All tomatoes were bought from a local supermarket and we put one tomato inside a zipper bag together with a banana for one week to overripe it. The calibration dummies with a height of 48 mm and diameters (in the horizontal plane) of 45, 50, 55, 60, 65 mm were printed with ABS filament on a Fused Filament Fabrication 3D printer (Ultimaker 3). The actuation system (Fig.~\ref{FigS_setup_ripeness}) was slightly different from that described in section \textbf{{Vacuum-Powered Commercial Soft Gripper}}. We used a 2 L air tank in between the proportional pressure regulator and the first valve in order to decrease the time it takes for the tank to fully stabilize at the set pressure using the pressure regulator. We put the flow resistor ($7.3 \times 10^9 \: \mathrm{Pa} \cdot \mathrm{s}/\mathrm{m}^3$) before the second valve instead of after it so that the venting doesn't go through the flow resistor and the gripper opens faster. The positions of the calibration dummies and tomatoes were preprogrammed in the python code. The experiment included one calibration round and five sensing rounds. A data acquisition frequency of 150 Hz was achieved by directly connecting the robotic arm to a laptop through Ethernet cable, while all other experiments with the robotic arm in this work were done with wifi connection. In the calibration round, we performed a reference measurement where the gripper closed in the air at the beginning and performed four gripping events on each dummy, the first of which was to align the dummy inside the gripper and the rest of which were used for calibration. In each sensing round, we performed a reference measurement and two gripping events on each tomato. Between sensing rounds, we shuffled the tomatoes randomly to test the robustness of the sensing strategy. For both calibration and sensing cycles, the tank pressure was set at -69.36 kPa.

\newpage
\noindent\textbf{Supplementary Video 1}

\noindent\textbf{Fluidic sensing of the soft robot-environment interaction}

\noindent \textbf{1. Fluidic response of a soft actuator inflated onto a rigid plate.} A PneuNet actuator is inflated with standard air volume of 28.7 ml onto a rigid plate from different heights $h$. Due to the compliance of the soft body, the interaction with the plate influences the fluidic response of the soft actuator. This fluidic response can be used to infer the interaction. \textbf{2. Size sensing.} A soft gripper is actuated by a 0.3 L air tank that is initially pressurized at 63 kPa. The final equilibrium pneumatic pressure of the gripper can be used to infer the size of the cylindrical object in the gripper.
\newline
\newline
\noindent\textbf{Supplementary Video 2}

\noindent\textbf{Time-enabled sensing versatility}

\noindent \textbf{1. Shape sensing.} A soft gripper is actuated by a 0.3 L air tank that is initially at a pressure of 63 kPa. The pneumatic pressure of the gripper is measured over time and compared to a reference response where the gripper closes without touching anything. Time of contacts can be determined from the pressure-time response and used to infer the aspect ratio of the rectangular object. \textbf{2. Stiffness sensing.} A PneuNet actuator is actuated by a 0.1 L air tank that is initially at a pressure of 49.9 kPa. The pressure of the actuator is measured over time and compared to a reference response where the actuator is inflated without touching anything. The time of contact can be determined from the pressure-time response and, together with the final equilibrium pressure, can be used to compare the stiffness values of different objects. \textbf{3. Profile scanning.} A PneuNet actuator is actuated by a 0.1 L air tank that is initially at a pressure of 49.9 kPa. The actuator is moved horizontally by a robotic arm for 100 mm at a speed of 2 mm/s. The pressure-time response of the actuator can be used to reconstruct the surface profile based on a calibration curve and a reference response where the actuator is inflated without touching the surface. 
\newline
\newline
\noindent\textbf{Supplementary Video 3}

\noindent\textbf{Retrofitting the fluidic sensing approach}

\noindent \textbf{1. Suction cup.} A suction cup is actuated onto silicone samples with different shore moduli by a 15 ml air tank that is initially at a pressure of -62 kPa. The surface stiffness of the sample influences the pressure-volume response of the suction cup through the amount of reduced internal geometric volume of the suction cup, which can be used in return to infer the surface stiffness. \textbf{2. TPU bending actuator.} A TPU bending actuator is actuated around cylindrical objects with different diameters by a 100 ml air tank that is initially at a pressure of 300 kPa. The size of the cylindrical object influences the pressure-volume response of the bending actuator. The equilibrium pneumatic pressure of the bending actuator can be used to infer the diameter of the cylindrical objects. \textbf{3. Filament actuator.} A filament actuator is actuated by a 6.3 ml air pipe that is initially at a pressure of 265 kPa, and acts as a muscle to rotate an arm towards a stopper. The pressure-volume response of the filament actuator changes with the total angular displacement $\theta$ of the joint. The equilibrium pneumatic pressure of the filament actuator can be used to infer the angular displacement $\theta$ of the joint. \textbf{4. McKibben actuator.} A McKibben actuator is actuated by a 6.3 ml air pipe that is initially at a pressure of 140 kPa, and acts as a muscle to rotate an arm towards a stopper. The pressure-volume response of the McKibben actuator changes with the total angular displacement $\theta$ of the joint. The equilibrium pneumatic pressure of the McKibben actuator can be used to infer the angular displacement $\theta$ of the joint. \textbf{5. Vacuum-powered commercial soft gripper.} A vacuum-powered commercial soft gripper is actuated by a 0.1 L air tank that is initially pressurized at -60 kPa. The final equilibrium pneumatic pressure of the gripper can be used to infer the size of the cylindrical object in the gripper. \textbf{6. Commercial soft PneuNet gripper.} A commercial soft PneuNet gripper is actuated by a 0.3 L air tank that is initially pressurized at 70 kPa. The final equilibrium pneumatic pressure of the gripper can be used to infer the size of the cylindrical object in the gripper.
\newline
\newline
\noindent\textbf{Supplementary Video 4}

\noindent\textbf{Closed-loop control with fluidic sensing: size sorting}

\noindent Four cylindrical objects with different diameters in a random input order are sorted by a robotic arm based on a modified insertion sort algorithm. The gripper is actuated by a 0.4 L air tank that is initially pressurized at 64 kPa. The equilibrium pneumatic pressure of the gripper provides a fluidic sensing feedback that is used to compare the size of the gripped cylindrical objects without using any calibration curve. The fluidic sensing feedback also works for sorting random objects.
\newline
\newline
\noindent\textbf{Supplementary Video 5}

\noindent\textbf{Closed-loop control with fluidic sensing: tomato picking}

\noindent Tomatoes with preprogrammed positions are picked and placed by the soft gripper. For each tomato, three attempts are made, in which the gripper is actuated by a 0.4 L air tank that is set to a pressure of 62 kPa, 70 kPa, 78 kPa, respectively. The equilibrium pneumatic pressure of the gripper after each attempt is compared with a corresponding reference response to determine whether the tomato is successfully picked or not.
\newline
\newline
\noindent\textbf{Supplementary Video 6}

\noindent\textbf{Closed-loop control with fluidic sensing: picking out the overripe tomato}

\noindent One overripe tomato, three ripe tomatoes and one dummy are placed at preprogrammed positions. The vacuum-powered commercial soft gripper grasps the objects one by one to pick out the overripe tomato. The experiment includes one calibration cycle and five sensing cycles. The objects are shuffled randomly between sensing cycles. The size of the object $D_{\mathrm{0}}$ upon gripping can be inferred by either $t_{\mathrm{c}}$ or $\Delta P_{\mathrm{max}}$. The final size of the object $D_{\mathrm{1}}$ after gripping can be inferred by $\Delta P_{\mathrm{eq}}$. The ripeness of tomato can be evaluate with $D_{\mathrm{0}}-D_{\mathrm{1}}$.
\newline
\newline
\noindent\textbf{References} \\

\noindent[1] O.~Byrne, F.~Coulter, E.T.~Roche, E.D~O'Cearbhaill, In silico design of additively manufacturable composite synthetic vascular conduits and grafts with tuneable compliance. \textit{Biomaterials Science}, \textbf{9}(12), 4343-4355 (2021).\newline

\noindent[2] A.~Iniguez-Rabago, Y.~Li, J.T.~Overvelde, Exploring multistability in prismatic metamaterials through local actuation, \textit{Nature Communications} \textbf{10}, 5577 (2019).\newline

\newpage 

\renewcommand\thetable{S\arabic{table}} 
\setcounter{table}{0}

\begin{table}[h]
        \centering
        \resizebox{160mm}{!}{\includegraphics{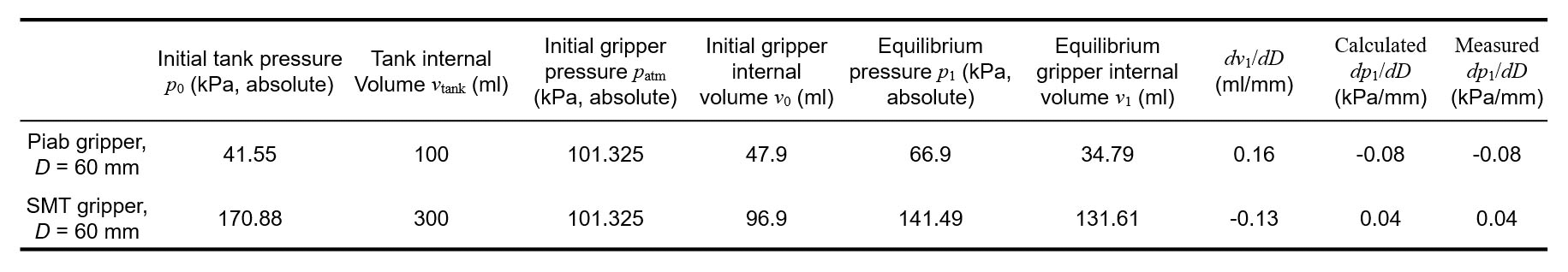}}
        \caption{\textbf{Comparison of the two commercial grippers.} The $dv_{\mathrm{1}}/dD$ is obtained from the linear fits in Fig.~\ref{FigS_commercialgrippers}c and f. The calculated $dp_{\mathrm{1}}/dD$ is based on equation \ref{eq:2}. The measured $dp_{\mathrm{1}}/dD$ is obtained from the linear fits of the data in Fig.~\ref{Fig_retrofit_2}h and k.}
        \label{TableS_commercialgrippers}
\end{table}

\begin{table}[h]
    \centering
    \resizebox{120mm}{!}{\includegraphics{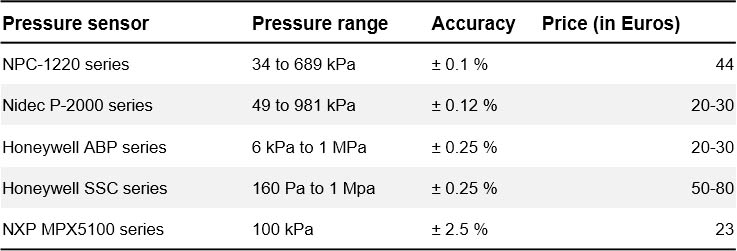}}
    \caption{\textbf{Comparison of selected commercial analog pressure sensors.} Prices are according to Mouser Electronics in 2023.}
    \label{TableS_pressuresensors}
\end{table}

\begin{table}[h]
    \centering
    \resizebox{110mm}{!}{\includegraphics{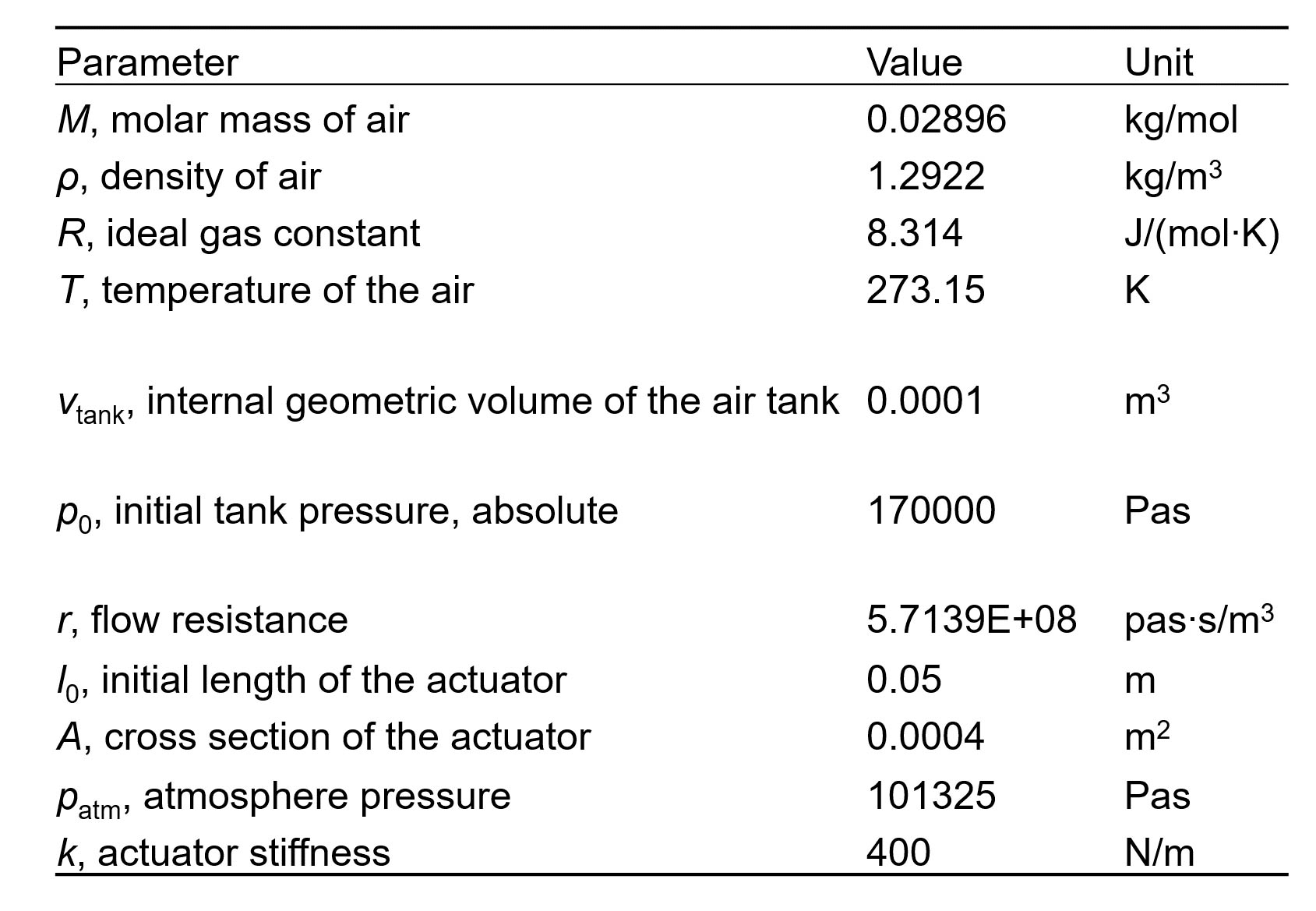}}
    \caption{\textbf{Parameter values used in the fluidic sensing model.}}
    \label{Table_modelparameters}
\end{table}

\begin{figure}[h]
    \centering
    \resizebox{160mm}{!}{\includegraphics{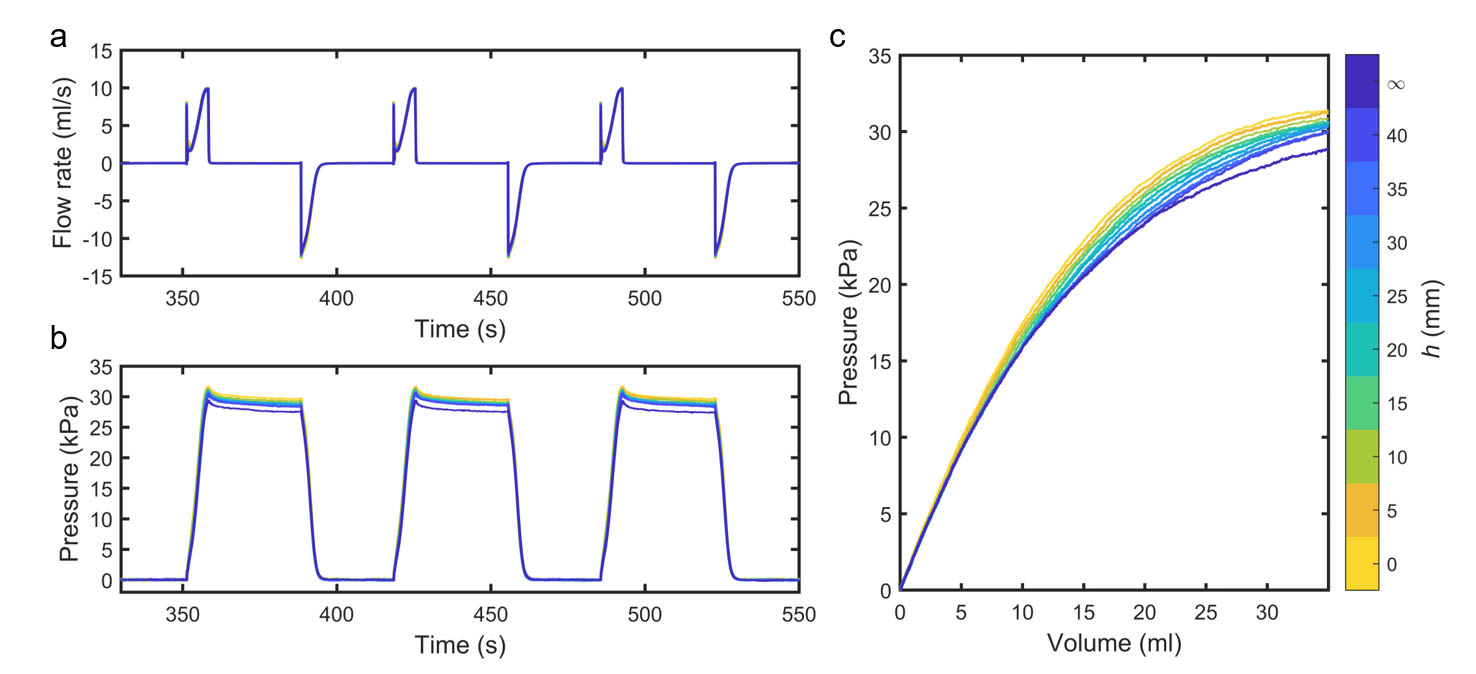}}
    \caption{\textbf{Characterization of pressure-volume curves of the soft PneuNet actuator bending onto a rigid plate from different heights \textit{h} through controlled flow input.} \textbf{a,} Flow rate of the flow in (positive) and out (negative) of the PneuNet actuator during the last three actuation cycles. \textbf{b,} Pressure of the PneuNet actuator during the last three actuation cycles. \textbf{c,} Pressure-volume curves of the PneuNet actuator from the last actuation cycle.}
    \label{FigS_pv_raw}
\end{figure}

\begin{figure}[h]
    \centering
    \resizebox{140mm}{!}{\includegraphics{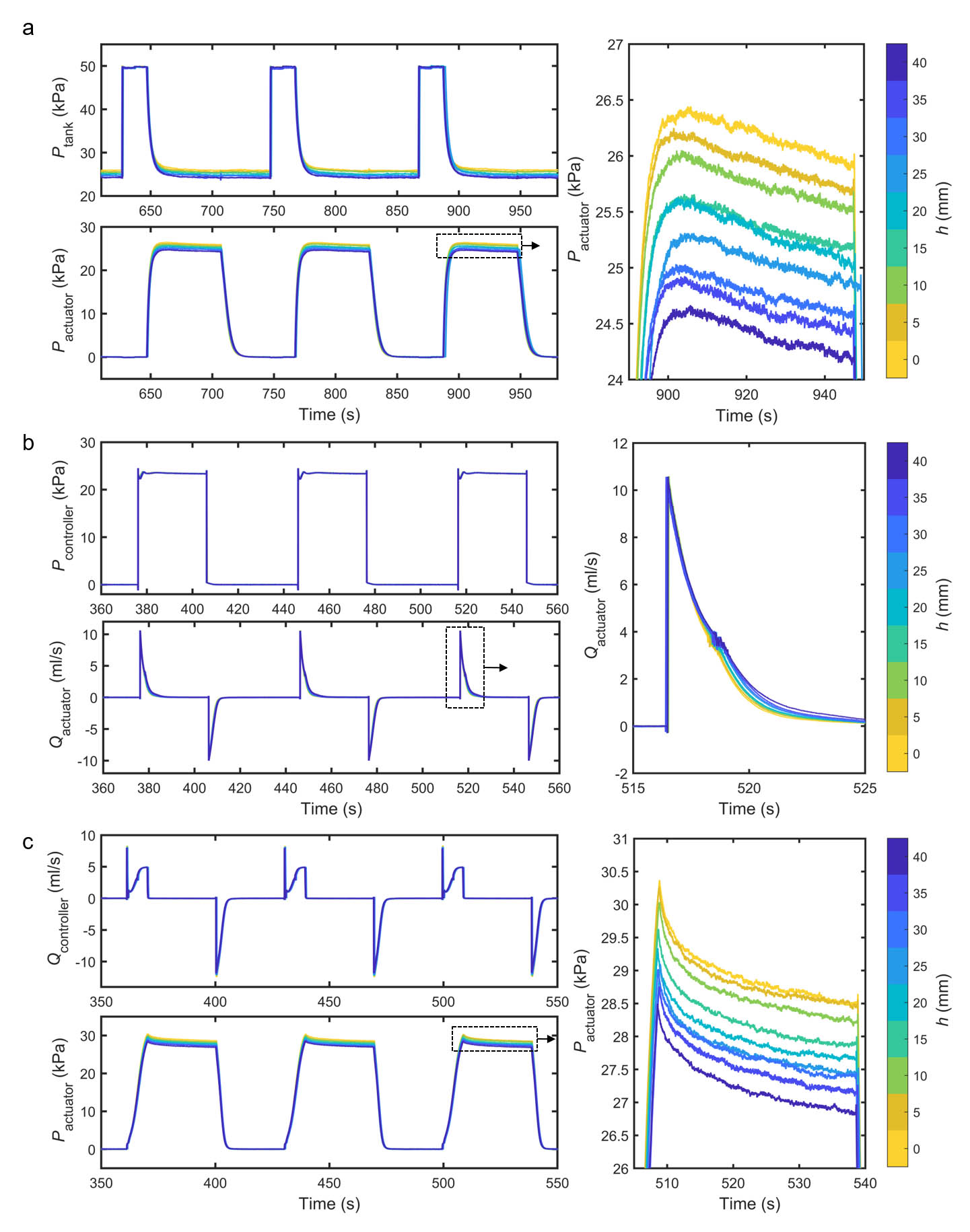}}
    \caption{\textbf{Experimental data for the fluidic response calibrations with the three fluidic sensing mehtods.} \textbf{a,} Control the total mass using a pressurized air tank and measure the pressure response. \textbf{b,} Control the pressure and measure the volume flow input. \textbf{c,} Control the volume flow input and measure the pressure. Positive and negative flow rate values represent the flow in and out of the actuator, respectively. For each method, a total of eight actuation cycles were performed at each \textit{h} and measurements from the last three cycles are shown here.}
    \label{FigS_pt_raw}
\end{figure}

\begin{figure}[h]
    \centering
    \resizebox{160mm}{!}{\includegraphics{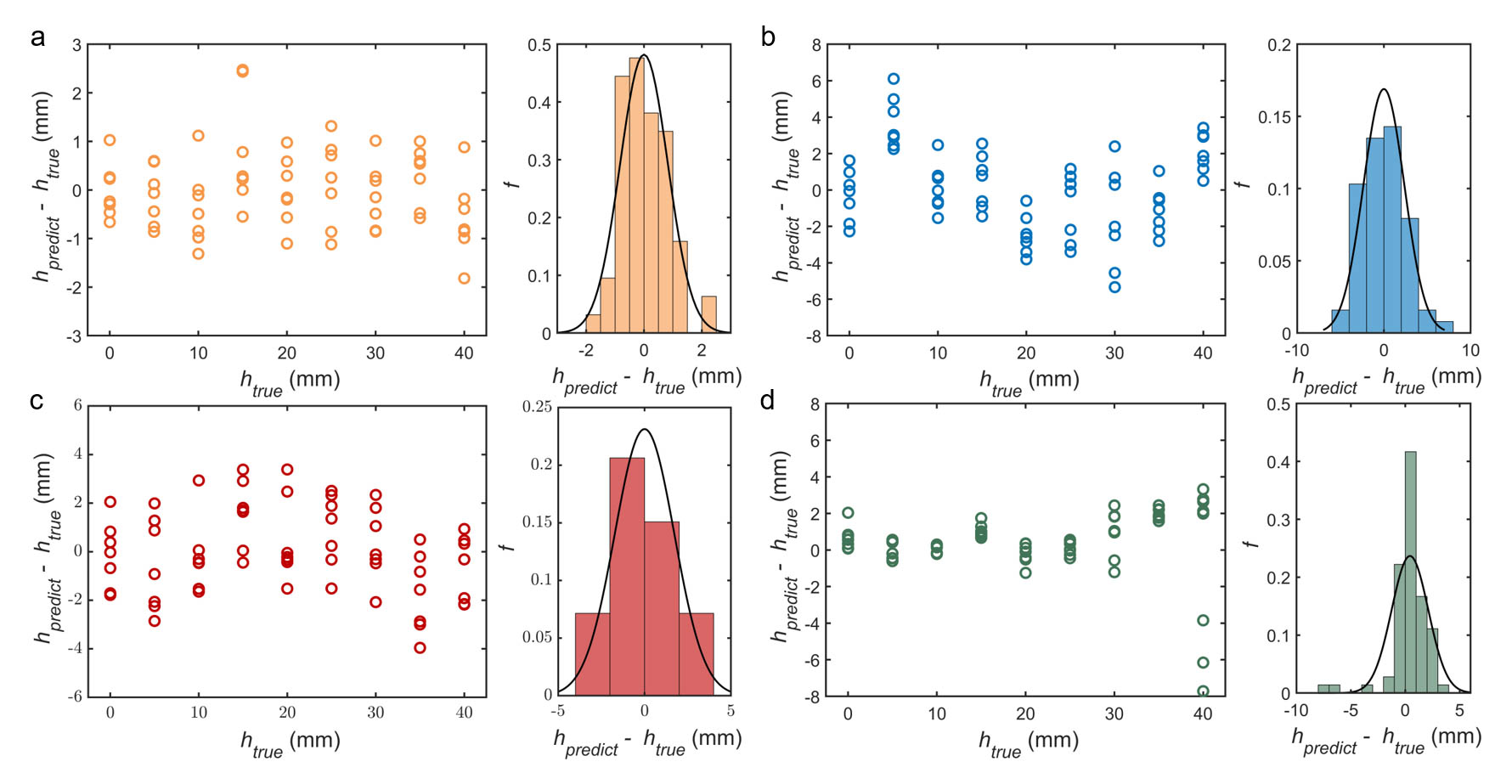}}
    \caption{\textbf{Accuracy of the various sensing approaches.} The sensing accuracy is determined by the difference between the ground truth (set distance $h$) and the predicted value of $h$ based on the calibration curves. The probability density function estimate $f$ is shown in histograms with a normal distribution. \textbf{a,} Accuracy obtained when controlling the total mass using a pressurized air tank and measuring the pressure response. The $h_{\mathrm{predict}}$ is calculated from the calibration curve in Fig.~\ref{Fig1}b at given pressure measurements. The sensing accuracy of this method is $\pm{1.7}$ mm with a 95\% confidence interval. \textbf{b,} Accuracy obtained when controlling the pressure and measuring the volume flow input. The $h_{\mathrm{predict}}$ is calculated from the calibration curve in Fig.~\ref{Fig1}c at given volume measurements. The sensing accuracy of this method is $\pm{4.7}$ mm with a 95\% confidence interval. \textbf{c,} Accuracy obtained when controlling the volume flow input and measuring the pressure. The $h_{\mathrm{predict}}$ is calculated from the calibration curve in Fig.~\ref{Fig1}d at given pressure measurements. The sensing accuracy of this method is $\pm{3.4}$ mm with a 95\% confidence interval. \textbf{d,} Accuracy obtained when controlling the total mass using a pressurized air tank and measure the pressure response over time. The $h_{\mathrm{predict}}$ is calculated based on the interpolation of vertical displacement curve of actuator in the case of free actuation in Fig.~\ref{Fig2}c at given time of contact measurements. The sensing accuracy of this method is -2.9 to 3.8 mm with a 95\% confidence interval.}
    \label{FigS_sensing_accuarcy}
\end{figure}

\begin{figure}[h]
    \centering
    \resizebox{160mm}{!}{\includegraphics{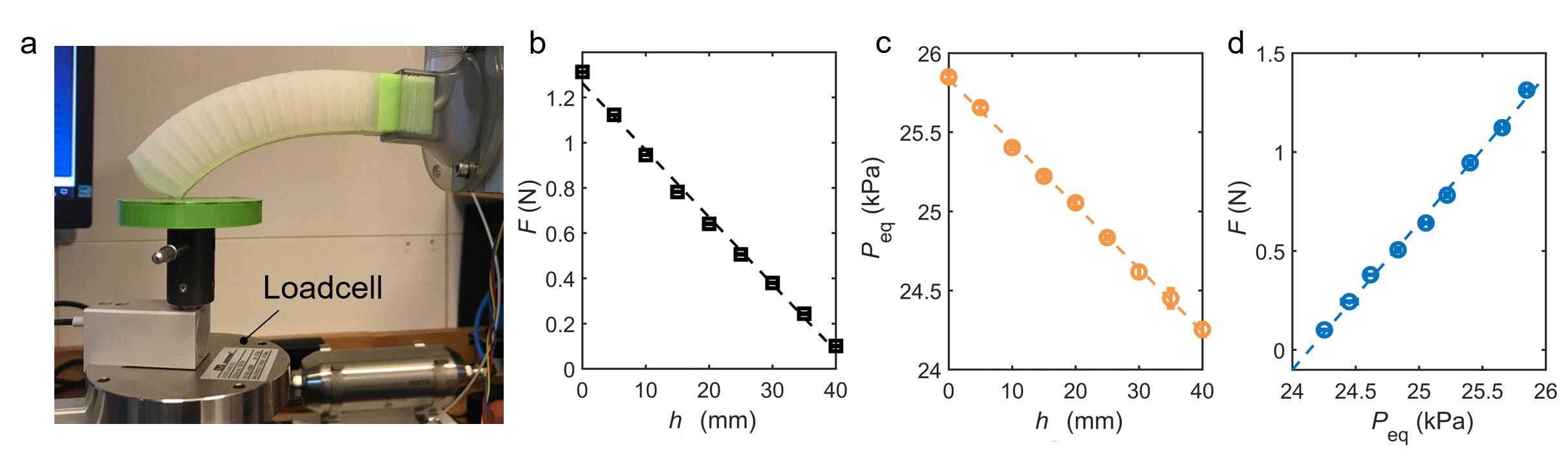}}
    \caption{\textbf{Force characterization.} \textbf{a,} Upon pressurization, the soft PneuNet actuator bends onto a load cell ($\pm$ 100 N, 132151, Instron) from a height $h$. \textbf{b,} Force measured for different initial $h$. \textbf{c,} Equilibrium pressure observed for different initial $h$. \textbf{d,} The obtained force-equilibrium pressure calibration curve. The dashed lines in \textbf{b-d,} represent a linear fitting. The error bars represent the standard deviation of four measurements.}
    \label{FigS_force_calibration}
\end{figure}

\begin{figure}[h]
    \centering
    \resizebox{160mm}{!}{\includegraphics{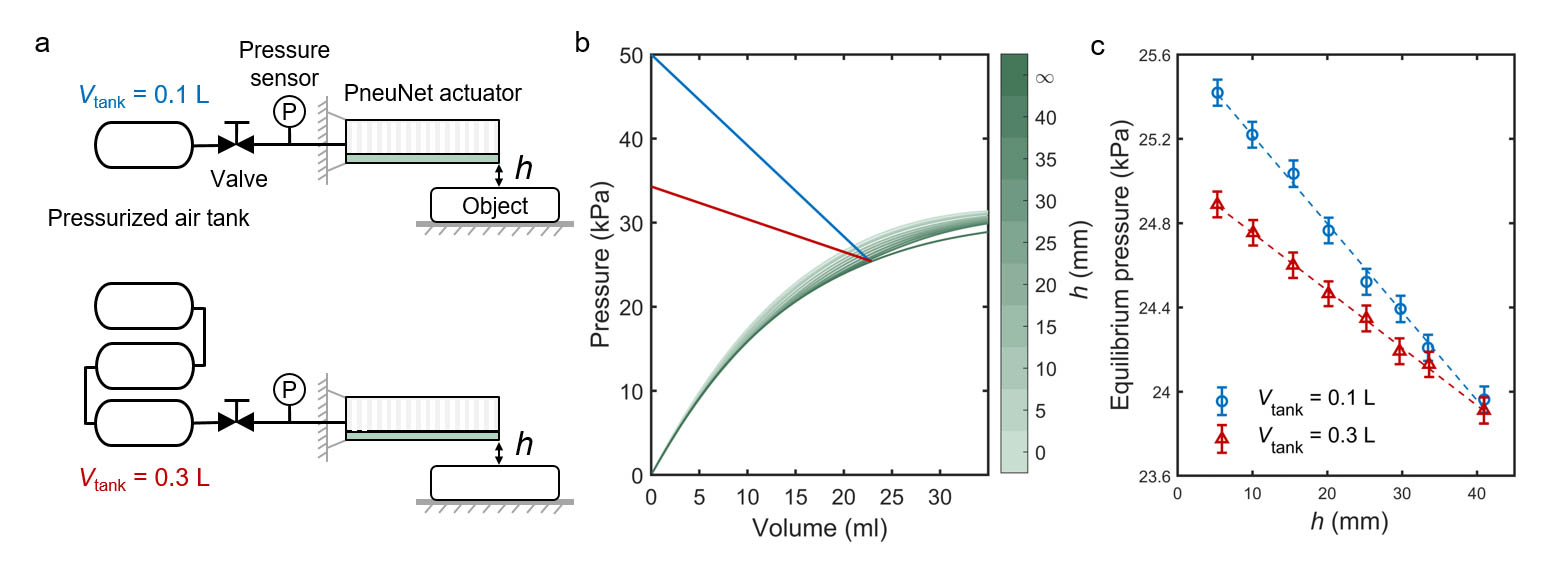}}
    \caption{\textbf{Tuning the sensing resolution by varying the air tank volume.} \textbf{a,} Schematics of the setup with 0.1 L and 0.3 L steel air tanks. \textbf{b,} Schematic of the intersections of pressure-volume curves between the actuator and air tanks. The blue and red lines represent the pressure-volume curves of the 0.1 L and 0.3 L air tanks, respectively.  \textbf{c,} \textit{h}-pressure calibration curves with 0.1 L and 0.3 L steel air tanks. One actuation cycle was performed at each \textit{h}, in which the actuator was inflated for 60s. The error bars represent the standard deviation of the non-smoothed pressure sensor (MPX5100DP, NXP) measurements (1000Hz) within the last 5s.}
    \label{FigS_resolution_2caps}
\end{figure}

\begin{figure}[h]
    \centering
    \resizebox{160mm}{!}{\includegraphics{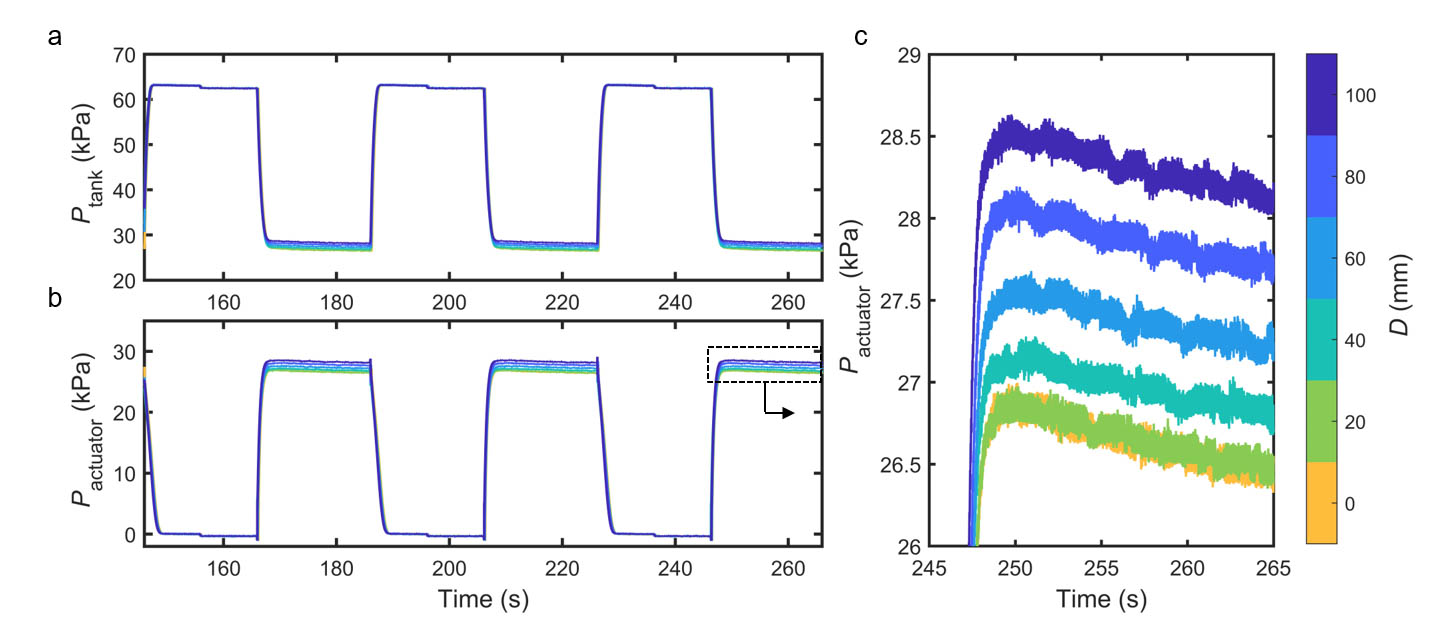}}
    \caption{\textbf{Experimental data for the \textit{D}-pressure calibrations of the soft gripper.} \textbf{a,} Pressure of the air tank during the last three actuation cycles. \textbf{b,} Pressure of the PneuNet actuator during the last three actuation cycles. \textbf{c,} Pressure of the PneuNet actuator during the last actuation cycle. The pressure-time curve at $D =$ 20 mm almost overlaps with the curve at $D =$ 0 mm because the gripper has a lower grasping limit slightly larger than 20 mm and the actuators are barely touching the object with $D =$ 20 mm. }
    \label{FigS_gripper_pt_raw}
\end{figure}

\begin{figure}[h]
    \centering
    \resizebox{160mm}{!}{\includegraphics{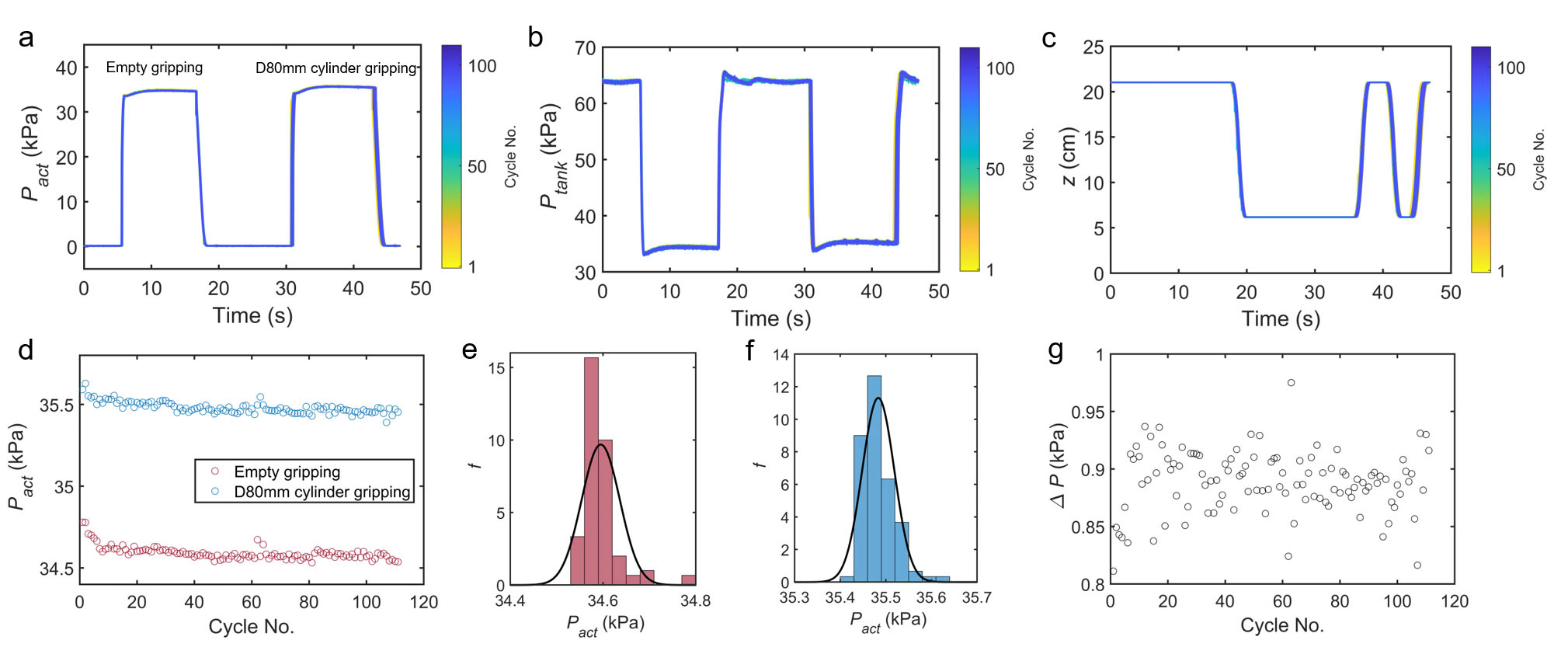}}
    \caption{\textbf{Cyclic testing of the sensing strategy.} We tested a total of 111 cycles of the soft PneuNet gripper (Fig.~\ref{Fig1}e) on the UR5 robotic arm. Each cycle contains two gripping events, the first one with nothing inside the gripper, the second one with a 80 mm diameter cylinder inside the gripper. In the second gripping event, after the gripper grasps the object, the robotic arm first moves up and holds the object in the air, then moves down and releases the object before going to the next cycle. For both gripping events, the equilibrium pressure inside the actuator is averaged between 8 s and 9 s after the opening of the valve, which corresponds to the time when the object was held in the air during the second gripping event. \textbf{a,} Pressure-time measurements of the actuator from 111 cycles superimposed on top of each other. \textbf{b,} Pressure-time measurements of the tank from 111 cycles superimposed on top of each other. \textbf{c,} Measurements of the TCP Z coordinate of the robotic arm from 111 cycles superimposed on top of each other. \textbf{d,} Equilibrium pressure of the two gripping events for all 111 cycles. \textbf{e,} Distribution of the equilibrium pressure of the first gripping event over 100 cycles. The mean value is $34.5951 \pm 0.0824$ kPa (95\% confidence interval). \textbf{f,} Distribution of the equilibrium pressure of the second gripping event over 100 cycles. The mean value is $35.4834 \pm 0.0706$ kPa (95\% confidence interval). Note that the soft gripper tested here is a different version from the one tested in Fig.~\ref{Fig1} because the original one was accidentally damaged. We slightly changed the actuator design, resulting in a higher actuation pressure. \textbf{g,} The pressure difference $\Delta P = P_{act\_D80} - P_{act\_empty}$ between the two gripping events over 111 cycles.}
    \label{FigS_cycling}
\end{figure}

\begin{figure}[h]
    \centering
    \resizebox{150mm}{!}{\includegraphics{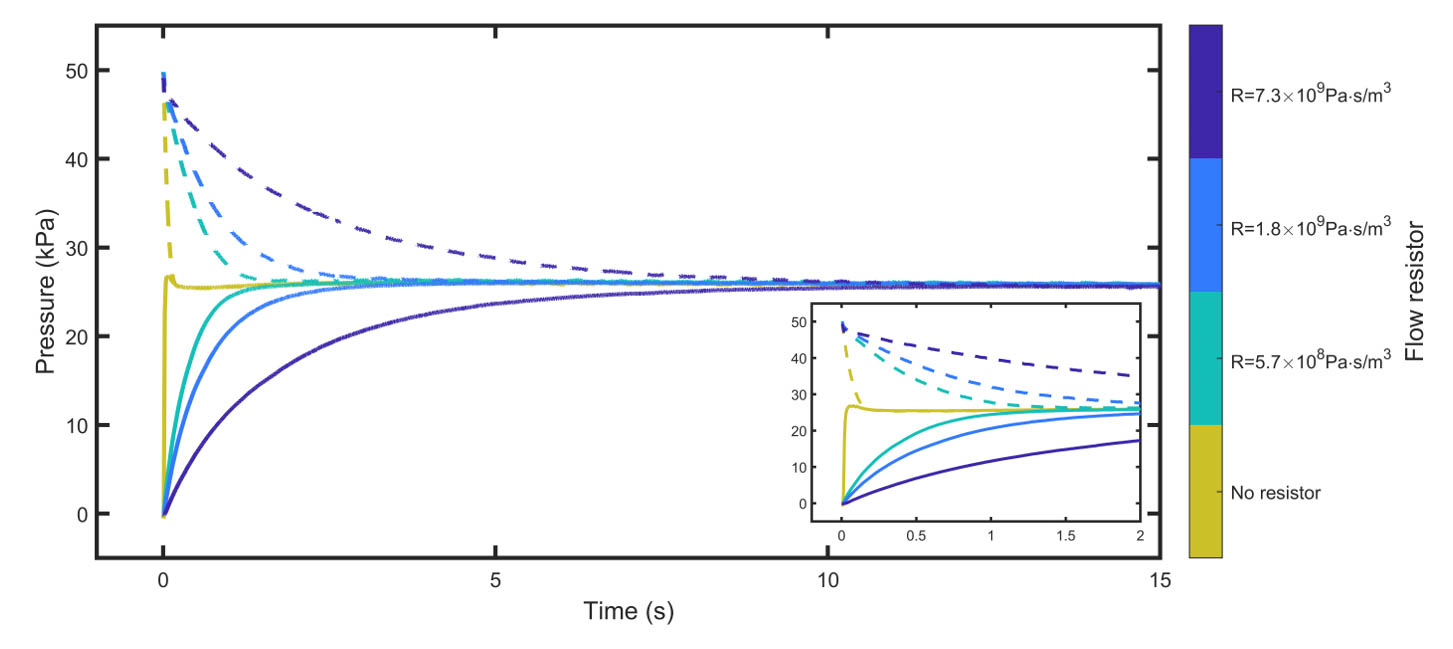}}
    \caption{\textbf{Tuning the response time with different flow resistors in between the valve and actuator.} The dashed curves represent the pressure in the 0.1 L air tank. The solid curves represent the pressure in the PneuNet actuator. The initial distance $h$ between the actuator and rigid plate was set at 20 mm. Tests were done by switching the flow resistor shown in Fig.~\ref{FigS_setup_masscontrol}. The fluidic circuit still has some flow resistances due to the valve and tube connections, even when no flow resistor was added to the circuit. Therefore, the response time can be further improved by reducing these flow resistances in the circuit.}
    \label{FigS_responsetime}
\end{figure}

\begin{figure}[h]
    \centering
    \resizebox{160mm}{!}{\includegraphics{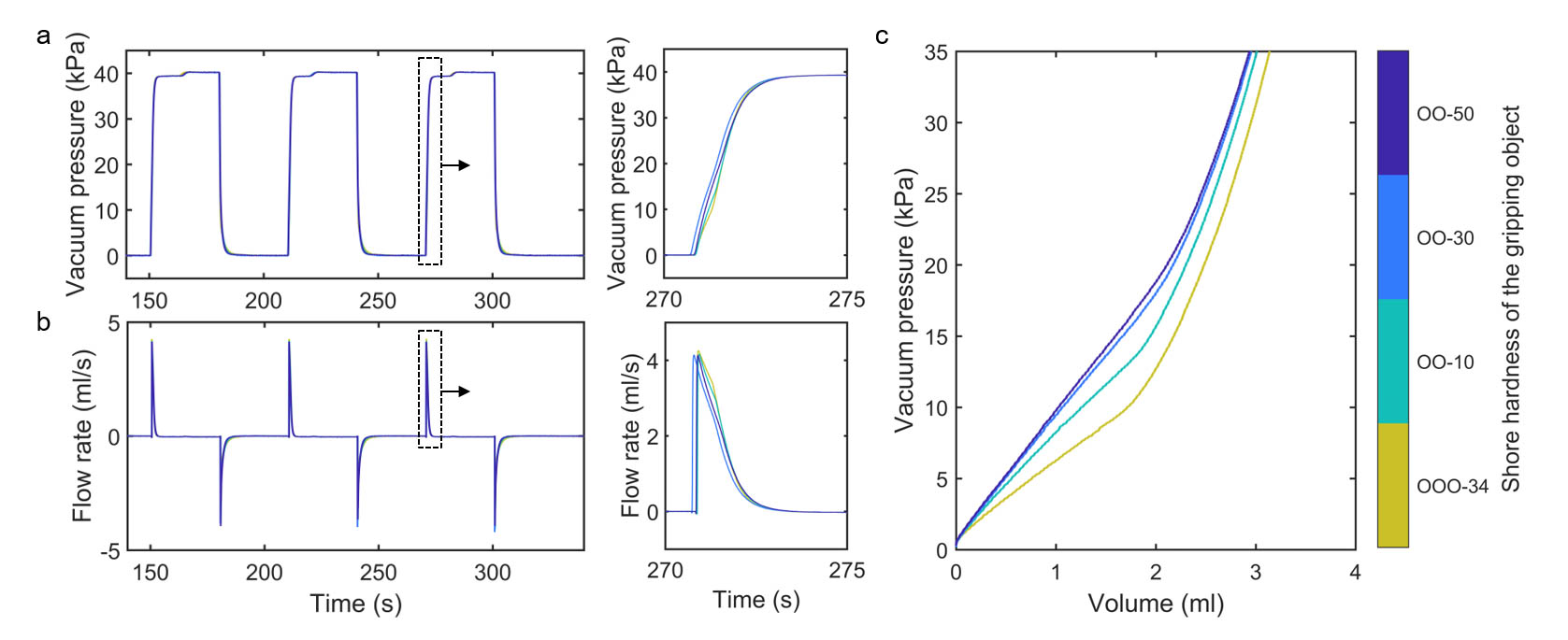}}
    \caption{\textbf{Characterization of pressure-volume curves of the suction gripper attaching to soft objects with different shore hardness through controlled pressure input.} \textbf{a,} Pressure of the suction gripper during the last three actuation cycles. \textbf{b,} Flow rate of the flow going out of (positive) and into (negative) the suction gripper during the last three actuation cycles.  \textbf{c,} Pressure-volume curves of the suction gripper from the last actuation cycle.}
    \label{FigS_suctioncup_pv}
\end{figure}

\begin{figure}[h]
    \centering
    \resizebox{160mm}{!}{\includegraphics{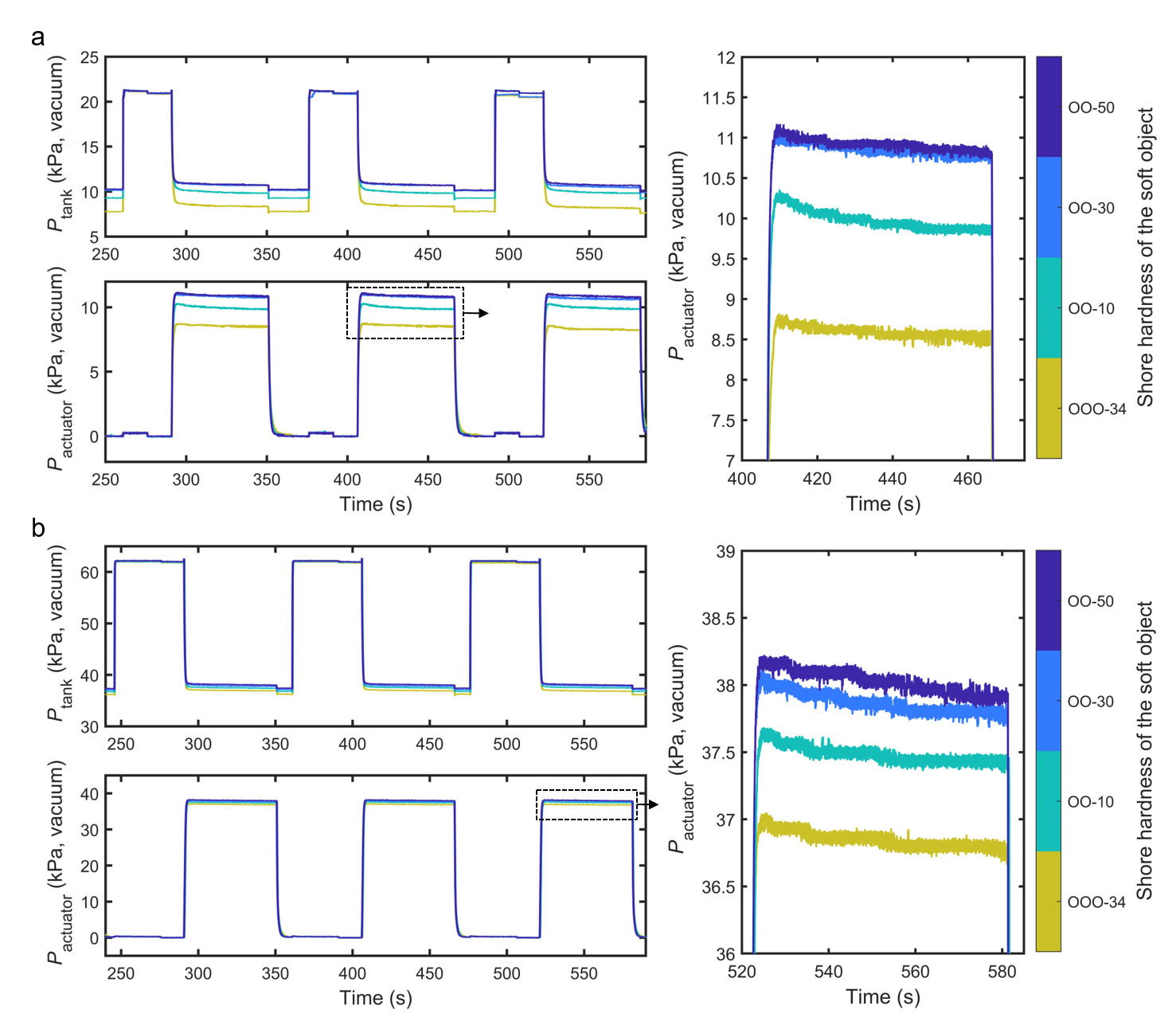}}
    \caption{\textbf{Experimental data for the stiffness-pressure calibrations of the suction gripper with two different vacuum pressures in the steel air tank.} \textbf{a,} $P_{\mathrm{0}}$ = 21 kPa (vacuum). \textbf{b,} $P_{\mathrm{0}}$ = 62 kPa (vacuum).}
    \label{FigS_suctioncup_softness_pt}
\end{figure}

\begin{figure}[h]
    \centering
    \resizebox{160mm}{!}{\includegraphics{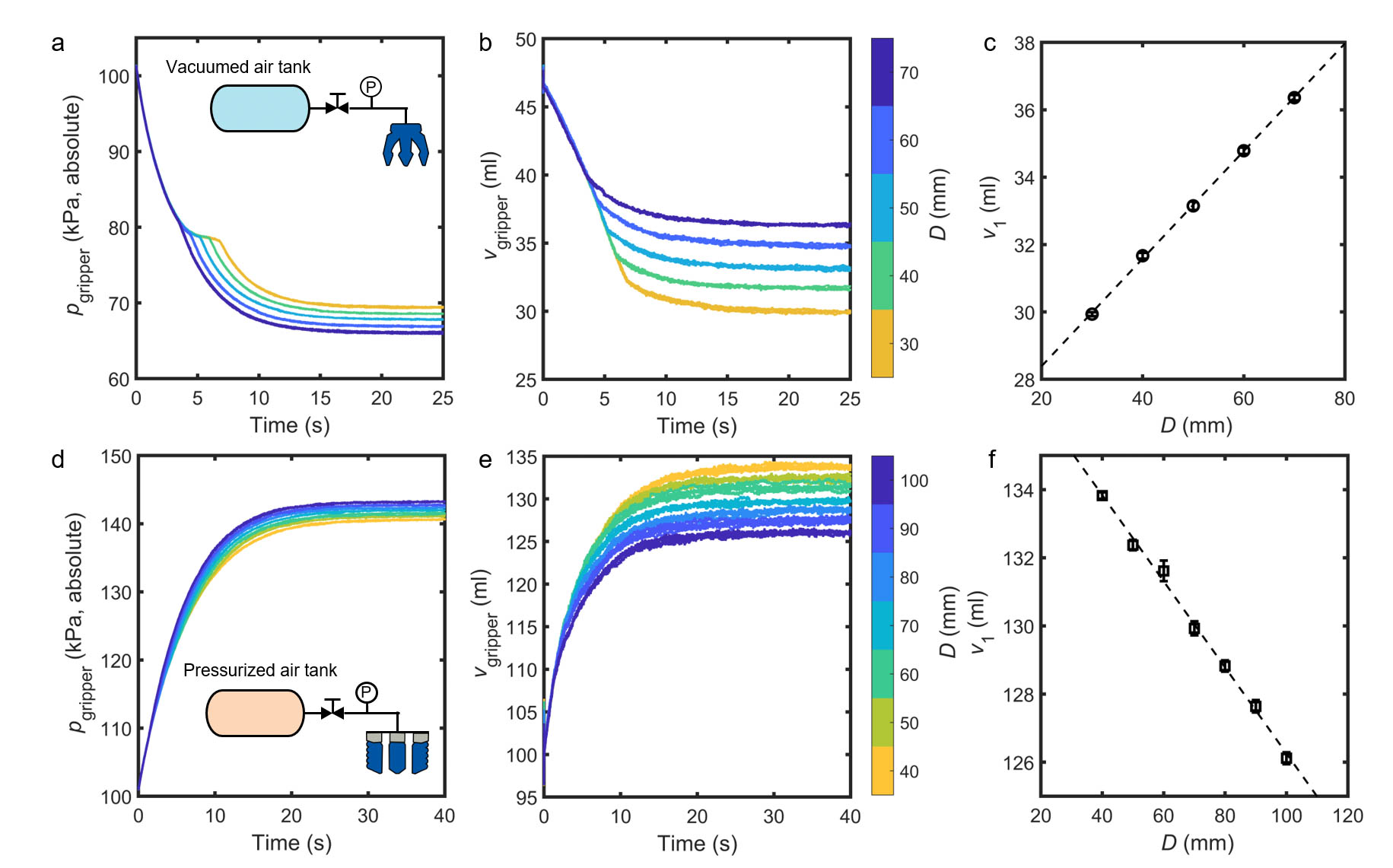}}
    \caption{\textbf{Internal volume estimation of the commercial grippers when gripping cylindrical objects with different diameters.} The dashed lines in \textbf{c and f} represent linear fits of the data.}
    \label{FigS_commercialgrippers}
\end{figure}

\begin{figure}[h]
\centering
\resizebox{160mm}{!}{\includegraphics{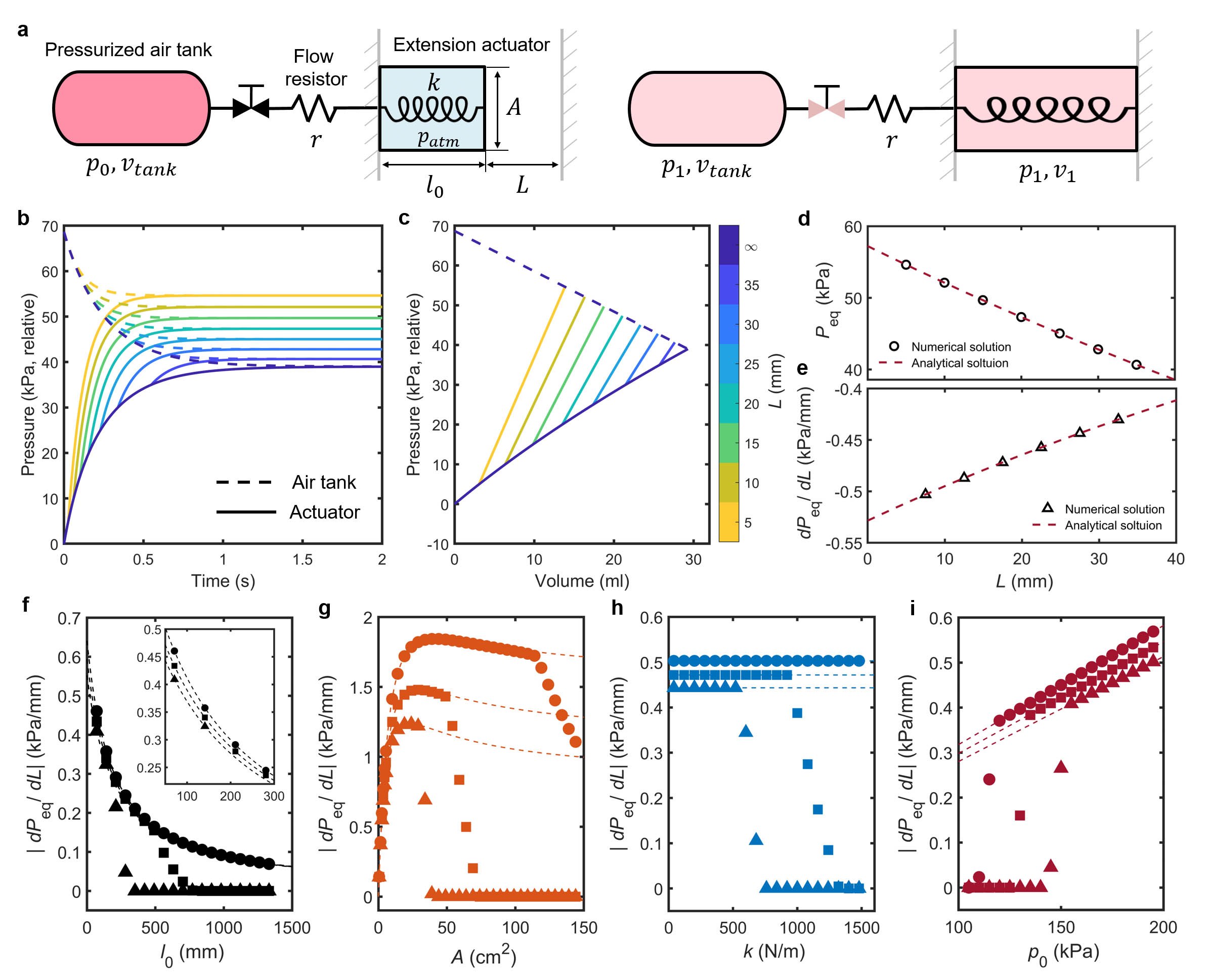}}
\caption{\textbf{Modeling the fluidic sensing approach. a,} Schematics of a simplified sensing scenario: an extension actuator with a linear stiffness $k$ comes into contact with a rigid wall from a distance $L$. \textbf{b-e,} Simulation results using the input values shown in Table S3: pressure-time (\textbf{b}) and pressure-volume (\textbf{c}) curves of the air tank and actuator; calibration curve of $P_{\mathrm{eq}}$ over $L$ (\textbf{d}); sensing resolution curve of $dP_{\mathrm{eq}}/dL$ over $L$ (\textbf{e}). \textbf{f-i} Simulation results of the effects of actuator's initial length $l_{\mathrm{0}}$ (\textbf{f}), cross section $A$ (\textbf{g}), stiffness $k$ (\textbf{h}) and initial tank pressure $p_{\mathrm{0}}$ (\textbf{i}) on the sensing resolution magnitude $|dP_{\mathrm{eq}}/dL|$. The makers and dashed lines in \textbf{f-i} represent numerical simulation results and analytical solutions, respectively. The circular, square and triangular markers in \textbf{f-i} represent $L = 7.5, 17.5, 27.5$ mm, respectively. The results in \textbf{f-i} are based on the input values in Table S3 except that an initial tank pressure of 300 kPa (absolute pressure) is used in \textbf{g}.}
\label{FigS_modelV4}
\end{figure}

\begin{figure}[h]
    \centering
    \resizebox{160mm}{!}{\includegraphics{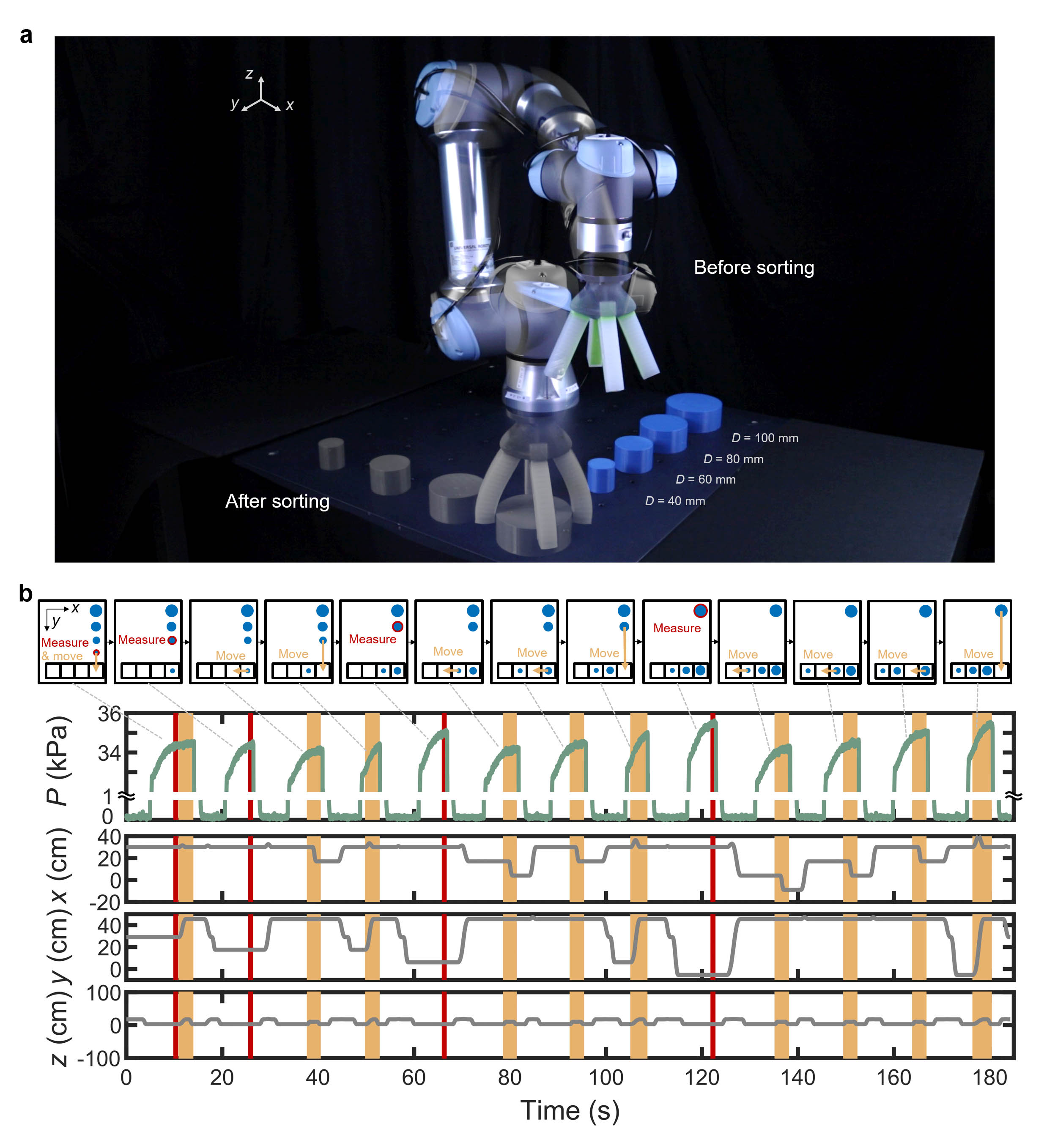}}
    \caption{\textbf{Sorting experiment from Fig.~\ref{Fig4}a with a different input order of cylindrical objects with a diameter of 40 mm, 60 mm, 80 mm, 100 mm.}}
    \label{FigS_sorting_1234}
\end{figure}

\begin{figure}[h]
    \centering
    \resizebox{160mm}{!}{\includegraphics{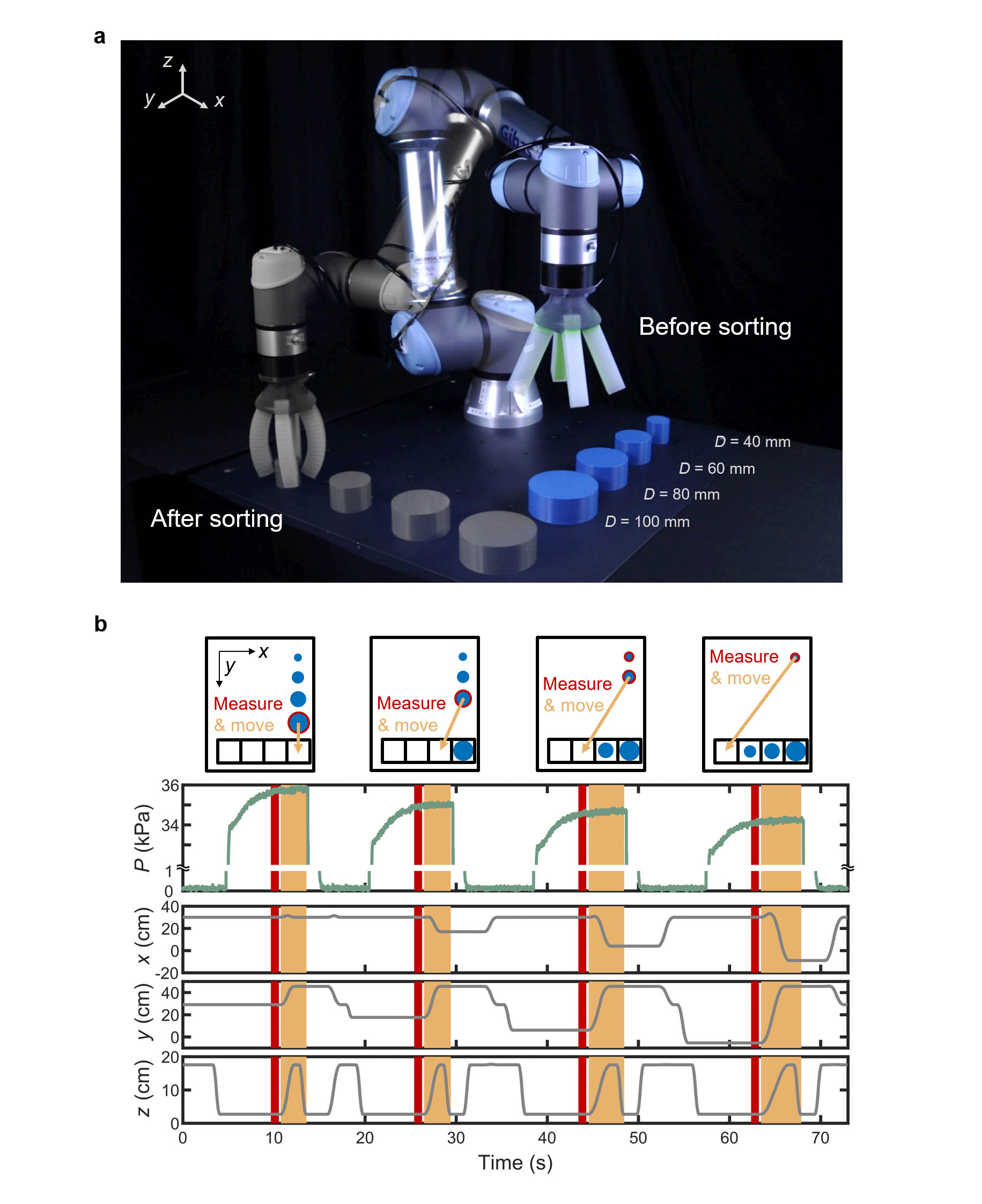}}
    \caption{\textbf{Sorting experiment from Fig.~\ref{Fig4}a with a different input order of cylindrical objects with a diameter of 100 mm, 80 mm, 60 mm, 40 mm.}}
    \label{FigS_sorting_4321}
\end{figure}

\begin{figure}[h]
    \centering
    \resizebox{150mm}{!}{\includegraphics{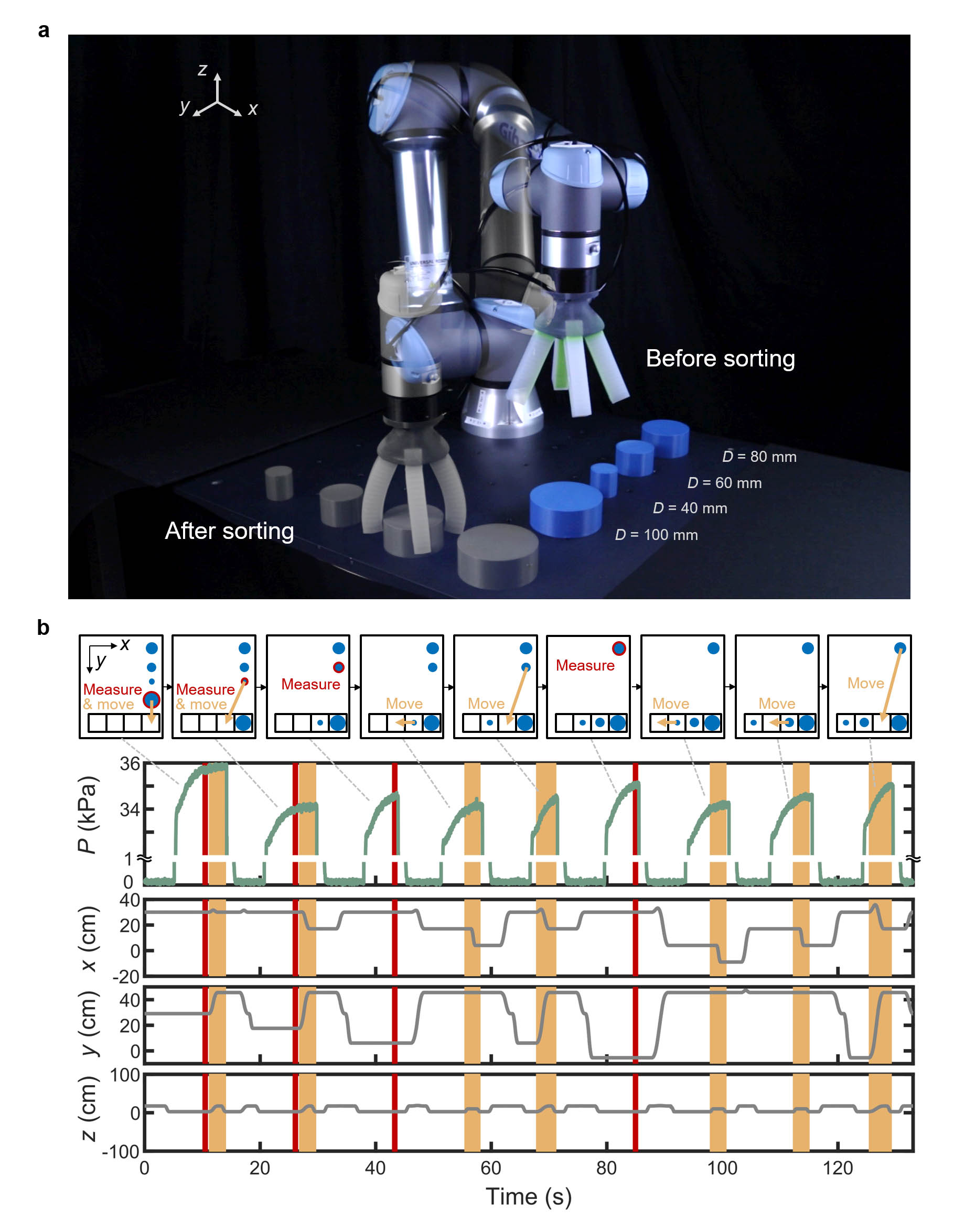}}
    \caption{\textbf{Sorting experiment from Fig.~\ref{Fig4}a with a different input order of cylindrical objects with a diameter of 100 mm, 40 mm, 60 mm, 80 mm.}}
    \label{FigS_sorting_4123}
\end{figure}

\begin{figure}[h]
    \centering
    \resizebox{150mm}{!}{\includegraphics{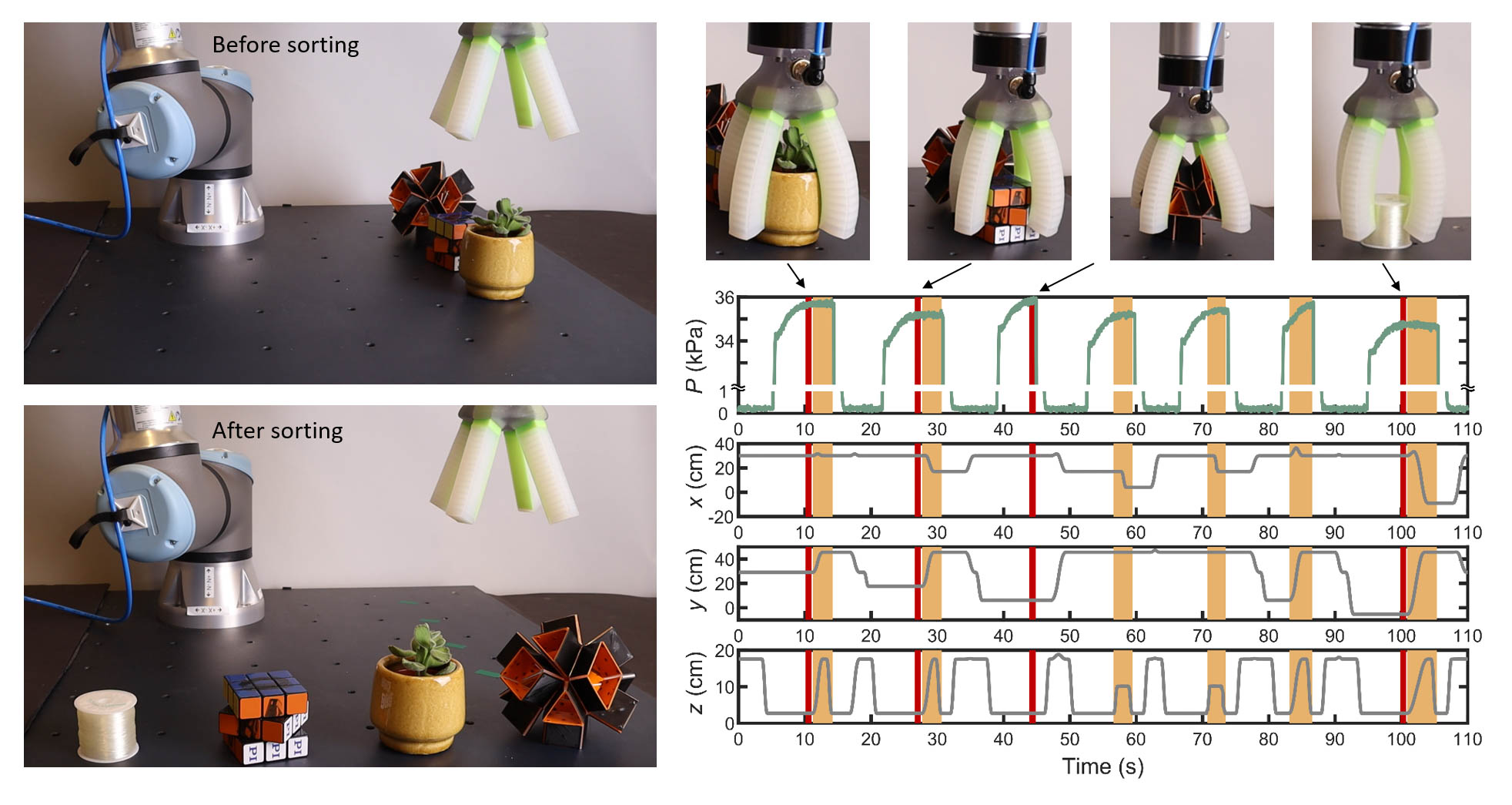}}
    \caption{\textbf{Sorting experiment with random objects: 1) Small plant. 2) Magic cube. 3) Prismatic metamaterial (\textit{2}). 4) Spool.}}
    \label{FigS_sorting_random1}
\end{figure}

\begin{figure}[h]
    \centering
    \resizebox{150mm}{!}{\includegraphics{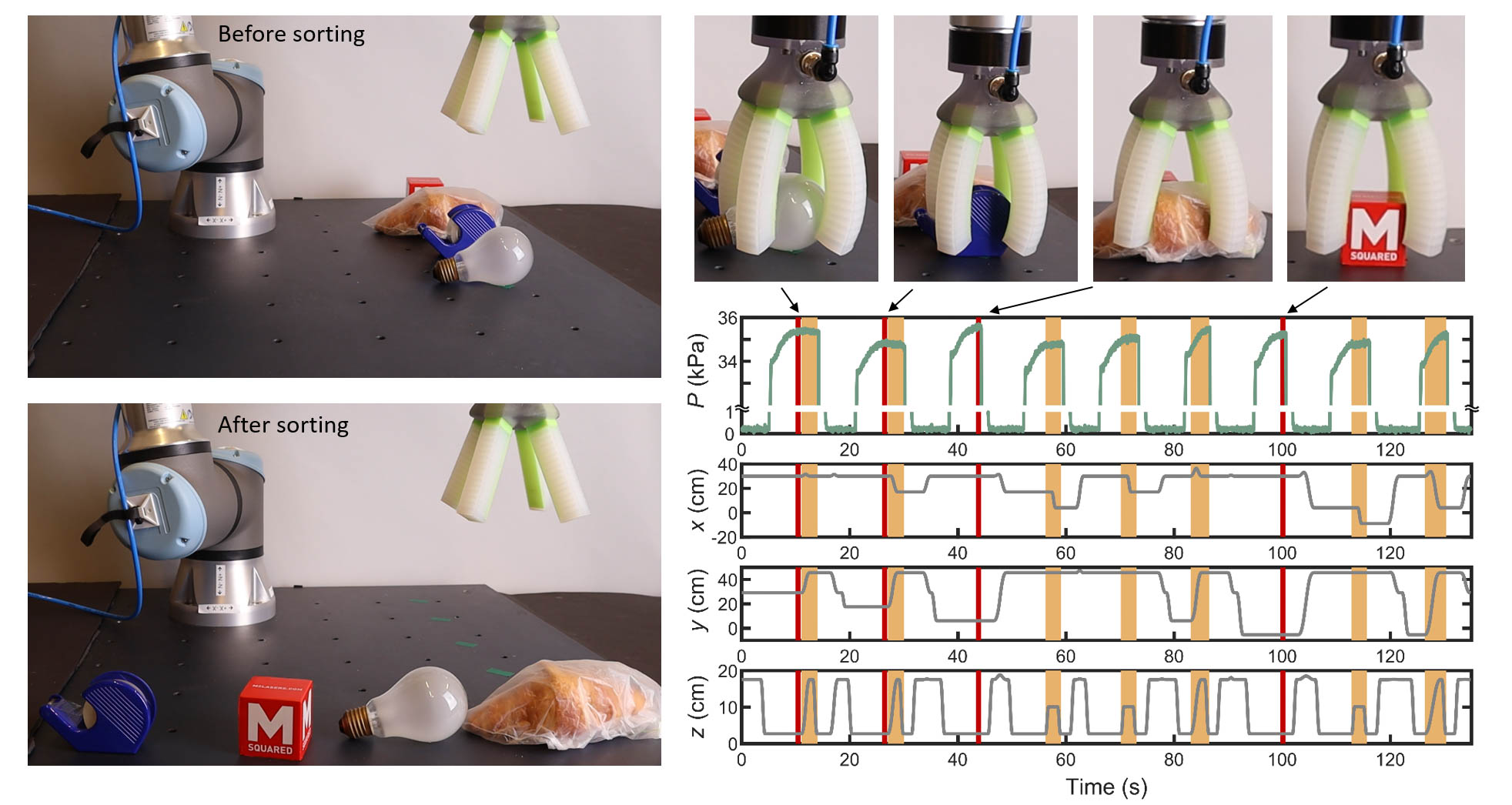}}
    \caption{\textbf{Sorting experiment with random objects: 1) Light bulb. 2) Tape cutter. 3) Croissant. 4) Stress relief cube.}}
    \label{FigS_sorting_random2}
\end{figure}

\begin{figure}[h]
    \centering
    \resizebox{160mm}{!}{\includegraphics{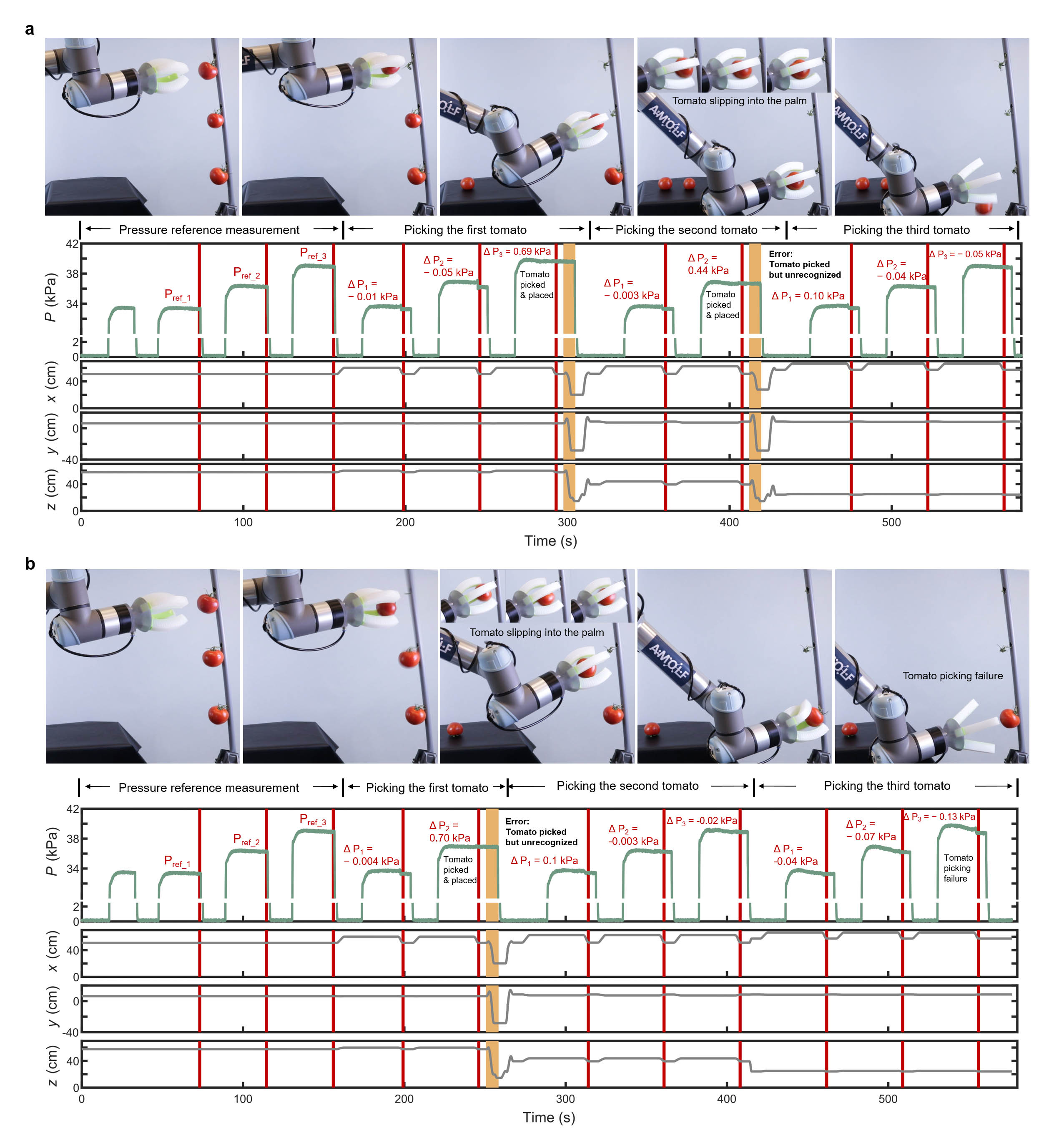}}
    \caption{\textbf{Snapshots of the tomato
picking experiments and gripper pressure and TCP coordinates over time during the picking experiments} (Demo 2 and 3 in Supplementary Video 5). The red band represents pressure feedback measurement, and the yellow band represents tomato placement. During the picking of the third tomato in \textbf{a} and the second tomato in \textbf{b}, the tomato slipped into the palm of the soft gripper, resulting in $\Delta P = 0.1$ kPa which is smaller than the threshold (0.2 kPa). The tomato picking was not recognized and the gripper dropped the tomato and started a new picking attempt. The gripper failed to pick the third tomato in \textbf{b}.}
    \label{FigS_tomatofailure}
\end{figure}

\begin{figure}[h]
    \centering
    \resizebox{160mm}{!}{\includegraphics{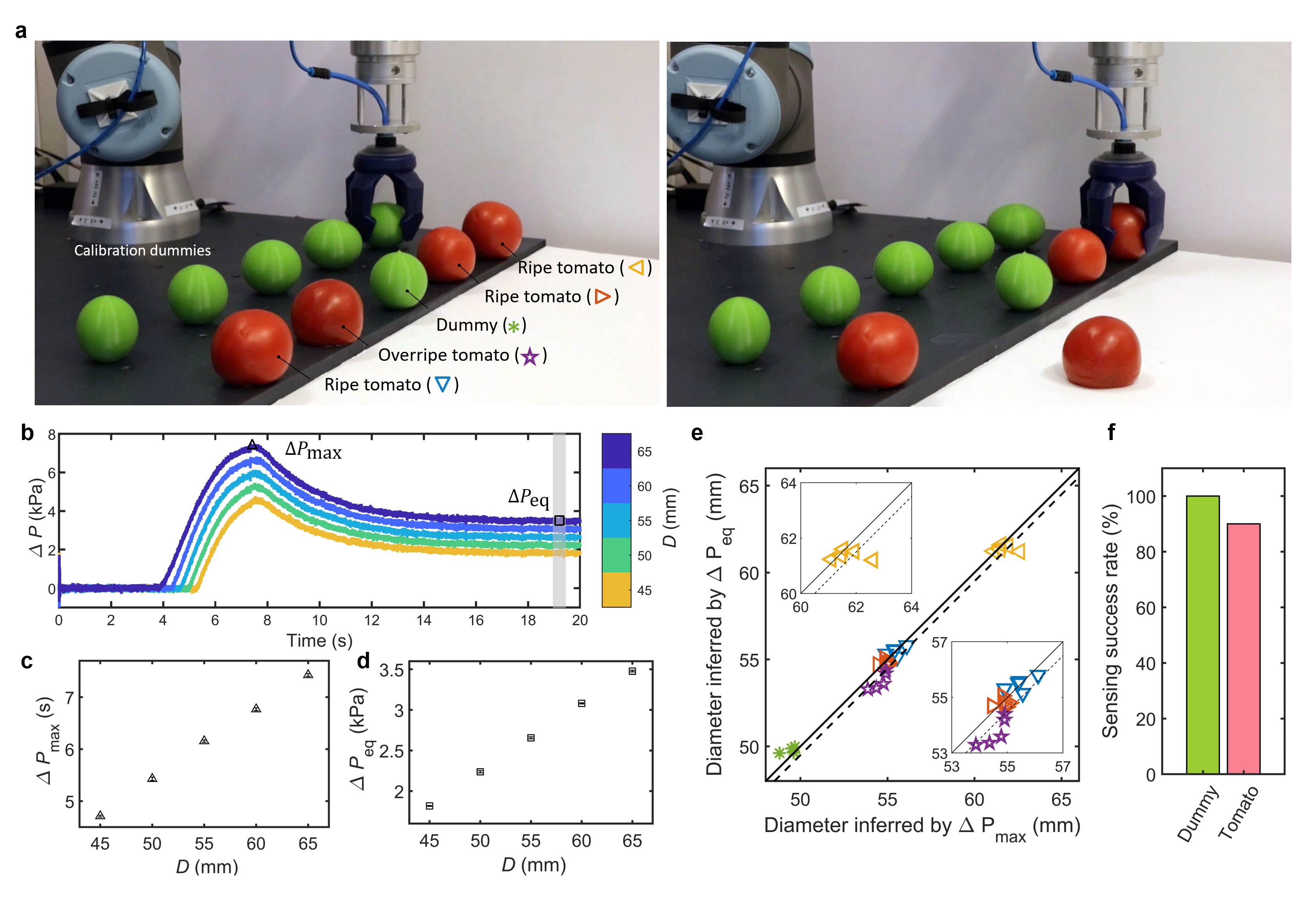}}
    \caption{\textbf{Picking out the overripe tomato with fluidic sensing based on calibrations with $\Delta P_{\mathrm{max}}$ and $\Delta P_{\mathrm{eq}}$.} \textbf{a,} Snapshots of the closed-loop control demonstration (Demo 2 in Supplementary Video 6). \textbf{b-d,} Calibration results. Experimental results from three measurements are plotted for each $D$ in \textbf{b}. The error bars in \textbf{c,d} represent the standard deviation of three measurements. \textbf{f-m,} Sensing results. The solid line in \textbf{e} represents that the object diameter inferred by $\Delta P_{\mathrm{max}}$ equals that inferred by $\Delta P_{\mathrm{eq}}$, the dashed line represents the object diameter inferred by $\Delta P_{\mathrm{eq}}$ is 0.5 mm smaller than that inferred by $\Delta P_{\mathrm{max}}$.}
    \label{FigS_tomato_ripeness_DPMAX}
\end{figure}

\begin{figure}[h]
    \centering
    \resizebox{150mm}{!}{\includegraphics{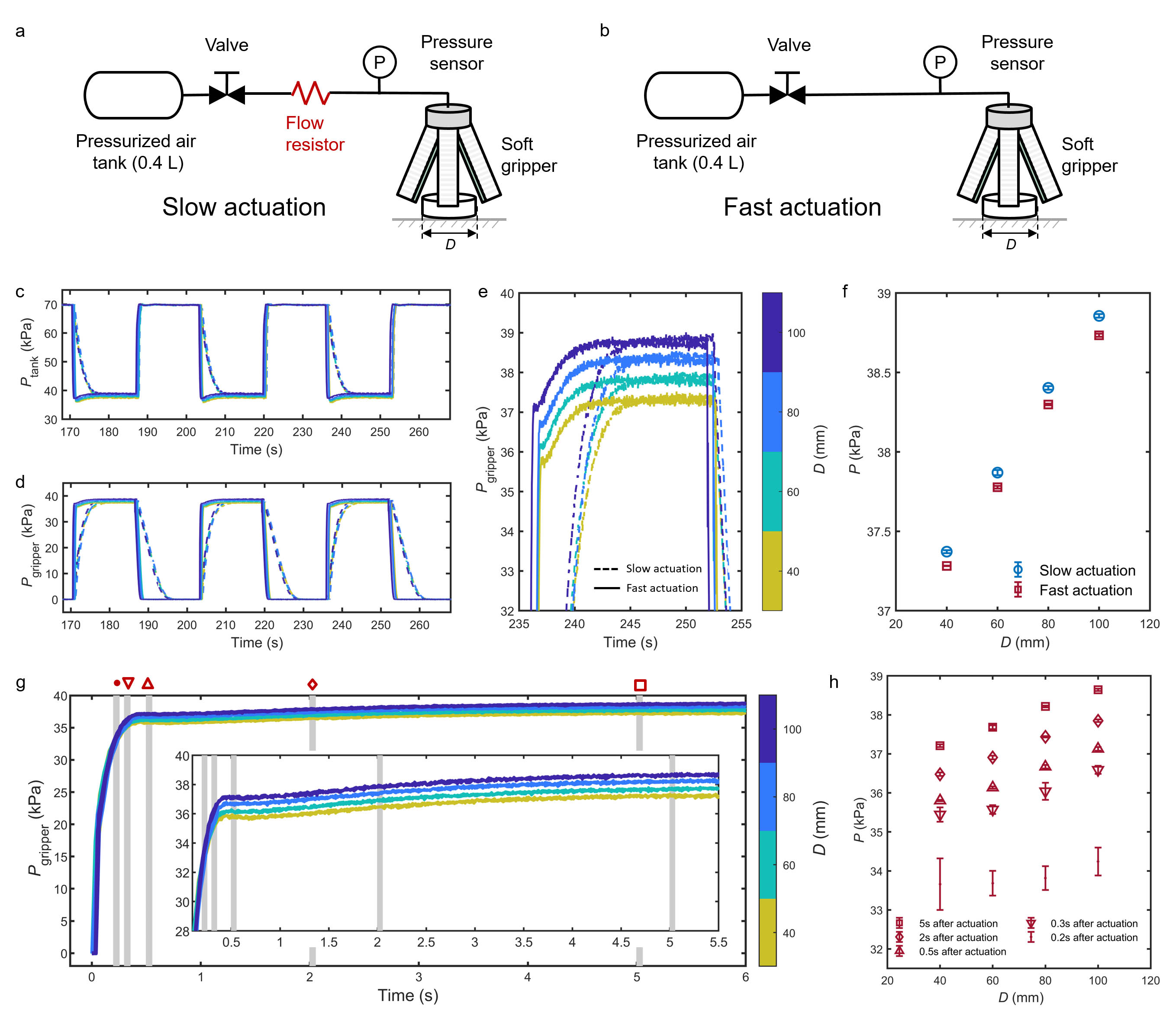}}
    \caption{\textbf{Influence of the actuation rate on the \textit{D}-pressure calibrations of the soft gripper.} \textbf{a, b,} We vary the actuation rate by placing a flow resistor ($R = 5.7 \times 10^8 Pa \cdot s/m^3$) between the solenoid valve and the gripper. \textbf{c-e,} Experimental results of the soft gripper gripping cylindrical objects with a diameter $D =$ 40 mm, 60 mm, 80 mm and 100 mm. The dashed and solid curves represent the tests in slow and fast actuation configurations, respectively. A total of eight actuation cycles were performed on each cylindrical object, and the last three cycles are shown in \textbf{c} and \textbf{d}. \textbf{f,} \textit{D}-pressure calibrations in slow and fast actuation configurations. Each data point represents the average value of the equilibrium pressure in the last five actuation cycles. For each actuation cycle, the equilibrium pressure is averaged over a 5 s period starting at 10 s after the actuation. \textbf{g, h,} Pressure measurements in fast actuation configuration and \textit{D}-pressure calibrations with different waiting time after the actuation. Results from the last five actuation cycles are superimposed in \textbf{g} for each gripping object. Each data point in \textbf{h} represents the average value of the gripper pressure in the last five actuation cycles. For each actuation cycle, the pressure in the gripper is averaged over a 0.05 s period starting at 0.2 s, 0.3 s, 0.5 s, 2 s, 5 s after the actuation (the opening of the valve in \textbf{b}). The error bar represents the standard deviation of the gripper pressure in the last five actuation cycles. }
    \label{FigS_actuationspeed}
\end{figure}

\begin{figure}[h]
    \centering
    \resizebox{150mm}{!}{\includegraphics{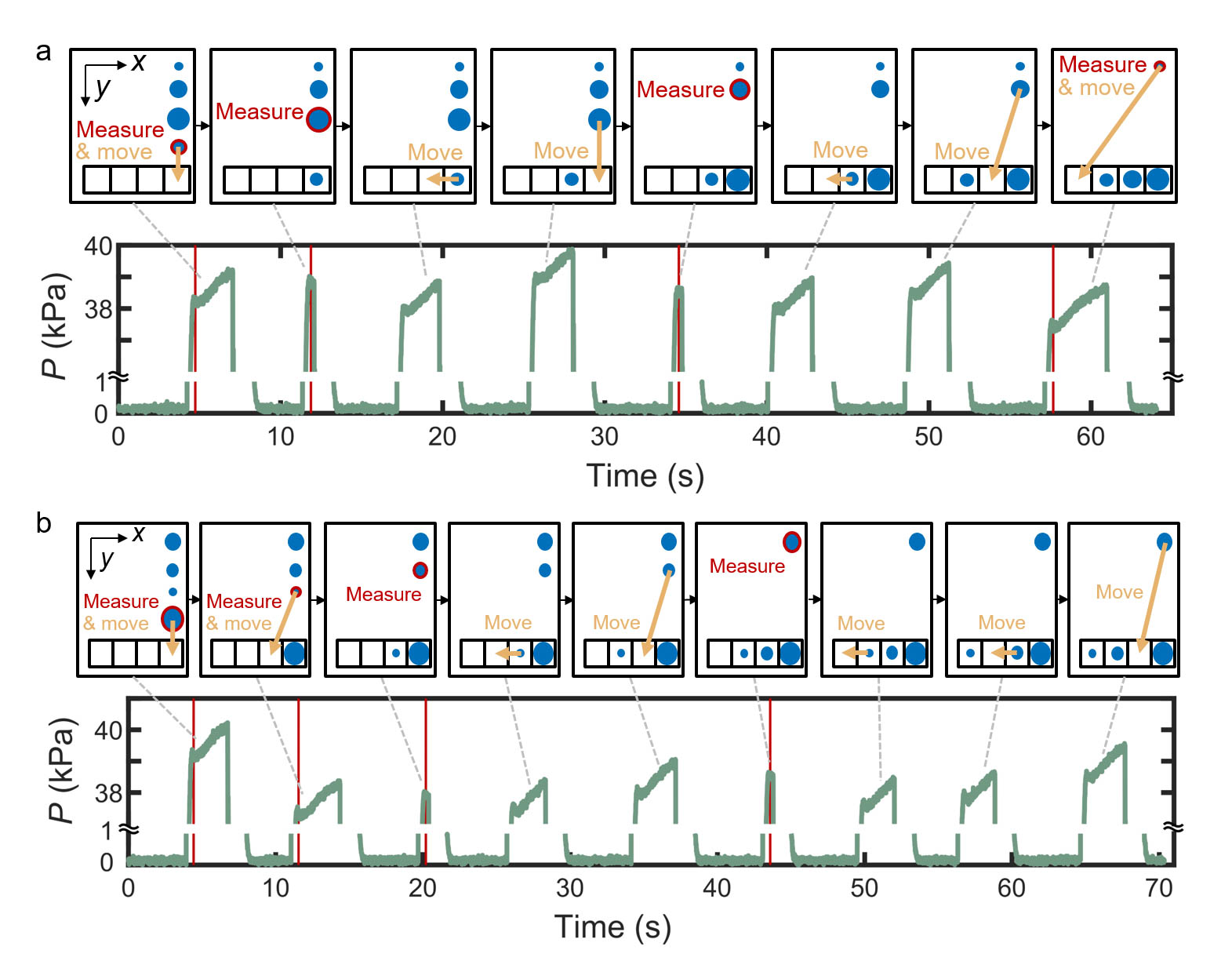}}
    \caption{\textbf{Sorting experiments with a faster speed.} The input order of cylindrical objects is 60 mm, 100 mm, 80 mm, 40 mm in \textbf{a} (Demo 7 in Supplementary Video 4) and 100 mm, 40 mm, 60 mm, 80 mm in \textbf{b} (Demo 8 in Supplementary Video 4). For each measurement event, the equilibrium pressure was averaged over a 0.05 s period starting at 0.5 s after the actuation, as indicated by the red bands in the plot. The four pressure feedback measurements (average $\pm$ standard deviation) in \textbf{a} are $38.21 \pm 0.11$ kPa, $38.84 \pm 0.07$ kPa, $38.59 \pm 0.07$ kPa, $37.44 \pm 0.07$ kPa, respectively. The four pressure feedback measurements (average $\pm$ standard deviation) in \textbf{b} are $39.17 \pm 0.11$ kPa, $37.33 \pm 0.06$ kPa, $37.92 \pm 0.07$ kPa, $38.52 \pm 0.06$ kPa, respectively. Note that the initial tank pressure was set at 74 kPa here to ensure successful gripping of each object, especially the one with a diameter of 40 mm. Note that the coordinates of the robotic arm were intentionally not saved in these experiments in order to obtain a high data frequency ($\sim 87$ Hz) that allows the fast sensing over a 0.05s period.}
    \label{FigS_sortingfaster}
\end{figure}

\begin{figure}[h]
    \centering
    \resizebox{160mm}{!}{\includegraphics{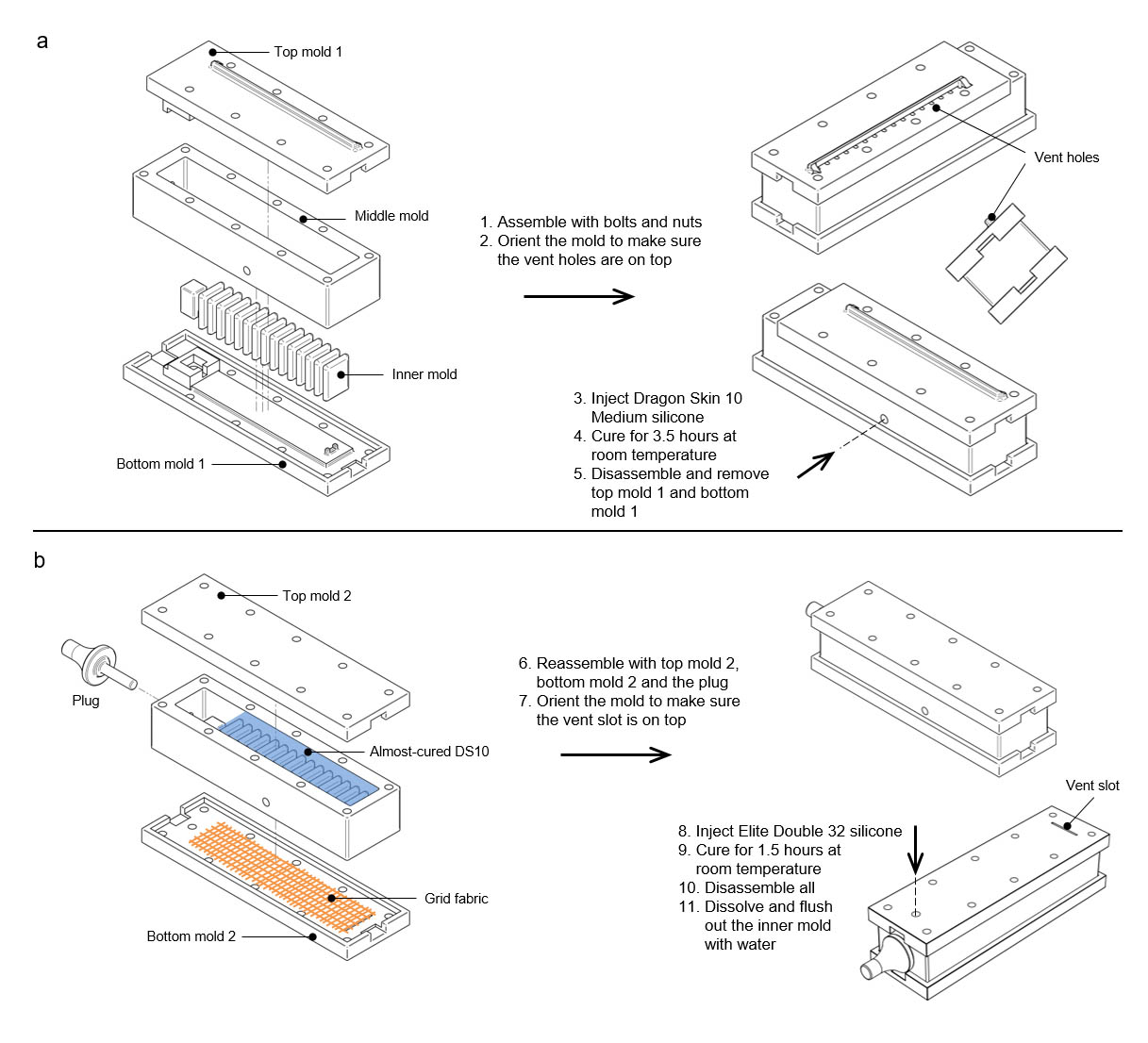}}
    \caption{\textbf{Fabrication of the PneuNet actuator.} \textbf{a,} Molding the extensible layer with Dragon Skin 10 Medium silicone. \textbf{b,} Molding the inextensible layer with Elite Double 32 silicone. The inner mold was printed with BVOH filament on a Fused Filament Fabrication 3D printer (Ultimaker 3) and the other molds were printed with VeroClear on a PolyJet 3D printer (Eden260VS, Stratasys). After the curing and disassembly, the inner mold was flushed for 24 hours using a water pump in parallel with a tunable flow resistor for venting.}
    \label{FigS_actuator_fabrication}
\end{figure}

\begin{figure}[h]
    \centering
    \resizebox{160mm}{!}{\includegraphics{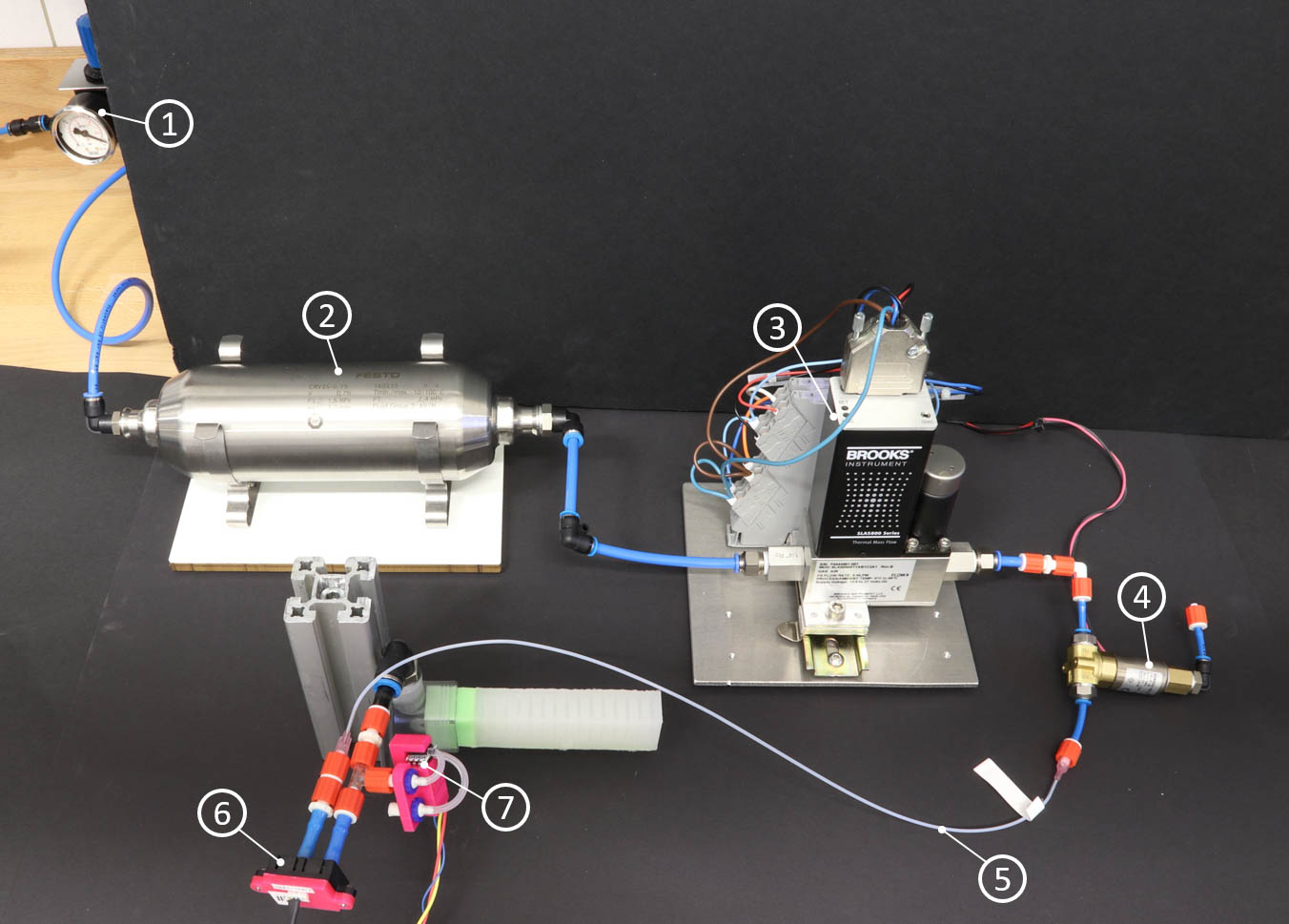}}
    \caption{\textbf{Experimental setup for flow control with pressure measurement.}  1) Wall-mounted pressure regulator. 2) 0.75 L air tank. 3) Mass flow controller. 4) Solenoid valve. 5) Flow resistor. 6) Bidirectional flow sensor. 7) Pressure sensor.}
    \label{FigS_setup_flowcontrol}
\end{figure}

\begin{figure}[h]
    \centering
    \resizebox{160mm}{!}{\includegraphics{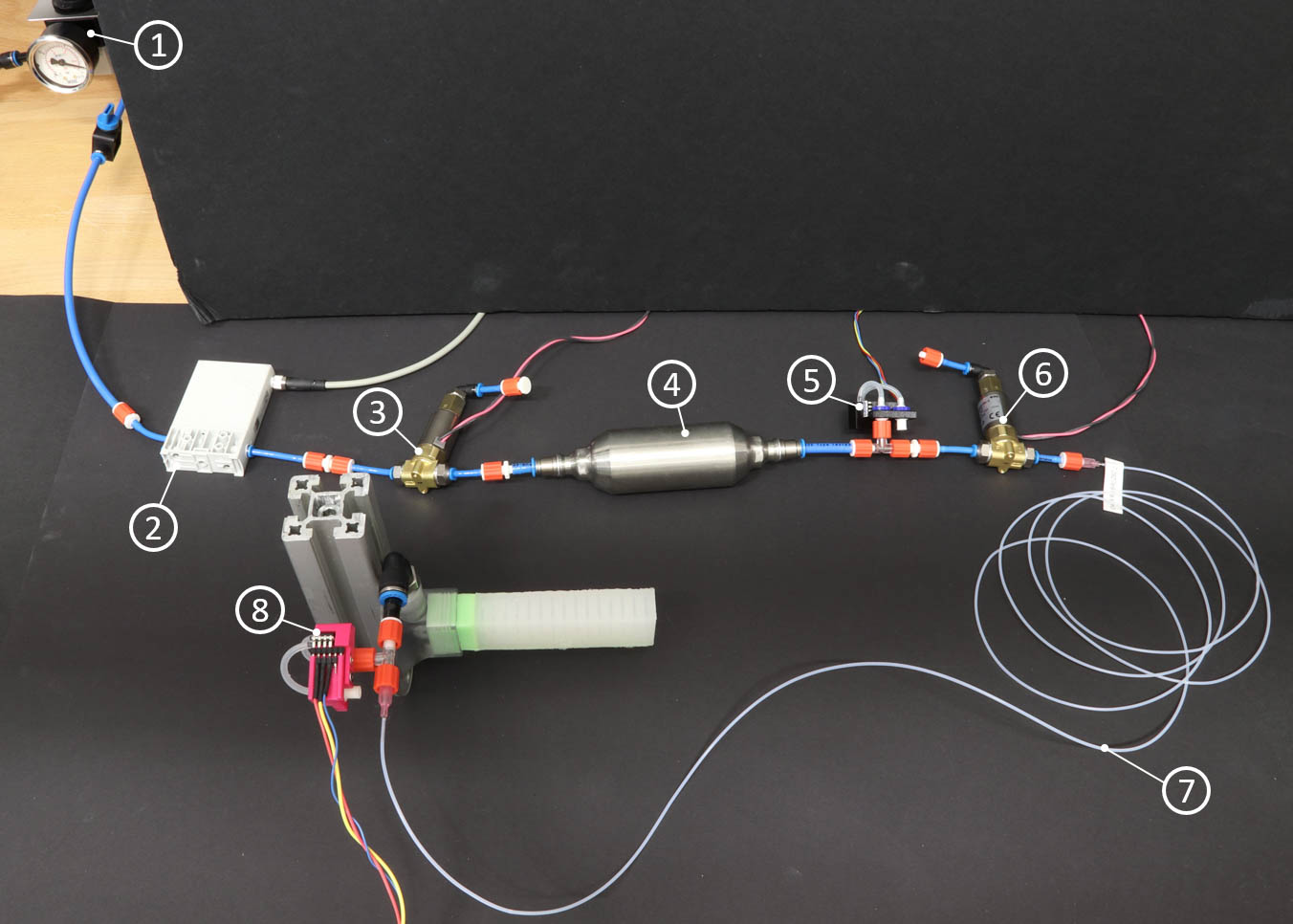}}
    \caption{\textbf{Experimental setup for pressure control with pressure measurement.} 1) Wall-mounted pressure regulator. 2) Proportional pressure regulator. 3) Solenoid valve. 4) 0.1 L air tank. 5) Pressure sensor. 6) Solenoid valve. 7) Flow resistor. 8) Pressure sensor.}
    \label{FigS_setup_masscontrol}
\end{figure}

\begin{figure}[h]
    \centering
    \resizebox{160mm}{!}{\includegraphics{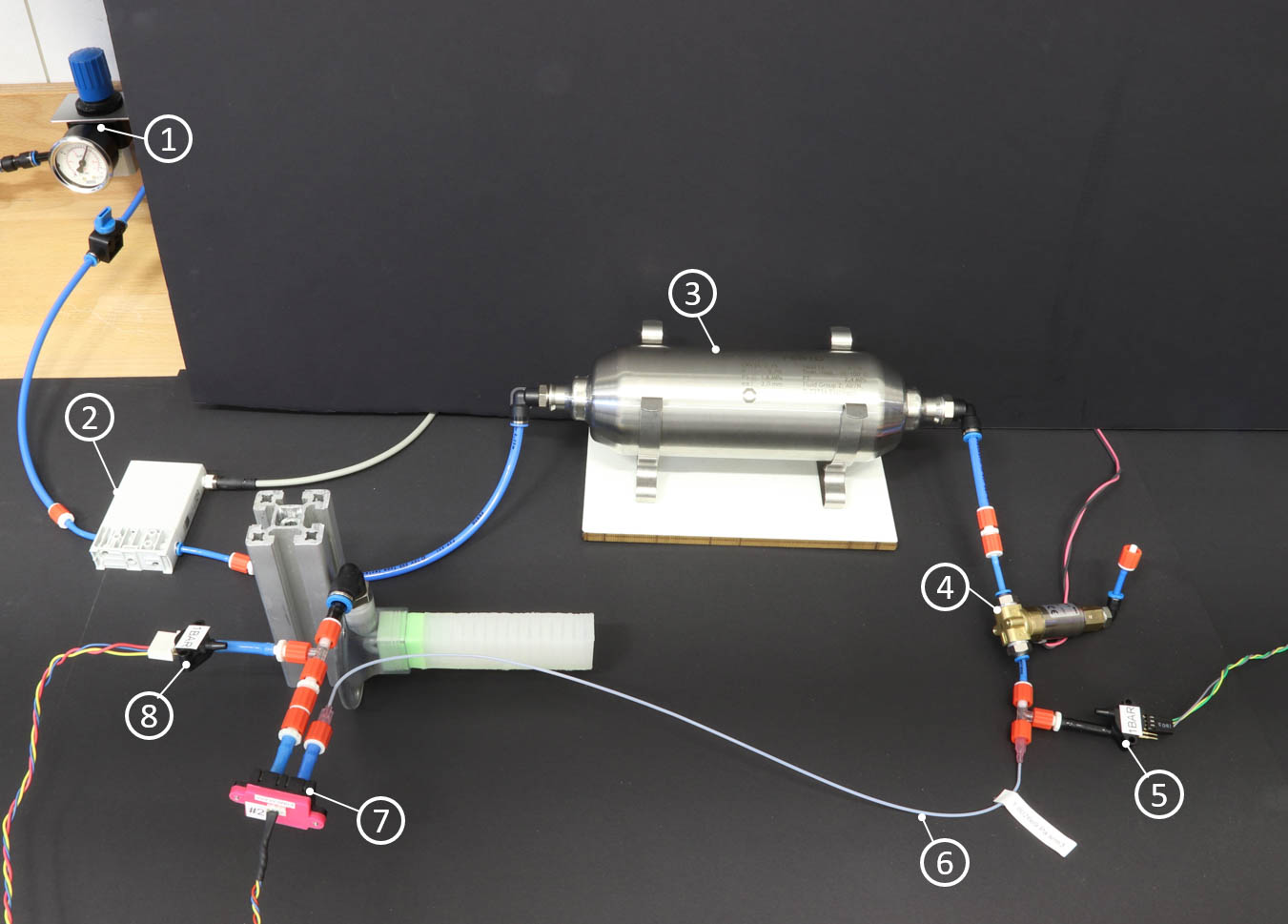}}
    \caption{\textbf{Experimental setup for pressure control with flow measurement.} 1) Wall-mounted pressure regulator. 2) Proportional pressure regulator. 3) 0.75 L air tank. 4) Solenoid valve. 5) Pressure sensor. 6) Flow resistor. 7) Bidirectional flow sensor. 8) Pressure sensor.}
    \label{FigS_setup_pressurecontrol}
\end{figure}

\begin{figure}[h]
    \centering
    \resizebox{160mm}{!}{\includegraphics{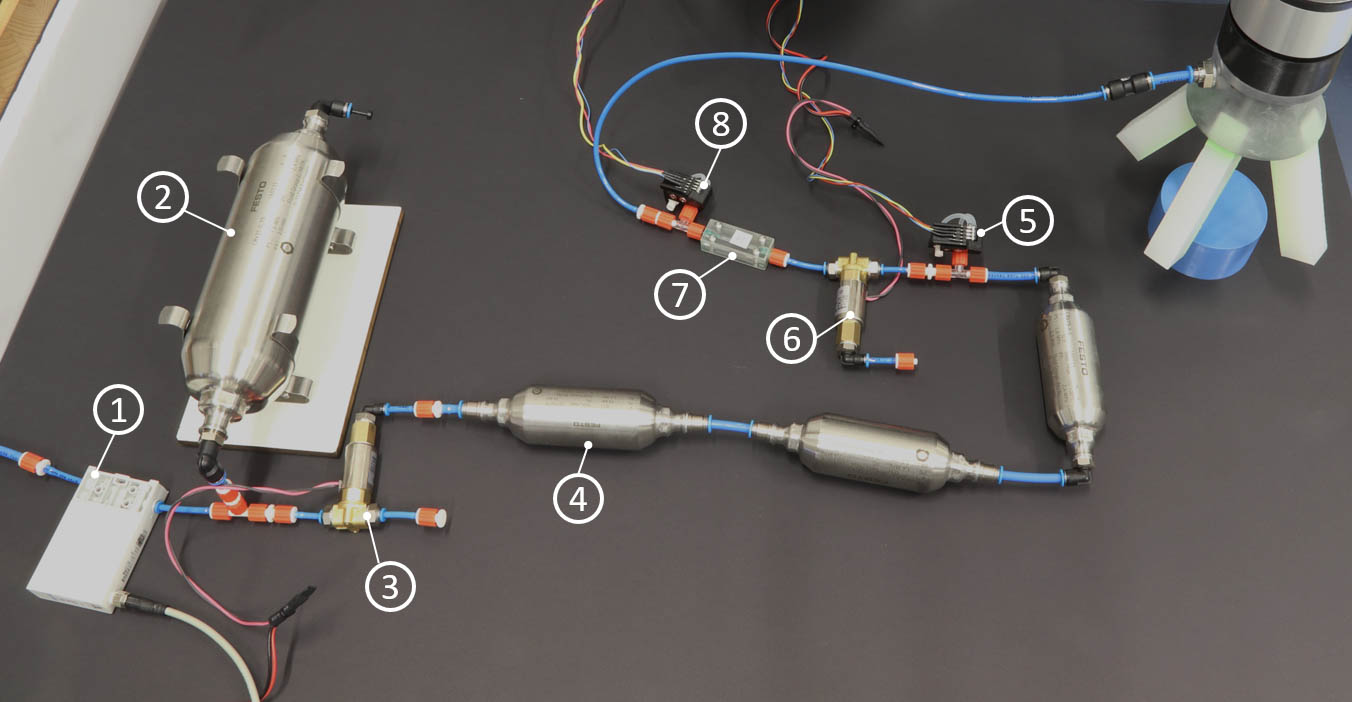}}
    \caption{\textbf{Experimental setup for size sensing with soft gripper.} 1) Proportional pressure regulator. 2) 0.75 L air tank. 3) Solenoid valve. 4) Three 0.1 L air tank in series. 5) Pressure sensor. 6) Solenoid valve. 7) Flow resistor. 8) Pressure sensor.}
    \label{FigS_setup_gripper}
\end{figure}

\begin{figure}[h]
    \centering
    \resizebox{160mm}{!}{\includegraphics{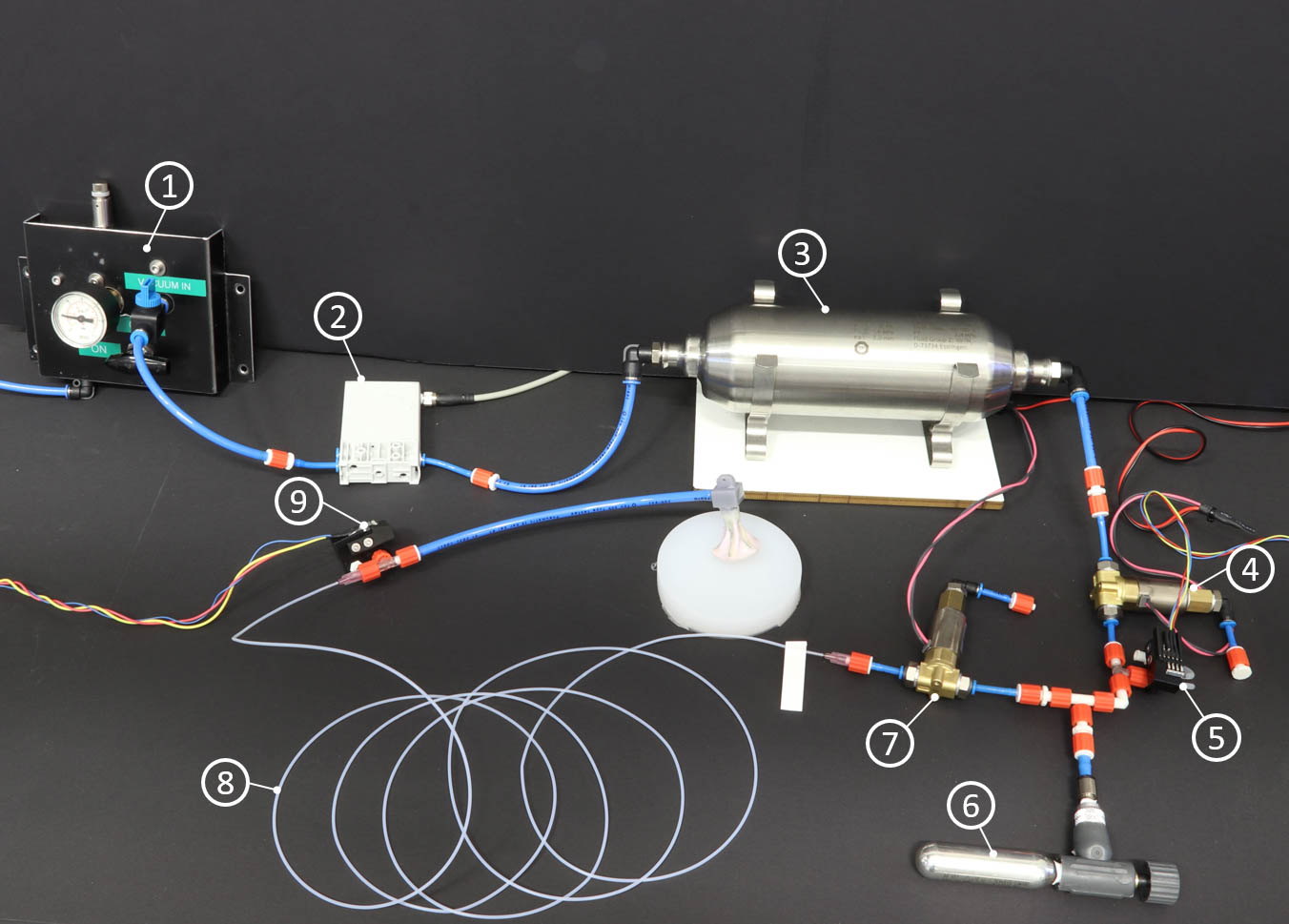}}
    \caption{\textbf{Experimental setup for the surface stiffness sensing with the suction-based soft gripper.} 1) Vacuum pressure regulator. 2) Proportional pressure regulator. 3) 0.75 L air tank. 4) Solenoid valve. 5) Pressure sensor. 6) 15 ml air tank. 7) Solenoid valve. 8) Flow resistor. 9) Pressure sensor. }
    \label{FigS_setup_suctioncup_softness}
\end{figure}

\begin{figure}[h]
    \centering
    \resizebox{160mm}{!}{\includegraphics{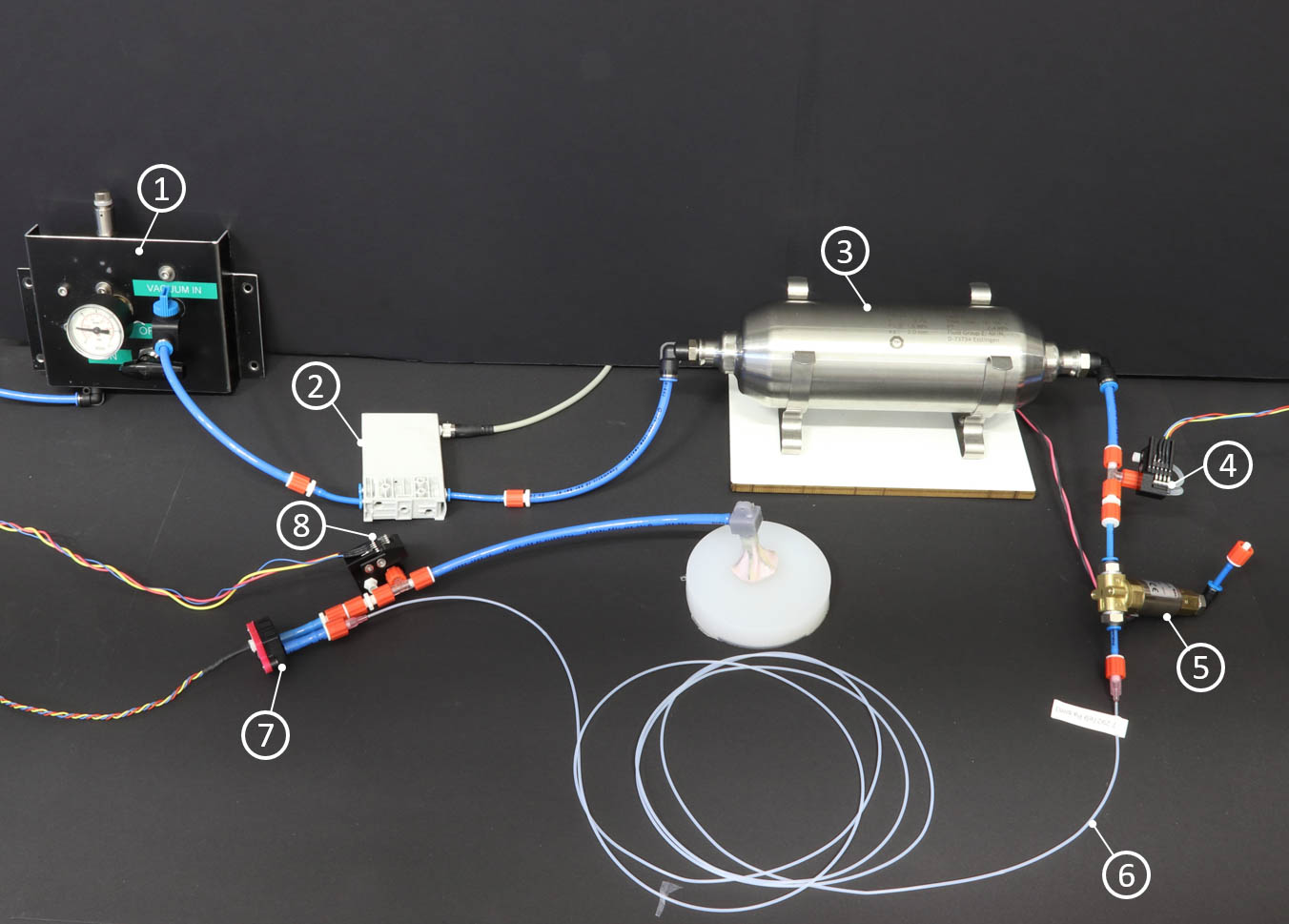}}
    \caption{\textbf{Experimental setup for the characterization of pressure-volume response of the suction-based soft gripper.} 1) Vacuum pressure regulator. 2) Proportional pressure regulator. 3) 0.75 L air tank. 4) Pressure sensor. 5) Solenoid valve. 6) Flow resistor. 7) Bidirectional flow sensor. 8) Pressure sensor.}
    \label{FigS_setup_suctioncup_PV}
\end{figure}

\begin{figure}[h]
    \centering
    \resizebox{160mm}{!}{\includegraphics{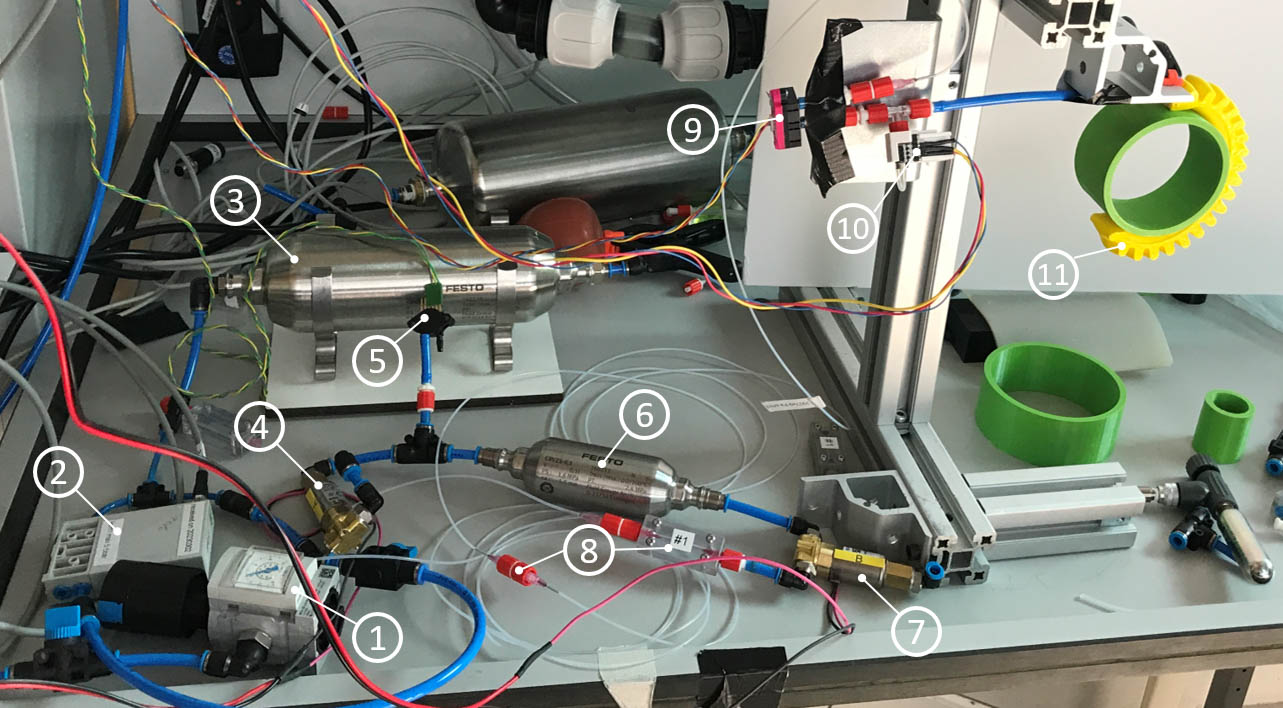}}
    \caption{\textbf{Experimental setup for the size sensing with TPU bending actuator.} 1) Wall-mounted pressure regulator. 2) Proportional pressure regulator. 3) 0.75 L air tank. 4) Solenoid valve (VDW250-5G-1-01F-Q, SMC). 5) Pressure sensor. 6) 0.1 L air tank. 7) Solenoid valve (VDW250-5G-1-01F-Q, SMC). 8) Flow resistor. 9) Bidirectional flow sensor. 10) Pressure sensor. 11) TPU bending actuator.}
    \label{FigS_setup_TPU_actuator}
\end{figure}

\begin{figure}[h]
    \centering
    \resizebox{160mm}{!}{\includegraphics{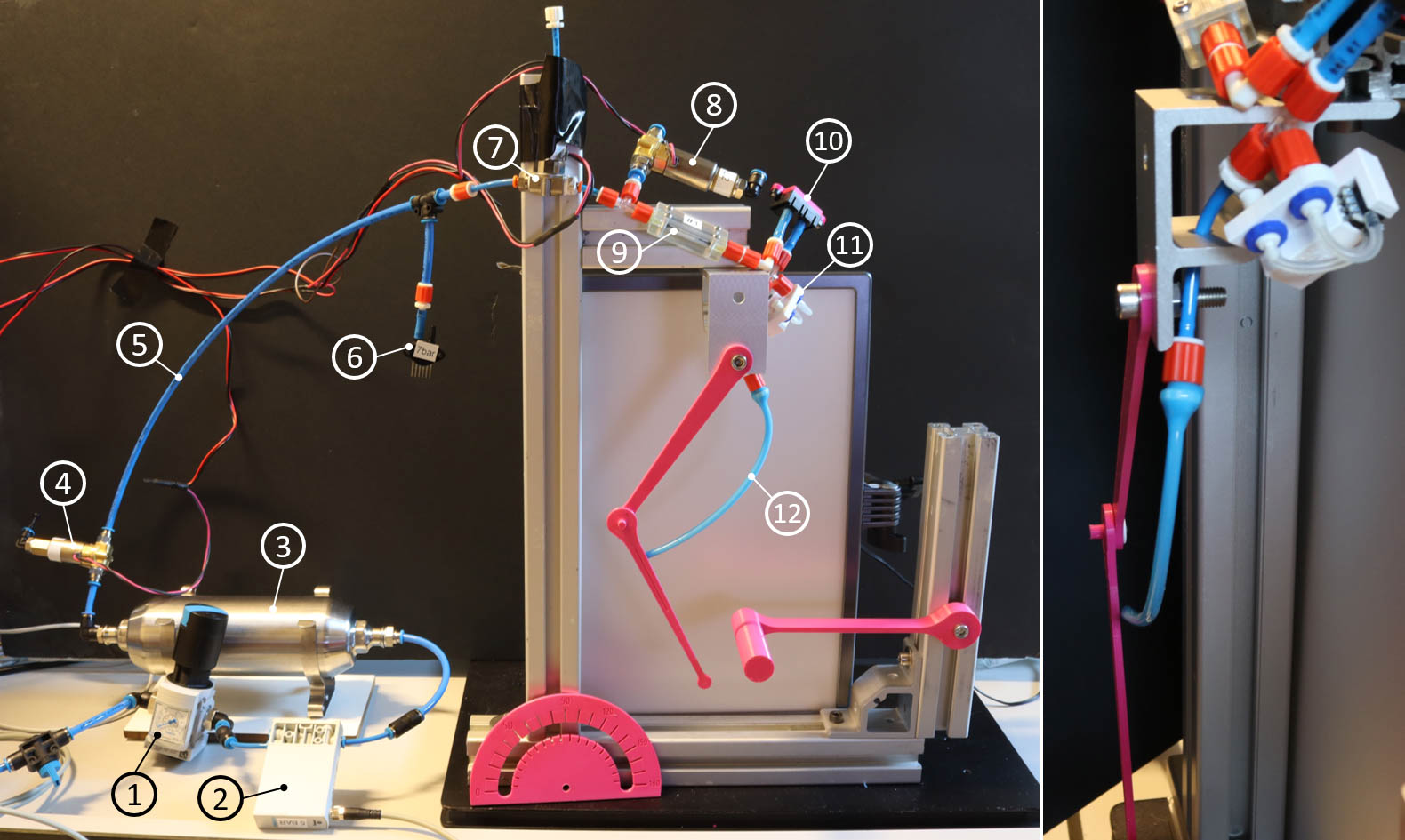}}
    \caption{\textbf{Experimental setup for the angular displacement sensing with filament actuator.} 1) Wall-mounted pressure regulator. 2) Proportional pressure regulator. 3) 0.75 L air tank. 4) Solenoid valve. 5) Pressurized air pipe. 6) Pressure sensor. 7) Solenoid valve. 8) Solenoid valve. 9) Flow resistor. 10) Flow sensor. 11) Pressure sensor. 12) Filament actuator. Note that this setup contains more than the minimum number of required components for the sensing strategy to work. Solenoid valves 7) and 8) were arranged in such way to avoid leakage through the valves at high actuation pressure required by the filament actuator. Flow sensor 10) was used to obtain the full pressure-volume responses of both the pressurized air pipe and filament actuator and the flow resistor 9) was used to restrict the flow within the range of the flow sensor.}
    \label{FigS_setup_muscledemo_filactuator}
\end{figure}

\begin{figure}[h]
    \centering
    \resizebox{160mm}{!}{\includegraphics{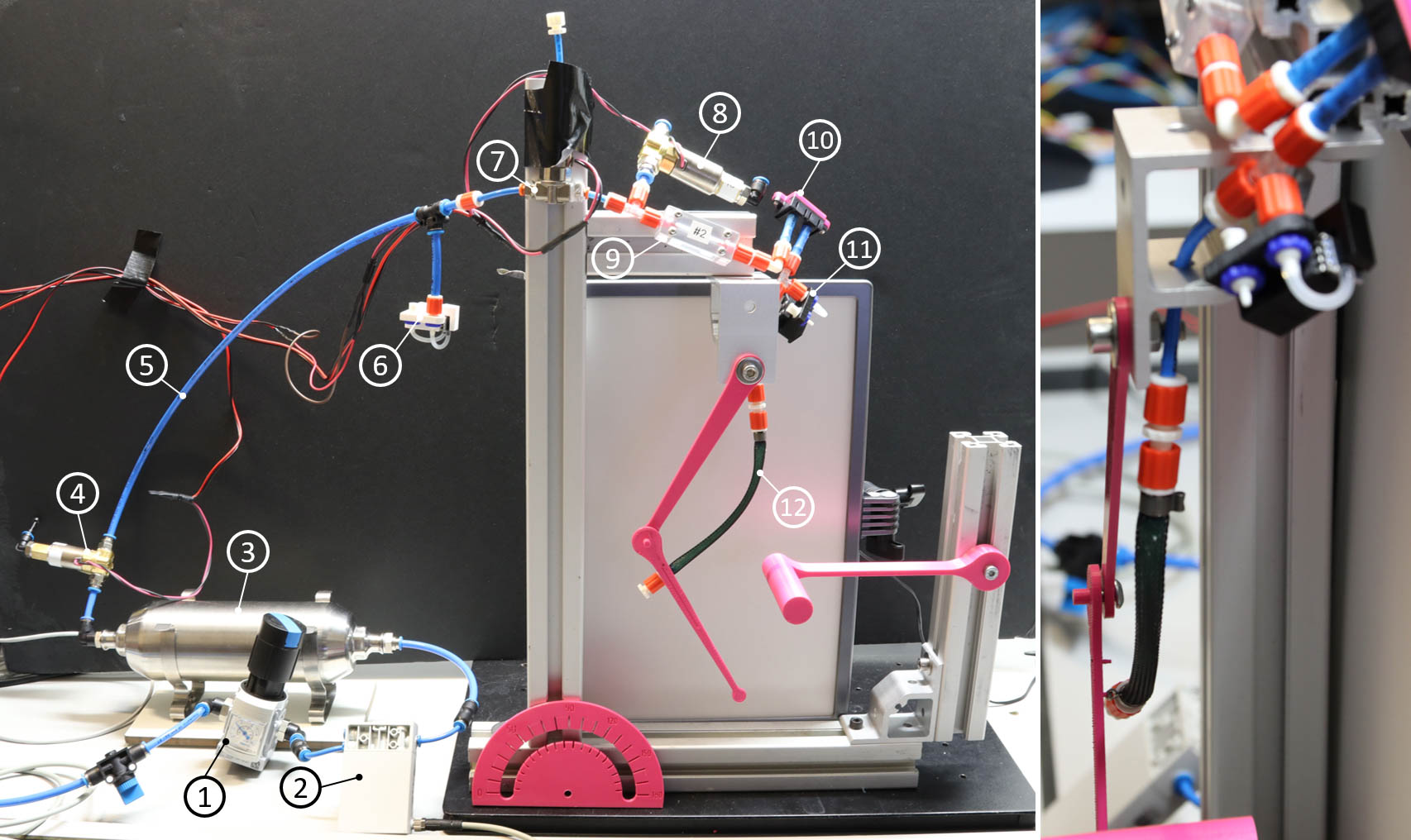}}
    \caption{\textbf{Experimental setup for the angular displacement sensing with McKibben actuator.} 1) Wall-mounted pressure regulator. 2) Proportional pressure regulator. 3) 0.75 L air tank. 4) Solenoid valve. 5) Pressurized air pipe. 6) Pressure sensor. 7) Solenoid valve. 8) Solenoid valve. 9) Flow resistor. 10) Flow sensor. 11) Pressure sensor. 12) McKibben actuator. The setup was kept the same as that for the filament actuator in Fig.~\ref{FigS_setup_muscledemo_filactuator} except for the pressure sensors 6) and 11) which were selected to suit the pressure range of the McKibben actuator.}
    \label{FigS_setup_muscledemo_McKibben}
\end{figure}

\begin{figure}[h]
    \centering
    \resizebox{160mm}{!}{\includegraphics{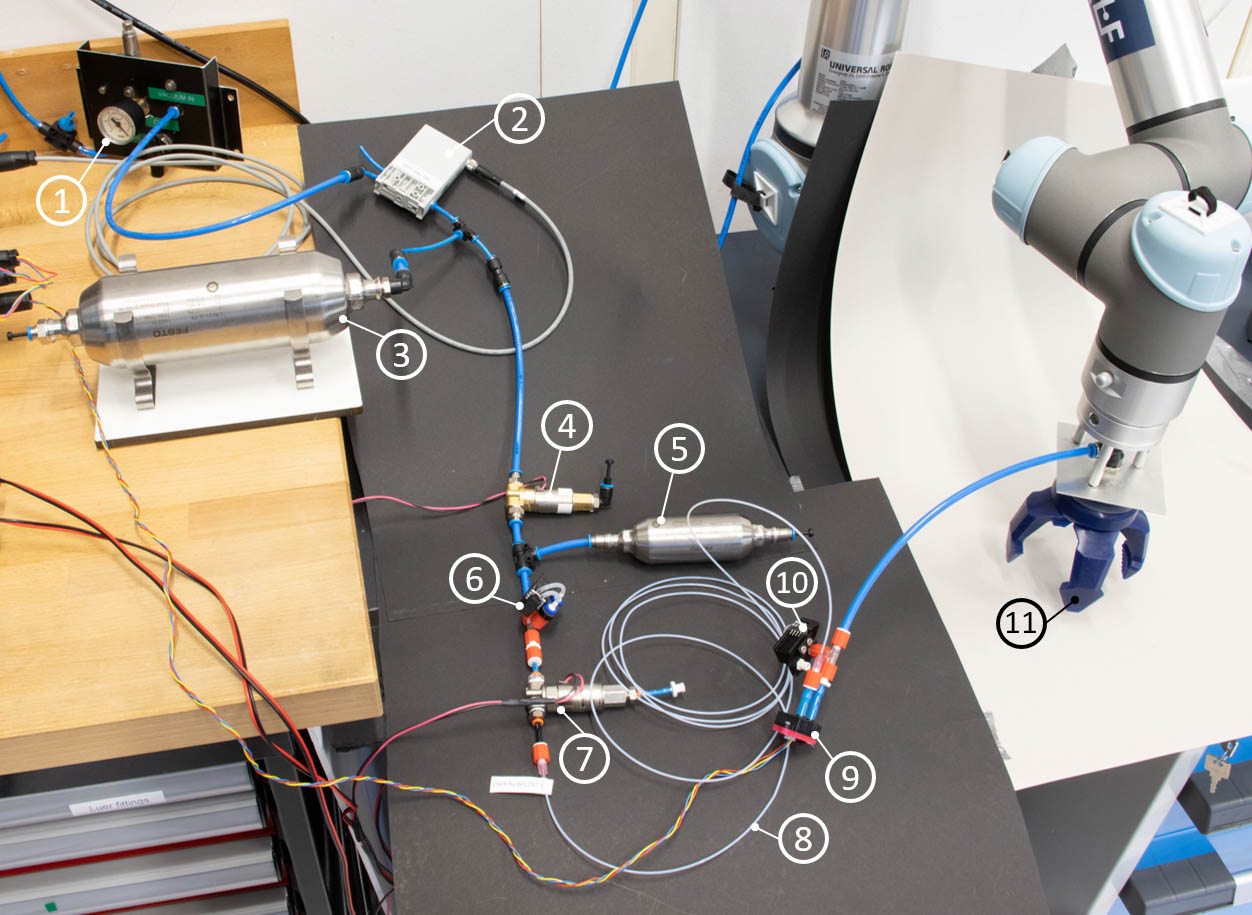}}
    \caption{\textbf{Experimental setup for the size sensing with vacuum-powered commercial soft gripper.} 1) Vacuum pressure regulator. 2) Proportional pressure regulator. 3) 0.75 L air tank. 4) Solenoid valve. 5) 0.1 L air tank. 6) Pressure sensor. 7) Solenoid valve. 8) Flow resistor. 9) Flow sensor. 10) Pressure sensor. 11) Commercial soft gripper. Note that the flow sensor 9) was used to obtain the full pressure-volume responses of both the air tank and the soft gripper and the flow resistor 8) was used to restrict the flow within the range of the flow sensor.}
    \label{FigS_setup_gripper_piab}
\end{figure}

\begin{figure}[h]
    \centering
    \resizebox{160mm}{!}{\includegraphics{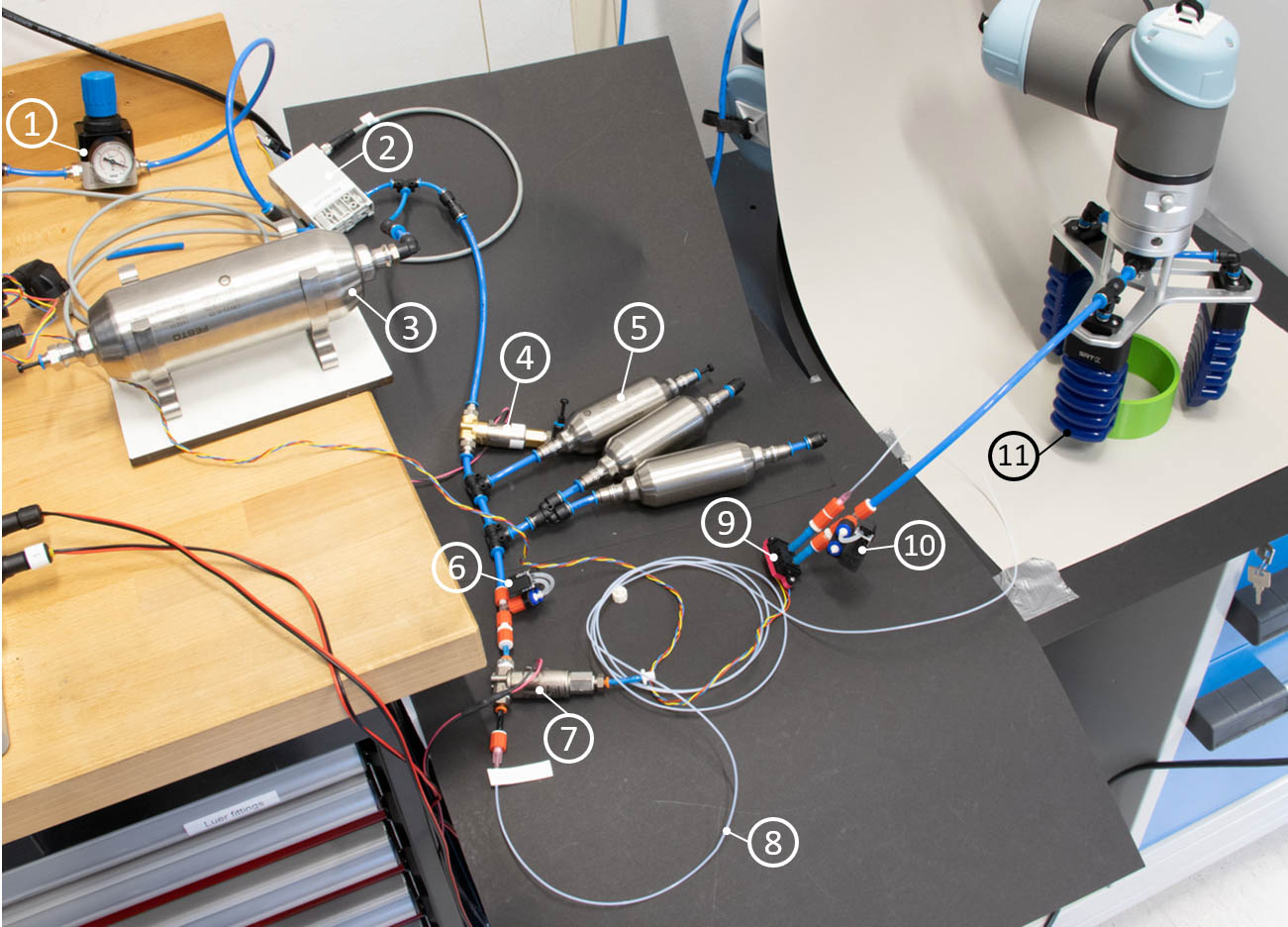}}
    \caption{\textbf{Experimental setup for the size sensing with commercial soft PneuNet gripper.} 1) Wall-mounted pressure regulator. 2) Proportional pressure regulator. 3) 0.75 L air tank. 4) Solenoid valve. 5) Three 0.1 L air tanks. 6) Pressure sensor. 7) Solenoid valve. 8) Flow resistor. 9) Flow sensor. 10) Pressure sensor. 11) Commercial soft PneuNet gripper. Note that the flow sensor 9) was used to obtain the full pressure-volume responses of both the air tank and the soft gripper and the flow resistor 8) was used to restrict the flow within the range of the flow sensor.}
    \label{FigS_setup_gripper_SRT}
\end{figure}

\begin{figure}[h]
    \centering
    \resizebox{160mm}{!}{\includegraphics{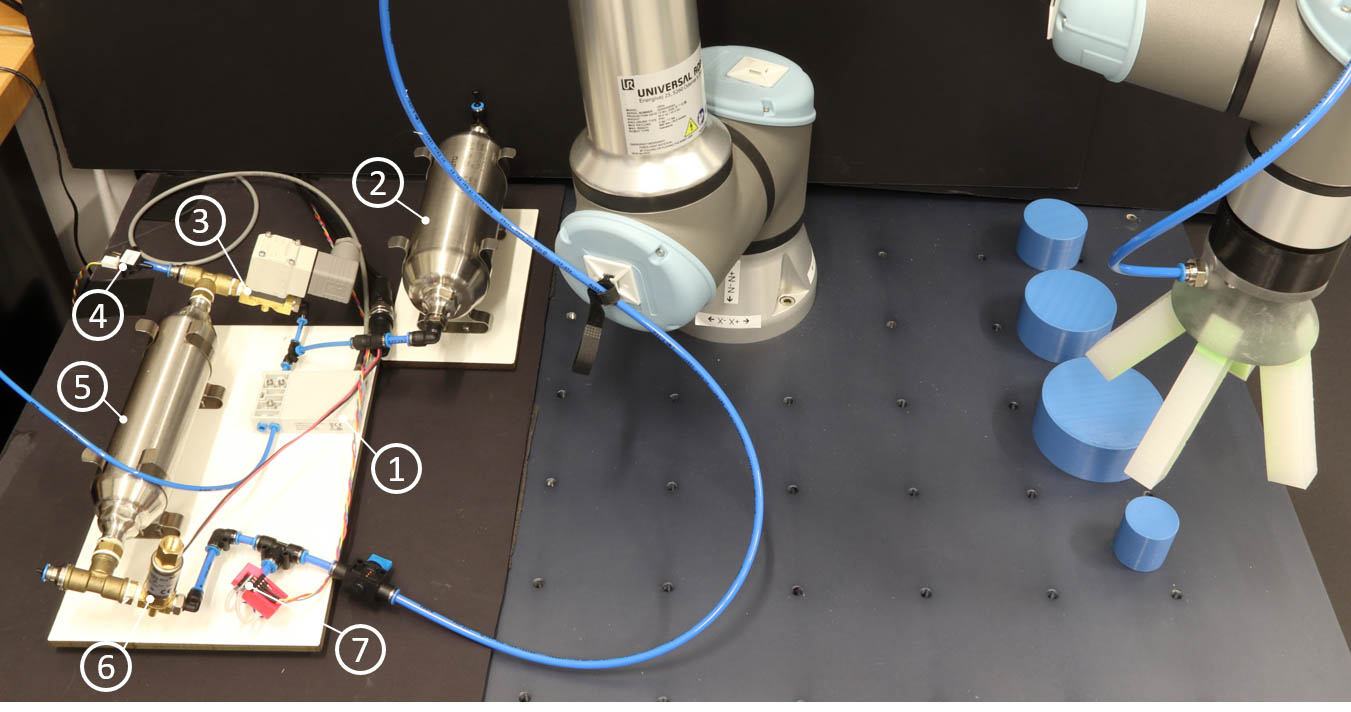}}
    \caption{\textbf{Experimental setup for the sorting experiment.} 1) Proportional pressure regulator. 2) 0.75 L air tank. 3) Solenoid valve. 4) Pressure sensor. 5) 0.4 L air tank. 6) Solenoid valve. 7) Pressure sensor.}
    \label{FigS_setup_sorting}
\end{figure}

\begin{figure}[h]
    \centering
    \resizebox{160mm}{!}{\includegraphics{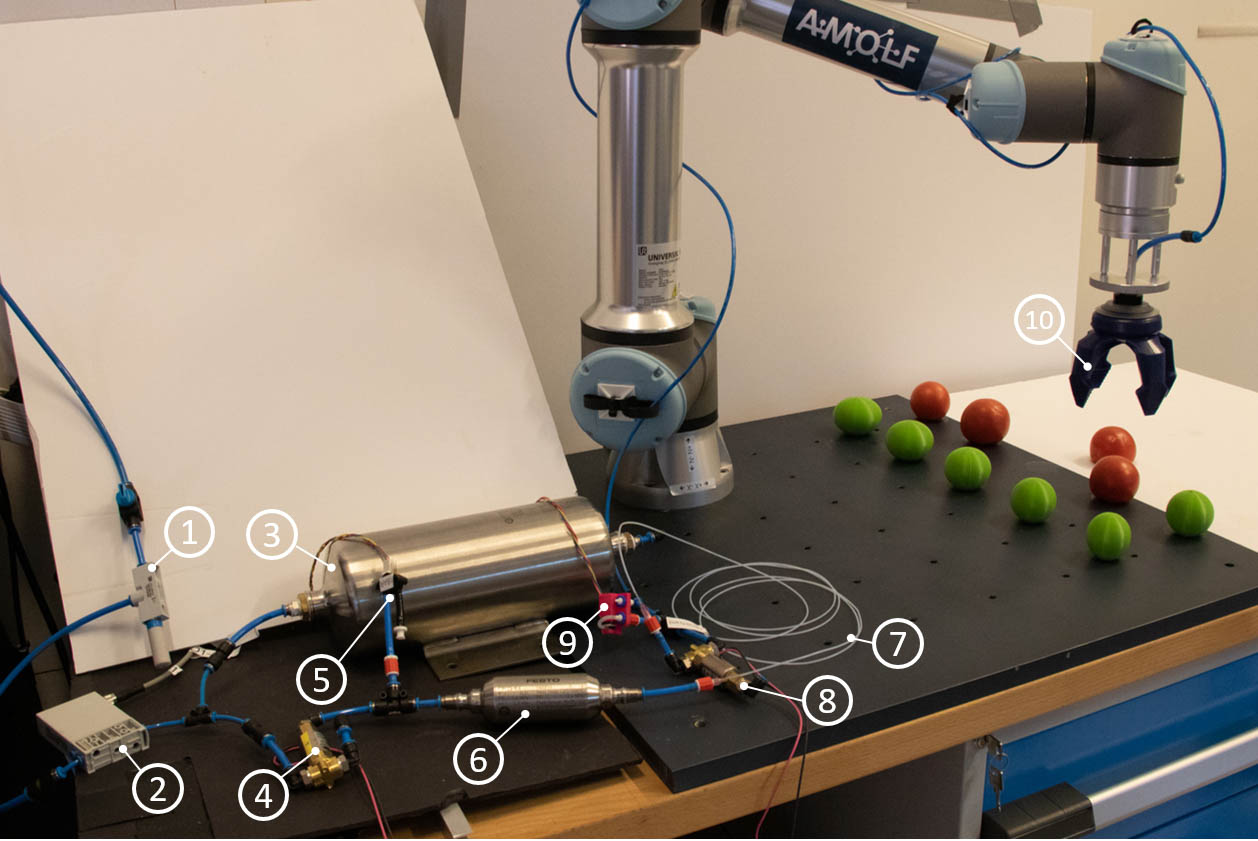}}
    \caption{\textbf{Experimental setup for the tomato ripeness sensing experiment.} 1) Vacuum generator (FVN-05-H-T3-PQ2-VQ2-RO1, Festo) connected to a 7 bar compressed air source. 2) Proportional pressure regulator. 3) 2 L air tank. 4) Solenoid valve (VDW250-5G-1-01F-Q, SMC). 5) Pressure sensor. 6) 0.1 L air tank. 7) flow resistor. 8) Solenoid valve (VDW250-5G-1-01F-Q, SMC). 9) Pressure sensor. 10) Commercial soft gripper.}
    \label{FigS_setup_ripeness}
\end{figure}

\clearpage

\end{document}